\documentclass[12pt]{article}
\usepackage{graphicx}
\usepackage[dvipsnames]{xcolor}
\usepackage[section]{placeins}
\usepackage[top=1in, left=1in, right=1in, bottom=1in]{geometry}
\usepackage[parfill]{parskip}	% Activate to begin paragraphs with an empty line rather than an indent
\usepackage{caption}
\usepackage{subcaption}
\usepackage[export]{adjustbox}
\usepackage{amssymb}
\usepackage{bbm}
\usepackage{bm}
\usepackage{dsfont}
\usepackage{amssymb}
\usepackage{amsmath}
\usepackage{amsfonts}
\newcommand{\indicator}[1]{\mathds{1}_{ {#1} }}

\usepackage{enumitem}
\usepackage{setspace}
\usepackage{booktabs}
\usepackage[compact]{titlesec}

\usepackage{lscape}

\usepackage{accents}

\usepackage[nocitation]{apacite}
\usepackage{natbib}
\bibpunct{(}{)}{;}{a}{}{,}
\usepackage{hyperref}
\usepackage{titling}
\usepackage{pifont}
\usepackage[capposition=top]{floatrow}
\newcommand{\subtitle}[1]{%
  \posttitle{%
    \par\end{center}
    \begin{center}\large#1\end{center}
    \vskip0.5em}%
}
\usepackage{amsthm}
\usepackage{amsmath} %maths
\usepackage{multirow}
\usepackage{multicol}
\usepackage{mathrsfs}

  {\end{smallmatrix}\right)}

\newcommand\independent{\protect\mathpalette{\protect\independenT}{\perp}}
\def\independenT#1#2{\mathrel{\rlap{$#1#2$}\mkern2mu{#1#2}}}

\usepackage{mathrsfs}

\usepackage{color}

\usepackage{array}
\newcolumntype{C}[1]{>{\centering\arraybackslash}m{#1}}

\begin{document}
\pagestyle{plain}

\newtheoremstyle{mystyle}% name
{\topsep}% Space above
{\topsep}% Space below
{\it}% Body font
{}% Indent amount
{\bf}% Theorem head font
{.}%Punctuation after theorem head
{.5em}%Space after theorem head
{}% theorem head spec
\theoremstyle{mystyle}
\newtheorem{assumptionex}{Assumption}
\newenvironment{assumption}
  {\pushQED{\qed}\renewcommand{\qedsymbol}{}\assumptionex}
  {\popQED\endassumptionex}
\newtheorem{assumptionexp}{Assumption}
\newenvironment{assumptionp}
  {\pushQED{\qed}\renewcommand{\qedsymbol}{}\assumptionexp}
  {\popQED\endassumptionexp}
\renewcommand{\theassumptionexp}{\arabic{assumptionexp}$'$}

\newtheorem{assumptionexpp}{Assumption}
\newenvironment{assumptionpp}
  {\pushQED{\qed}\renewcommand{\qedsymbol}{}\assumptionexpp}
  {\popQED\endassumptionexpp}
\renewcommand{\theassumptionexpp}{\arabic{assumptionexpp}$''$}

\newtheorem{assumptionexppp}{Assumption}
\newenvironment{assumptionppp}
  {\pushQED{\qed}\renewcommand{\qedsymbol}{}\assumptionexppp}
  {\popQED\endassumptionexppp}
\renewcommand{\theassumptionexppp}{\arabic{assumptionexppp}$'''$}

\renewcommand{\arraystretch}{1.3}

\newcommand\carl[1]{\cmnt{#1}{Carl}}
\newcommand\ambarish[1]{\cmnt{#1}{Ambarish}}
\newcommand\jose[1]{\cmnt{#1}{Jose}}

\newcommand{\argmin}{\mathop{\mathrm{argmin}}}
\makeatletter
\newcommand{\grande}{\bBigg@{2.25}}
\newcommand{\enorme}{\bBigg@{5}}

\newcommand{\blind}{0}

\newcommand{\tit}{Targeted Quality Measurement of Health Care Providers}

\if0\blind

%{\title{\tit\thanks{For comments and suggestions, we thank...}}
{\title{\tit
\thanks{The authors thank Ambarish Chattopadhyay, Michael Chernew, Richard Frank, Avi Feller, Luke Keele, Bijan Niknam, and Alan Zaslavsky for helpful comments and conversations.
This study used the linked SEER-Medicare database. The interpretation and reporting of these data are the sole responsibility of the authors. The authors acknowledge the efforts of the National Cancer Institute; the Office of Research, Development and Information, CMS; Information Management Services (IMS), Inc.; and the Surveillance, Epidemiology, and End Results (SEER) Program tumor registries in the creation of the SEER-Medicare database.\\
\indent \indent The collection of cancer incidence data used in this study was supported by the California Department of Public Health as part of the statewide cancer reporting program mandated by California Health and Safety Code Section 103885; the National Cancer Institute's Surveillance, Epidemiology and End Results Program under contract HHSN261201000140C awarded to the Cancer Prevention Institute of California, contract HHSN261201000035C awarded to the University of Southern California, and contract HHSN261201000034C awarded to the Public Health Institute; and the Centers for Disease Control and Prevention's National Program of Cancer Registries, under agreement \# U58DP003862-01 awarded to the California Department of Public Health. The ideas and opinions expressed herein are those of the author(s) and endorsement by the State of California Department of Public Health, the National Cancer Institute, and the Centers for Disease Control and Prevention or their Contractors and Subcontractors is not intended nor should be inferred. The authors acknowledge the efforts of the National Cancer Institute; the Office of Research, Development and Information, CMS; Information Management Services (IMS), Inc.; and the Surveillance, Epidemiology, and End Results (SEER) Program tumor registries in the creation of the SEER-Medicare database.\\
\indent \indent \emph{Funding and support}: This research was supported by Arnold Ventures. 
Jos\'{e} Zubizarreta also acknowledges support by award ME-2019C1-16172 from the Patient-Centered Outcomes Research Institute (PCORI).
The views presented here are those of the authors and not necessarily those of Arnold Ventures, PCORI, its directors, officers, or staff.
}\vspace*{.3in}}
\author{
Yige Li
\thanks{Department of Health Care Policy, Harvard University.
email: \url{yige_li@hcp.med.harvard.edu}.},
\and 
Nancy L. Keating
\thanks{Department of Health Care Policy, Harvard University; Division of General Internal Medicine, Brigham and Women's Hospital. email: \url{keating@hcp.med.harvard.edu}.},
\and
Mary Beth Landrum
\thanks{Department of Health Care Policy, Harvard University. 
email: \url{landrum@hcp.med.harvard.edu}.},
\and
Jos\'{e} R. Zubizarreta\thanks{Departments of Health Care Policy, Biostatistics, and Statistics, Harvard University, 180 Longwood Avenue, Office 307-D, Boston, MA 02115; email: \url{zubizarreta@hcp.med.harvard.edu}.}
}

\date{} 

\maketitle
}\fi

\if1\blind
\title{\tit}
\date{} 
\maketitle
\fi

\begin{abstract}

\indent \hspace{.5cm} Assessing the quality of cancer care administered by US health providers poses numerous challenges {due to} meaningful heterogeneity in patient populations.
Patients undergoing oncology treatment exhibit substantial variation in disease presentation among other crucial characteristics.
In this paper, we present a framework for institutional quality measurement that {addresses} this patient heterogeneity. 
Our framework follows recent advancements in health outcomes research, conceptualizing quality measurement as a causal inference problem. 
This approach enables us to use flexible covariate profiles to target specific patient populations of interest{.} 
%To our knowledge, this application of covariate profiles is novel in the quality measurement literature.
{We} use different clinically relevant covariate profiles and evaluate methods for {case-mix} adjustments.
These adjustments integrate weighting and regression modeling approaches in a progressive manner in order to reduce model extrapolation and allow for provider effect modification.
We evaluate these methods in an extensive simulation study{, comparing their performance in terms of point estimates and estimated rankings. We} highlight the practical utility of weighting methods that {can generate stable weights when covariate overlap is limited and} alert investigators when case-mix adjustments are infeasible without some form of extrapolation that goes beyond the support of the {observed} data.
In our study of cancer-care outcomes, we assess the performance of oncology practices for different profiles that correspond to important types of patients who may receive cancer care. 
We describe how the methods examined may be particularly important for high-stakes quality measurement, such as public reporting or performance-based payments. 
These methods have the potential to help inform individual patient health care decisions and contribute to progress toward more personalized quality measurement. 
%They may also be important for individual patients seeking practices that provide high-quality care to patients like them. 
%This approach applies to other settings besides health care, including business and education, where instead of cancer practices, we have companies and schools.
\end{abstract}

%\vspace*{.3in}

\begin{center}
\end{center}
\clearpage
\doublespacing

%\singlespacing
%\pagebreak
%\tableofcontents
%\pagebreak
%\doublespacing
\section{Introduction}
\label{sec_introduction}

In recent decades, a primary focus of statistical and public policy research has been the development of reliable and valid measures to evaluate the performance of institutions and organizations \citep{normand2016league}. 
These metrics are crucial for promoting accountability and improving the performance of these entities. 
Moreover, in sectors such as education and healthcare, such measures enable informed individual decision-making around vital life choices through their ability to measure the quality of schools and hospitals, often in league tables \citep{goldstein1996league}. 
However, quality measurement is a complex task that demands careful consideration of several factors \citep{ash2013statistical}.

Firstly, the methods must be fair and account for systematic differences in the case-mixes of the populations the institutions serve. 
Additionally, the methods ought to be transparent and promote an open understanding of their case-mix adjustments by stakeholders, such as patients and physicians.  
Finally, there must be accountability for the great heterogeneity of the individuals in the population thinly spread across numerous organizations in the data.   
Finally, careful consideration must be given to the heterogeneity and sparsity of the data, where highly diverse individuals are thinly distributed across numerous organizations.
For example, in this paper we study the performance of cancer care providers in the US.
In our data, there are 600 cancer practices, with a median of 112 patients, each characterized by 20 covariates.
In nearly 50\% of the practices, there is at least one (constant) zero count in a relevant covariate dimension, further complicating case-mix adjustments because there is no variation on that dimension.
Appropriately handling sparse, high-dimensional data is key to reliable measurement. 

Often, these performance or quality assessments are conducted using parametric outcome regression models.
For instance, Medicare's Hospital Compare evaluates hospital performance across the U.S. based on a random-effects logit model, which includes random hospital indicators and covariates that account for patient risk factors \citep{krumholz2011administrative}. 
However, this approach does not clearly reveal the exact form of the case-mix adjustments applied to the data. 
In other words, it is unclear how individual patient-level data are weighted to generate aggregate quality estimates.
Also, the characteristics of the target population to which inferences apply are not manifest and may be distorted by the implied weighting of the regression procedure.
In particular, if some of the covariates are sparse, such as racial minority indicators across health care providers, the implied weights of such procedures can result in a weighted population that omits the minority (see Section 7.1 of \citeauthor{chattopadhyay2023implied}, 2023, for a general discussion of this and related matters in linear regression adjustments).
Finally, it is not clear to what extent the quality assessments for a given health care provider are based on data from the provider in question as opposed to other health providers that see different patients, involving some form of extrapolation based on a model that can be misspecified.
More generally, as discussed by \cite{george2017mortality}, this approach can produce rankings of provider organizations (e.g., hospitals) that are not checkable against data.

In this study, we build upon the contributions of \cite{silber2014template}, \cite{longford2020performance}, and \cite{keele2023hospital}, viewing quality measurement through the lens of causal inference (see also \citealp{varewyck2014shrinkage}; \citealp{daignault2017doubly}; \citealp{tang2020construct}).
This is important because it helps to clearly define the estimands and to state and evaluate the assumptions needed for their identification.
Also, it helps to decouple the estimands from the estimation methods and to envision alternative approaches that can be used for case-mix adjustment (for example, approaches that combine new weighting and regression methods).
Importantly, this helps to recognize when these methods produce quality assessments based on extrapolations for which there is little or no data for each health provider to support them.
Unlike some hierarchical models \citep{jones2011identification}, this framework also helps to delimit the design and analysis stages of a study and to use methods for case-mix adjustment that do not require information about the outcomes \citep{rubin2008objective}. 
Finally, it provides a more coherent framework for discussing hospital or practice effects, which is something that is routinely done in quality measurement, although often without a formal justification.

A central concept in this paper is the evaluation of healthcare provided based on a \emph{covariate profile} of a target population. 
This profile characterizes the population of study interest in terms of summary statistics of the joint distribution of its observed variables (see also \citeauthor{chattopadhyay2020balancing}, 2020, and \citeauthor{han2024privacy}, 2024).
As we illustrate, this profile may correspond to the entire patient population across all health care providers, the population of a specific hospital or practice, or a target population with a similar covariate profile to that of an \emph{individual} patient.
The latter may be of particular interest to patients selecting the best providers for treating patients like them.
This covariate profile helps to recognize from the outset that patients are heterogeneous and that some health care providers may specialize in treating different types of patients.
In addition, it allows us to understand the problem of sparsity in terms of a profile (in this sense, we provide a characterization of the distribution of the covariates relative to the profile) and to use methods that ground the analysis as much as possible on the available data and identify when the results are an extrapolation beyond the support of the data of a given health provider.

Utilizing this profile and borrowing from the causal inference literature, we evaluate weighting methods for quality measurement which emphasize transparency, flexibility, and robustness in the case-mix adjustments.
First, in terms of transparency, these weighting methods are based on constrained convex optimization problems that are computationally tractable and warn the investigator when some form of extrapolation beyond the support of the observed data is needed to perform the adjustments.
Also, it is straightforward to conduct covariate balance checks, which are intuitive for stakeholders.
Second, in terms of flexibility, the profiles can represent varied target populations.
Finally, in terms of robustness, the methods examined conduct the case-mix adjustments in a layered way, adjusting for as many covariates as possible by weighting, without extrapolating, and leaving the rest of the covariates to modeling, borrowing strength across providers from the parametric form of the outcome model.
As a result, these methods are less sensitive to model misspecification than traditional regression approaches used in quality measurement.

Starting from a target population of interest, our goal is to make the case-mix covariate adjustments transparent, and, to the greatest extent possible, ground the comparisons of practices or hospitals on the available data. 
This paper is organized as follows.
In Section \ref{sec_data}, we delineate our data, comprised of SEER cancer registry and Medicare administrative records.
In Section \ref{sec_framework}, we describe a framework from causal inference where target estimands, identification assumptions, and covariate profiles are key.
In Section \ref{sec_methods}, we outline methods that use balancing weights and regression modeling to perform the adjustments.
In Section \ref{sec_simulation_study}, we evaluate the performance of the methods in a simulation study.
In Section \ref{sec_case_study}, we present the results from our case study.
Finally, in Section \ref{sec_discussion}, we provide a discussion and remarks.

%%%%%%%%%%%%%%%%%%%%
%%%%%%%%%%%%%%%%%%%%
%%%%%%%%%%%%%%%%%%%%

%%%%%%%%%%%%%%%%%%%%%%%%%%%%%%%%%%%%%%%%%%%
%%%%%%%%%%%%%%%%%%%%%%%%%%%%%%%%%%%%%%%%%%%
%%%%%%%%%%%%%%%%%%%%%%%%%%%%%%%%%%%%%%%%%%%
\section{SEER cancer registry and Medicare administrative data}
\label{sec_data}

We use cancer registry data from the Surveillance, Epidemiology, and End Results (SEER) Program for individuals diagnosed with cancer in 2011-2013 linked to administrative data from Medicare for 2010-2014. 
The SEER-18 data include demographic and clinical information for cancer patients from areas covering 28\% of the total US population. Medicare data include inpatient, outpatient, provider carrier, durable medical equipment, and hospice/home health files. 
Our study population includes individuals aged 65 years and older who were enrolled in parts A and B of fee-for-service Medicare through at least 6 months after their cancer diagnosis.  
Patients were attributed to medical oncology practices based on outpatient evaluation and management visits to a medical oncology practice \citep{gondi2021assessment}. 

The main outcome is survival one year after the first visit to the practice (we also studied survival through six months after diagnosis and through 18 months diagnosis). The covariates we used for case-mix adjustment are: age (continuous), sex, race/ethnicity (White, Black, other/unknown), marital status (unmarried, married, unknown), census-tract median household income (quartiles), Census-level proportion of residents without a high school education (quartiles), Klabunde modification of the Charlson Comorbidity Index (0, 1, 2, $\geq$3), cancer type (breast, colorectal, lung, ovary, pancreas, prostate), and cancer stage (1, 2, 3, 4, missing). We excluded practices with less than 30 patients.

%%%%%%%%%%%%%%%%%%%%
%%%%%%%%%%%%%%%%%%%%
%\vspace{-0.2cm}

\section{Framework}
\label{sec_framework}

%%%%%%%%%%%%%%%%%%%%%%%%%%%%%%%%%%%%%%%%%%%
%%%%%%%%%%%%%%%%%%%%%%%%%%%%%%%%%%%%%%%%%%%
\subsection{Notation}
\label{sec_notation}

Let $i = 1, ..., n$ index the patients across different practices and write $\boldsymbol{X}_{i} \in \mathbb{R}^K$ for their vector of $K$ observed covariates.
Put $\boldsymbol{x}^*$ for the target covariate profile.
This profile can characterize various target populations, such as the entire patient population across all practices or the subpopulation of patients similar to an individual patient of interest.
The profile $\boldsymbol{x}^*$ summarizes the target population in terms of its covariate means, or more, more generally in terms of transformations $\widetilde{\boldsymbol{X}}_{i}$ of its covariates, such as higher moments. 

Now, let $p = 1, ..., P$ index the practices and let $\indicator{p,i}$ stand for the practice assignment indicator of patient $i$.
Write $Y_i(p)$ for the potential outcome \citep{neyman1923application, rubin1974estimating} under practice assignment $\indicator{p,i}$, $p = 1, ..., P$.
Denote $Y_i$ for their observed outcome, $Y_i = \sum_{p = 1}^{P} \indicator{p,i} Y_i(p)$.
Finally, define $\boldsymbol{U}$ as the vector unobserved covariates.
See Table \ref{tab_notation} in the Online Supplementary Materials for a summary of this notation.

%%%%%%%%%%%%%%%%%%%%%%%%%%%%%%%%%%%%%%%%%%%
%%%%%%%%%%%%%%%%%%%%%%%%%%%%%%%%%%%%%%%%%%%
\subsection{Estimands}
\label{sec_estimands}

%But what about the patient who is looking to identify how a patient like him/her will do in the practice?
%
%And it is not only patients that may want to compare one practice to another. A payer may want to do this to determine which practice to contract with. Or as you say later, one practice may want to compare itself with another.
%
%It seems that this section might be better off ignoring the ÒperspectiveÓ (patient vs. system, which I think is incomplete) and instead focusing on the purpose of the quality measurement:
%(1)	Comparing one practice with one or more others and (2) ranking practices for the purpose of tiering or performance-based payments.
%---
%As I got into the simulation discussion, I see that you come back to ÒindividualÓ vs. ÒsystemÓ. I think that is okÉ but maybe you could still use the (1) and (2) above, but just label them Òindividual practice-level comparisonsÓ and Òsystem-level comparisonsÓ?

We are fundamentally interested in estimating the mean potential outcome under health care provider $p$ for patients with covariate values equal to the profile $\boldsymbol{x}^*$; that is,
\begin{equation}
\label{eq_msm}
\mu_p(\boldsymbol{x}^*) := \mathrm{E} [Y_i(p) | \boldsymbol{X}_i = \boldsymbol{x}^*].
\end{equation}
More generally, considering that a covariate profile may summarize a distribution, we can write $\mu_p(\boldsymbol{x}^*) = \int \mathrm{E} [Y_i(p) | \boldsymbol{X}_i = \boldsymbol{x}]f^*(\boldsymbol{x})d\boldsymbol{x}$, where $f^*(\boldsymbol{x})$ is the density of such a distribution.
We take two perspectives: the first one, perhaps of an individual patient, which for the purpose of selecting a care provider is primarily interested in the average effect of receiving care in practice $p'$ instead of practice $p''$, $\mu_{p'}(\boldsymbol{x}^*) - \mu_{p''}(\boldsymbol{x}^*)$, where $\boldsymbol{x}^*$ are the observed covariates of the target patient; and the second one, perhaps of the system of health providers, which for performance payment purposes is mainly interested in the ranking of all the practices for a given target population of patients described by
$\boldsymbol{x}^*$, $\mu_{(1)}(\boldsymbol{x}^*), \mu_{(2)}(\boldsymbol{x}^*), ..., \mu_{(P)}(\boldsymbol{x}^*)$, where the subscript $(p)$ enclosed in parentheses denotes the $p$-th order statistics of the mean potential outcome function conditional on covariates.
In other words, the first perspective is primarily interested in contrasts whereas the second one is mainly interested in rankings.
Arguably, there is a third important perspective which is interested in the two previous quantities.
For instance, this can be the perspective of a given practice, which is interested the average treatment effect for benchmarking itself to another practice, and the ranking for understanding its relative standing in the wider concert of all practices.
In all these cases, the core objective is reduced to identifying and estimating the conditional potential outcome $\mu_{p}(\boldsymbol{x}^*)$, grounded in the assumptions outlined below.

%%%%%%%%%%%%%%%%%%%%%%%%%%%%%%%%%%%%%%%%%%%
%%%%%%%%%%%%%%%%%%%%%%%%%%%%%%%%%%%%%%%%%%%
\subsection{Assumptions}
\label{sec_assumptions}

Our notation implies the Stable Unit Treatment Value Assumption (SUTVA; \citealp{rubin1980randomization}) which in our setting states that there is no interference between patients and that there are no additional versions of the ``treatments'' beyond receiving care in one of the $P$ practices. 
Again, our problem boils down to estimating the conditional mean potential outcome function $\mu_p(\boldsymbol{x}^*)$ for all $p = 1, ..., P$.
For identification, or \emph{causal} quality measurement, we also require the unconfoundedness assumption, $\indicator{p,i} \independent Y_i(p) \mid X_i$, and the positivity assumption, $0 < \Pr(\indicator{p,i} = 1 | \boldsymbol{X}_i = \boldsymbol{x}^*) < 1$, for all $\boldsymbol{x}^*$,  $p = 1, ..., P$ \citep{imbens2000role}. Hence, $\mu_p(\boldsymbol{x}^*) := \mathrm{E} [Y(p) | \boldsymbol{X} = \boldsymbol{x}^*] = \mathrm{E} [Y | \boldsymbol{X} = \boldsymbol{x}^*,p]$.
The unconfoundedness assumption states that observed covariates determine selection into practices, whereas the positivity assumption states that for every value of the target $\boldsymbol{x}^*$ there is a positive probability of receiving care in all the practices.
To our knowledge, these assumptions are implicit in traditional forms of quality measurement (e.g., \citealp{krumholz2011administrative}).

%%%%%%%%%%%%%%%%%%%%%%%%%%%%%%%%%%%%%%%%%%%
%%%%%%%%%%%%%%%%%%%%%%%%%%%%%%%%%%%%%%%%%%%
\subsection{Profiles}
\label{sec_targets}

The covariate profiles $\boldsymbol{x}^*$ can represent different target populations.
Examples of target populations of interest are the population of patients across all practices, the population of patients in the top 10\% best performing practices, and in the bottom 10\% worst performing practices.
In other settings, one may be interested in the population of patients that go to large or small practices, that receive care in a given geographic area, or that correspond to a group of policy interest, such as vulnerable patients as defined by race ethnicity, socioeconomic status, or clinical complexity.
For personalization, the profile $\boldsymbol{x}^*$ can also represent a target population with similar covariate values as those of a specific individual who may be seeking care.
Throughout, we consider the profile to be fixed.

Based on $\boldsymbol{x}^*$, we distinguish between three types of covariates: those whose sample covariate values contain the target profile $\boldsymbol{x}^*$ inside its convex hull and for which case-mix adjustments toward $\boldsymbol{x}^*$ can be conducted by interpolation (i.e., by a convex combination of the observations in a given practice); those practices for which $x^*$ is outside its convex hull and for which the adjustments can only be performed by extrapolation (i.e., by an affine transformation of the observations in the practice); and those covariates for which $\boldsymbol{x}^*$ is outside its convex hull and for which at least one practice has a covariate with constant value different from that in $\boldsymbol{x}^*$ for all its patients.
These are covariates that cannot be weighted toward the target $\boldsymbol{x}^*$, even when extrapolation is allowed.
These covariates require borrowing information across the practices.
We call these null-case covariates $\boldsymbol{X}_{\textrm{null}}$ and denote their complement by $\boldsymbol{X}_{\textrm{null}}^{\mathsf{c}}$, such that $\boldsymbol{X}_{\textrm{null}} \cup \boldsymbol{X}_{\textrm{null}}^{\mathsf{c}} = \boldsymbol{X}$.
In other words, null-case covariates take a constant value different to the corresponding values of $\boldsymbol{x}^*$ for all patients in a practice.
In our analysis, we focus on null-case covariates alongside the first type of covariates which can be adjusted for by interpolation.

%%%%%%%%%%%%%%%%%%%%
%%%%%%%%%%%%%%%%%%%%
%%%%%%%%%%%%%%%%%%%%

\section{Methods for layered case-mix adjustments}
\label{sec_methods}

We consider imputation estimators of the form 
\begin{equation}
\label{eq_imputation}
\widehat{Y}^{\textrm{imp}}_{i: \indicator{p,i} = 1} := \frac{1}{n_p} \sum_{i: \indicator{p,i} = 1} \widehat{\mu}_{p,i}(\boldsymbol{x}^*),
\end{equation}
where $\widehat{\mu}_{p,i}(\boldsymbol{x}^*)$ represents the estimated conditional mean potential outcome for patient $i$ under practice $p$, and $n_p$ denotes the sample size for practice $p$.
In particular, we focus on linear estimators of the form
\begin{equation}
\label{eq_hajek}
\widehat{Y}^{\textrm{imp, lin}}_{i: \indicator{p,i} = 1} := \sum_{i: \indicator{p,i} = 1} \omega_i Y_i,
\end{equation}
where $\omega_i$ are suitable weights that add up to one.
Equation $(\ref{eq_hajek})$ is a H\'ajek estimator \citep{hajek1971comment}.
Many common estimators admit this representation; for example, ordinary least squares and ridge regression imputation approaches (see, e.g., \citealp{ben2018augmented, chattopadhyay2021implied}).

For the most, we consider such linear weighting estimators for their simplicity and because the weights can be computed without the outcomes (that is, as part of the design of the study, \citealp{rubin2008objective}), a valuable feature for practical and objective quality measurement.
To adjust for the null-case covariates, the estimators have to leverage information from other practices, so
\begin{equation}
\widehat{Y}^{\textrm{imp, lin}}_{i: \indicator{p,i} = 1} = \sum_{i = 1}^n \omega_i Y_i.
\end{equation}

For continuous and binary outcomes, we utilize linear and linear probability models, respectively \citep{Hellevik2009Linear}. 
One challenge in fitting these outcome models is the potential for overfitting or encountering singularity issues when attempting to account for all possible effect modifiers. For example, in outcome models of the form $\mathrm{E}[Y|\boldsymbol{X},p] =  \alpha_p+ \boldsymbol{\beta}_p^\top \boldsymbol{X}$, the coefficient vector $\boldsymbol{\beta}_p$ may not be identifiable for some practices if there are null-case covariates. 

To address such null-case covariates, we posit the following outcome model,
\begin{equation*}
     \mathrm{E}[Y|\boldsymbol{X}, p] = \alpha_p + 
 \boldsymbol{\alpha}^\top \widetilde{\boldsymbol{X}}_{\text{null}} + \boldsymbol{\beta}_p^\top \widetilde{\boldsymbol{X}}^{\mathsf{c}}_{\text{null}}, 
\end{equation*}
where only the covariates in $\widetilde{\boldsymbol{X}}^{\mathsf{c}}_{\text{null}}$ can be effect modifiers. 
As a result, $\mu_p(\boldsymbol{x}^*)$ becomes $\alpha_p +\boldsymbol{\alpha}^\top \widetilde{\boldsymbol{x}}^*_{\text{null}}+\boldsymbol{\beta}_p^\top \widetilde{\boldsymbol{x}}^{*\mathsf{c}}_{\text{null}}$, where $\widetilde{\boldsymbol{x}}^*_{\text{null}}$ is the component of the profile for $\widetilde{\boldsymbol{X}}_{\text{null}}$, while $\widetilde{\boldsymbol{x}}^{*\mathsf{c}}_{\text{null}}$ corresponds to $\widetilde{\boldsymbol{X}}^{\mathsf{c}}_{\text{null}}$.

When estimating $\mu_p(\boldsymbol{x}^*)$ for quality measurement, three key questions are the method for adjustment, the covariate functions for adjustment, and the degree of adjustment.
We study these questions through the following methods that combine weighting and modeling approaches.
In what follows, the categorization $\{weighting, modeling\}$ is more procedural than conceptual as estimators of the form $(\ref{eq_hajek})$ can be collapsed to a single weighting or modeling approach.
We choose this procedural categorization because it maps more closely to common practice.

The methods we consider are summarized in Table \ref{tab_methods} in the Online Supplementary Materials.
They combine weighting and modeling approaches or use one of them.
First, we describe traditional modeling approaches.
Afterward, we consider approaches that start by weighting and increasingly impose stronger parametric modeling assumptions where there is less or no data.  

For modeling-only approaches, we consider different forms of linear regression imputation.
First, with fixed-effects (FE) model, we regress the outcome on the covariates $\boldsymbol{X}$ and the practice indicators $\indicator{1},...,\indicator{P}$, then predict the patient outcomes and average them within each practice.
We implement this approach with the original covariates $\boldsymbol{X}$ and transformations $\widetilde{\boldsymbol{X}}$ thereof.
The transformations that we consider aim to balance the covariance matrices of the covariates; specifically, in $\widetilde{\boldsymbol{X}}$ we consider the original covariates and the square of their principal component scores of $\boldsymbol{X}_{\text{null}}^{\mathsf c}$.
% (\textcolor{black}{see Section X for details}).
Second, we consider stratified regression (SR) of the outcome on the covariates, where we fit separate models for each practice, to then predict the patient outcomes and average them within each of the practices.
Here again, we fit the models on the original covariates $\boldsymbol{X}$ and the transformations $\widetilde{\boldsymbol{X}}$. We also consider pooled regression (PR) with interactions on $\boldsymbol{X}_{\text{null}}^{\mathsf c}$, where we regress the outcome on $\boldsymbol{X}_{\text{null}}$, $(\indicator{1},...,\indicator{P})$, and interactions of $\boldsymbol{X}_{\text{null}}^{\mathsf c}$ and $(\indicator{1},...,\indicator{P})$.

For weighting-only approaches, we consider inverse probability weighting using logistic regression (LW)
and the stable balancing weighting (SBW; \citealp{zubizarreta2015stable}).
Unlike modeling approaches to weighting, the latter balancing approach is better suited for comparing hundreds of treatments (i.e., practices in our case study).
Specifically, SBW finds weights of minimum variance that approximately balance functions of the covariates.
See \cite{wang2020minimal} and \cite{chattopadhyay2020balancing} for theoretical properties and empirical performance of the SBW in causal inference settings.
Ideally, the weights will take non-negative values and balance the covariates by interpolating within the convex hull of the available data.
Of course, this non-extrapolating form of adjustment is feasible only for practices that contain $\boldsymbol{x}^*$ inside its convex hull; otherwise, the optimization problem does not admit a solution indicating that the investigator needs to extrapolate to adjust.
In Table \ref{tab_methods} (see the Online Supplementary Materials), we denote this approach by $\{\textrm{SBW}(\mathbb{R}^+_0, \boldsymbol{X}), \cdot\}$ or $\{\textrm{SBW}(\mathbb{R}^+_0, \widetilde{\boldsymbol{X}}), \cdot\}$ depending on whether the original covariates $\boldsymbol{X}$ or their transformations $\widetilde{\boldsymbol{X}}$ are used (in our empirical studies, $\widetilde{\boldsymbol{X}}$ includes the square of principal component scores of the covariates in order to approximately balance their covariance matrix).
Since some covariates are null for some of the practices (that is, when some covariates take a constant value different to the corresponding values of $\boldsymbol{x}^*$ for all patients in some of the practices), alternatively, we combine the SBW with a modeling step as follows.

In order to control extrapolation and allow for effect modification, the idea is to balance as many covariates as possible (i.e., $\boldsymbol{X}_{\textrm{null}}^{\mathsf{c}}$) through weighting and leave the rest (i.e., $\boldsymbol{X}_{\textrm{null}}$) for modeling, borrowing strength across practices from the parametric form of the outcome model.
Here we adjust for $\boldsymbol{X}_{\textrm{null}}^{\mathsf{c}}$ using constrained SBW allowing approximate balance and then use these weights in weighted linear regression.
Specifically, in $\{\textrm{SBW}(\mathbb{R}_0^+, \boldsymbol{X}_{\textrm{null}}^{\mathsf{c}}), \textrm{WR}(\boldsymbol{X})\}$ we perform weighted regression of the outcome on $\boldsymbol{X}$ and the practice indicators, where the weights are from $\textrm{SBW}(\mathbb{R}_0^+, \boldsymbol{X}_{\textrm{null}}^{\mathsf{c}})$; then we predict the patient outcomes and average them within each of the practices.
To facilitate comparisons with other viable approaches, we include two additional methods: $\{\textrm{LW}(\mathbb{R}^+, \boldsymbol{X}_{\textrm{null}}^{\mathsf{c}}), \textrm{WR}(\boldsymbol{X})\}$ and $\{\textrm{LW}(\mathbb{R}^+, \boldsymbol{X}_{\textrm{null}}^{\mathsf{c}}), \textrm{FE}(\boldsymbol{X})\}$.
The first method uses inverse propensity scores estimated from a logistic regression model as weights in the weighted linear regression.
The second method is a bias-corrected (BC) estimator where the outcomes are regressed on covariates $\boldsymbol{X}$ and the practice indicators $\indicator{1},...,\indicator{P}$. More detailed explanations of these methods are provided in the Online Supplementary Materials. 

%%%%%%%%%%%%%%%%%%%%
%%%%%%%%%%%%%%%%%%%%
%%%%%%%%%%%%%%%%%%%%

\section{Simulation study}
\label{sec_simulation_study}

To assess the performance of the previous methods, we extend the simulation study design by \cite{yang2016propensity} to (i) encompass higher-dimensional ``treatments,''  corresponding to $P = 100$ {and $P = 200$} practices, (ii) include {$K = 30$ covariates in the practice assignment and outcome models}, (iii) and examine both linear and non-linear outcome models.
This simulation design {reflects the performance of methods in} the real-world data by incorporating a similar number of covariates, degrees of sparsity, and outcomes generated under models of increasing complexity.

%%%%%%%%%%%%%%%%%%%%%%%%%%%%%%%%%%%%%%%%%%%
%%%%%%%%%%%%%%%%%%%%%%%%%%%%%%%%%%%%%%%%%%%
\subsection{Study design}
\label{sec_study_design}

In our simulation, we consider a vector of {30} observed covariates, $\boldsymbol{X}_i = (X_{i1}, X_{i2}, ..., X_{i30})$, distributed as follows: 
$X_{i1}, X_{i2}, X_{i3} \sim \mathcal{N}_{3}(\boldsymbol{\mu}_3, \boldsymbol{\Sigma}_3)$, with $\boldsymbol{\mu}_3 = (0, 0, 0)^\top$ and $\boldsymbol{\Sigma}_3 = [ (2, 1, -1), $ $(1, 1, -0.5), $ $(-1, -0.5, 2) ]$; 
$X_{i4} \sim \mathrm{Unif}[-3, 3]$; 
$X_{i5} \sim \chi_1^2$; 
$X_{i6} \sim \operatorname{Bern}(0.5)$; 
$X_{i7}, X_{i8}, X_{i9},X_{i10} \sim \mathcal{N}_{4}(\boldsymbol{\mu}_4, \boldsymbol{\Sigma}_4)$, with $\boldsymbol{\mu}_4 = (0, 0, 0, 0)^\top$, $\boldsymbol{\Sigma}_4 = [ (2, 1, -1, -1), $ $(1, 1, -0.5, -0.5), $ $(-1, -0.5,$ $2, .5), $ $(-1, -0.5, 0.5, 1)]$; and 
{$X_{ik} \sim \operatorname{Bern}(0.5)$ for $k = 11,...,30$}.
The practice assignment indicators are generated in the following way: 
$(\indicator{1,i}, \indicator{2,i}, ..., \indicator{P,i}) \sim \operatorname{Multinom} \{\Pr(1|\boldsymbol{X}_i), \Pr(2|\boldsymbol{X}_i), ..., \Pr(P|\boldsymbol{X}_i) \}$, where $\Pr(\indicator{p,i} | \boldsymbol{X}_i) = \frac{\exp(\boldsymbol{X}_i^\top\eta_{p})}{\sum_{\zeta=1}^{P} \exp(\boldsymbol{X}_i^\top\eta_{\zeta})}$, with $\eta_p = (1 - \frac{p}{P}) \times (1, 1, 1, -1, 1, 1,{0, 0, 0, 0, \eta_0})^\top$ and {$\eta_0 = (1,-1,1,-1,....,1,-1)$}.
Using these covariates and practice assignment indicators, we generate the outcomes under four models of increasing complexity:
\begin{itemize}
\item Setting 1, linear model: $Y_i(p) = (1 + {2p/P})X_{i1} + \{(-1)^p + 2 \} X_{i2} + \{ (-1)^p + 2 \} X_{i3} + (-1)^{p+1}X_{i4} + (-1)^p(X_{i5} - 1) + (-1)^p(X_{i6} - 0.5) {+ \sum_{k = 11}^{20}} 0.5(X_{ik}-0.5) - {\sum_{k = 21}^{30}} 0.5(X_{ik}-0.5) + 0.1p + \varepsilon_i$, where $\varepsilon_i \sim \mathcal{N}(0,1)$. 

\item Setting 2, weakly non-linear model: $Y_i(p) = (1 + {2p/P})(X_{i1} {+ 0.25 X_{i1}^2 - 0.5}) + \{ (-1)^p + 2 \} X_{i2} + \{ (-1)^p + 2 \} X_{i3} + (-1)^{p+1}X_{i4} + (-1)^p(X_{i5} - 1) + (-1)^p(X_{i6} - 0.5) + {\sum_{k = 11}^{20}} 0.5(X_{ik}-0.5) - {\sum_{k = 21}^{30}} 0.5(X_{ik}-0.5) + 0.1p + \varepsilon_i$, where $\varepsilon_i \sim \mathcal{N}(0,1)$.

\item Setting 3, moderately non-linear model: $Y_i(p) = (1 + {2p/P})(X_{i1}{+0.5X_{i1}^2 - 1}) + \{ (-1)^p + 2 \} X_{i2} + \{ (-1)^p + 2 \} X_{i3} + (-1)^{p+1}X_{i4} + (-1)^p(X_{i5} - 1) + (-1)^p(X_{i6} - 0.5){+ \sum_{k = 11}^{20}} 0.5(X_{ik}-0.5) - {\sum_{k = 21}^{30}} 0.5(X_{ik}-0.5) + 0.1p + \varepsilon_i$, where $\varepsilon_i \sim \mathcal{N}(0,1)$. 

\item Setting 4, highly non-linear model: $Y_i(p) = (1 + {2p/P})(X_{i1}+X_{i1}^2 - 2) + \{ (-1)^p + 2 \} X_{i2} + \{ (-1)^p + 2 \} X_{i3}  + (-1)^{p+1}X_{i4} + (-1)^p(X_{i5} - 1) + (-1)^p(X_{i6} - 0.5) +{\sum_{k = 11}^{20}} 0.5(X_{ik}-0.5) - {\sum_{k = 21}^{30}} 0.5(X_{ik}-0.5) + 0.1p + \varepsilon_i$, where $\varepsilon_i \sim \mathcal{N}(0,1)$.
\end{itemize}

In this manner, the true mean potential outcome function for practice $p$ is $\operatorname{E}\left[Y_i(p)\right] = $ $ \operatorname{E}\left[\operatorname{E}\left[Y_i(p)\mid \boldsymbol{X}_i\right]\right] = 0.1 p$ for all $p = 1, ..., {P}$.
Thus, the effect of practice $p = 1$ in place of practice $p = 10$ is $0.1 - 1 = -0.9$.
Under this data generating mechanism, {for $P = 100$ and $P = 200$,} we generate 1000 data sets of size $n = 10000$, with corresponding average practice sizes of {100 and 50. For all the approaches, we let $\boldsymbol{X}_{\textrm{null}} := (X_{1},...,X_{10})$, $\boldsymbol{X}_{\textrm{null}}^{\mathsf c} := (X_{11},...,X_{30})$}.

%%%%%%%%%%%%%%%%%%%%%%%%%%%%%%%%%%%%%%%%%%%
%%%%%%%%%%%%%%%%%%%%%%%%%%%%%%%%%%%%%%%%%%%
\subsection{Comparative performances}
\label{sec_comparative_performances}

We evaluate the performance of all the methods from two perspectives: an {absolute} perspective, focusing on bias and root mean square error (RMSE) {of the estimated survival rates (Tables \ref{tab_the_system_perspective} and \ref{tab_the_system_perspective_200})}, and a {relative} perspective, based on the errors in the estimated practice rankings {(Tables \ref{tab_the_practice_perspective} and \ref{tab_the_practice_perspective_200})}.
For both perspectives, we consider three target profiles: patients across the overall system of practices, {patients in one practice near the center of covariate distribution, and patients in one practice at the boundary of covariate distribution}.
We evaluate the performance of the methods described in Table \ref{tab_methods}.

{Tables \ref{tab_the_system_perspective} and \ref{tab_the_system_perspective_200} present the biases and RMSEs for $P = 100$ and $P = 200$, respectively.}
{For $P =100$ with a linear outcome model, practice-pooled regression (PR) exhibits relatively small biases and RMSEs, outperforming both fixed-effects regression (FE) and practice-stratified regression (SR). However, at $P = 200$, PR become more sensitive to model misspecification, exhibiting the highest RMSEs among all methods when similar higher-order terms are adjusted. Across both scenarios with $P = 100$ and $P = 200$, as well as for both linear and non-linear outcomes, the proposed approach (\{SBW, WR\}) consistently shows stable and robust performance.  It effectively handles different forms of covariate adjustment, whether the specification is parsimonious ($\{\textrm{SBW}(\mathbb{R}_0^+, \boldsymbol{X}_{\textrm{null}}^{\mathsf{c}}), \textrm{WR}(\boldsymbol{X})\}$) or more comprehensive ($\{\textrm{SBW}(\mathbb{R}_0^+, \widetilde{\boldsymbol{X}}_{\textrm{null}}^{\mathsf{c}}), \textrm{WR}(\widetilde{\boldsymbol{X}})\}$). 
Even when a parsimonious specification results in underfitting, the approach remains robust to misspecification. Conversely, when the more comprehensive specification leads to overfitting, the estimator still maintains its stability.

In general, when $P$ is large relative to the sample size $n$, the combined weighting and modeling approaches (\{LW, FE\}, \{LW, WR\}, and \{SBW, WR\}), which incorporate forms of weighting that adjust only for $X_{\textrm{null}}^{\mathsf{c}}$, exhibit more accurate performance than the one-step modeling approaches (\{$\cdot$, FE\}, \{$\cdot$, SR\}, \{$\cdot$, PR\}, \{LW, $\cdot$\}). 
In extended simulation study results (not included in the paper due to space constraints but available upon request), approaches that incorporate LW weights and additionally adjust for the null-case binary covariates $X_{\textrm{null}}$ in the treatment assignment models, increase biases and RMSE, despite the relevance of these covariates.

In Tables \ref{tab_the_practice_perspective} and \ref{tab_the_practice_perspective_200}, two metrics are presented: the mean and the maximum absolute difference between the true and estimated rankings. For both metrics, the findings are consistent with the results observed previously.}

\section{Case study}
\label{sec_case_study}

The methods that exhibited the most accurate and robust performance in the simulation study are $\{\textrm{SBW}(\mathbb{R}_0^+, \boldsymbol{X}_{\textrm{null}}^{\mathsf{c}}), \textrm{WR}(\boldsymbol{X})\}$, hence we will now use them to assess survival rates across oncology practices. 
First, we will describe the covariate profiles of the target populations, including all patients in the system and three individual patients. 
In the case of the individual patients, we envision a population of patients with the same covariate values as the individual patient. 
Second, we will find the practices with null-case covariates for each profile. 
Again, these are practices for which quality estimates can be provided only by extrapolation using data from other practices. 
Third, we will report quality estimates using $\textrm{SBW}(\mathbb{R}^+_0, \boldsymbol{X})$ and $\{\textrm{SBW}(\mathbb{R}_0^+, \boldsymbol{X}_{\textrm{null}}^{\mathsf{c}}), \textrm{WR}(\boldsymbol{X})\}$. 
Finally, we will devise rankings using $\{\textrm{SBW}(\mathbb{R}_0^+, \boldsymbol{X}_{\textrm{null}}^{\mathsf{c}}), \textrm{WR}(\boldsymbol{X})\}$.

Table \ref{tab_profiles} describes the four covariate profiles under study.
The first profile is the average ``patient'' across all practices (the system).
The next three profiles are examples of hypothetical patients that often require cancer care.
In particular, Patient 1 represents a White man in his early 70s with stage 3 lung cancer and generally low socioeconomic status.
Patient 2 corresponds to a White woman in her late 70s with stage 4 pancreatic cancer, high comorbidity, and higher socioeconomic status.
Patient 3 is identical to Patient 2, except for race which is now Black.
As we will see, just changing this covariate from Patient 2 to Patient 3 will have important implications for quality measurement.

%%\singlespacing
%\begin{table}[htb]
%\small
%\caption{Covariate profiles for the system and three hypothetical patients}
%\begin{center}
%\label{tab_profiles}
%\begin{tabular}{lrrrrrr}
%\hline
%& System & Patient 1 & Patient 2 & Patient 3\\ 
%\hline
%Age & 76.10 & 70.77 & 78.34 & 78.34\\ 
%Sex (Male) & 0.34 & 1 & 0 & 0 \\ 
%Sex (Female) & 0.66 & 0 & 1 & 1\\
%Race (White) & 0.82 & 1 & 1 & 0\\ 
%Race (Black) & 0.08 & 0 & 0 & 1\\ 
%Race (Other) & 0.10 & 0 & 0 & 0\\  
%Marital status (Married) & 0.52 & 0 & 1 & 1\\ 
%Marital status (Unmarried)  & 0.44 & 1 & 0 & 0\\ 
%Percentage of residents without a high school education & 13.09 & 27.75 & 13.57 &  13.57\\
%Census-tract median household income & 1.08 & 0.60 &  0.89 &  0.89\\ 
%Charlson Comorbidity Index (0) & 0.34 & 0 & 0 & 0\\ 
%Charlson Comorbidity Index (1) & 0.26 & 0 & 0 & 0\\ 
%Charlson Comorbidity Index (2) & 0.16 & 0 & 0 & 0\\ 
%Charlson Comorbidity Index ($\geq$3) & 0.24 & 1 & 1 & 1\\ 
%Cancer type (Breast) & 0.35 & 0 & 0 & 0\\ 
%Cancer type (Colorectal) & 0.17 & 0 & 0 & 0\\ 
%Cancer type (Lung) & 0.34 & 1 & 0 & 0\\ 
%Cancer type (Ovary) & 0.02 & 0 & 0 & 0\\ 
%Cancer type (Pancreas) & 0.07 & 0 & 1 & 1\\ 
%Cancer type (Prostate)& 0.05 & 0 & 0 & 0\\ 
%Cancer stage (1) & 0.27 & 0 & 1 & 1\\ 
%Cancer stage (2) & 0.22 & 0 & 0 & 0\\ 
%Cancer stage (3) & 0.20 & 1 & 0 & 0\\ 
%Cancer stage (4) & 0.27 & 0 & 1 & 1\\ 
%\hline
%\end{tabular}
%\end{center}
%\end{table}
%%\doublespacing

%Can you make the labels larger? (including legend)

For each of these profiles, we inspect the data to understand the extent to which data are available to assess the survival outcomes for each of the different practices. 
Figure \ref{fig_null_covariates} 
in the Online Supplementary Materials counts the number of practices with null-case covariates for each profile.
Again, these are practices for which it is not possible to carry out the case-mix adjustments (and therefore, estimate the mean outcome) using data solely from the practice in question; that is, without leveraging the functional form of a model fitted on data from patients in other practices.   
For the system profile, across all 600 practices, there are 306 practices with no null-case covariates and 294 practices with at least one null-case covariate.
Among these, 83 have three or more null-case covariates.
The covariates that produce the largest number of null cases are race (Black and other), education (quartile 1), income (quartile 4), and cancer type (ovary and prostate).
For patients 1, 2, and 3, the numbers of practices with null-case covariates are reduced substantially, but there are still sizable numbers of practices that did not treat a single patient with a certain covariate value of the target profiles. 
This is particularly true for Patient 3, because many practices did not treat any Black women with pancreatic cancer, a cancer that is relatively infrequently diagnosed.

When dealing with null-case covariates, $\textrm{SBW}(\mathbb{R}^+_0, \boldsymbol{X})$ does not produce estimates for every practice because its case-mix optimization problem is infeasible.
By contrast, $\{\textrm{SBW}(\mathbb{R}_0^+, \boldsymbol{X}_{\textrm{null}}^{\mathsf{c}}), \textrm{WR}(\boldsymbol{X})\}$ borrows information from data across practices, specifically for the null-case covariates sex, race, cancer type, and cancer stage. 
While $\textrm{SBW}(\mathbb{R}^+_0, \boldsymbol{X})$ is sample bounded \citep{robins2007comment} and allows for effect modification for all the covariates $X$, $\{\textrm{SBW}(\mathbb{R}_0^+, \boldsymbol{X}_{\textrm{null}}^{\mathsf{c}}), \textrm{WR}(\boldsymbol{X})\}$ admits effect modification for $X_{\textrm{null}}^{\mathsf{c}}$, which generally constitutes a smaller subset of covariates.
Consequently, using $\textrm{SBW}(\mathbb{R}^+_0, \boldsymbol{X})$ we can inspect covariate balance as follows.

\begin{figure}[!htbp]
\begin{center}
\caption{Covariate balance relative to the system profile before and after weighting. After weighting adjustments, imbalances, as measured by standardized mean differences, are close to zero.} 
\label{fig_balance}
\includegraphics[scale=0.5]{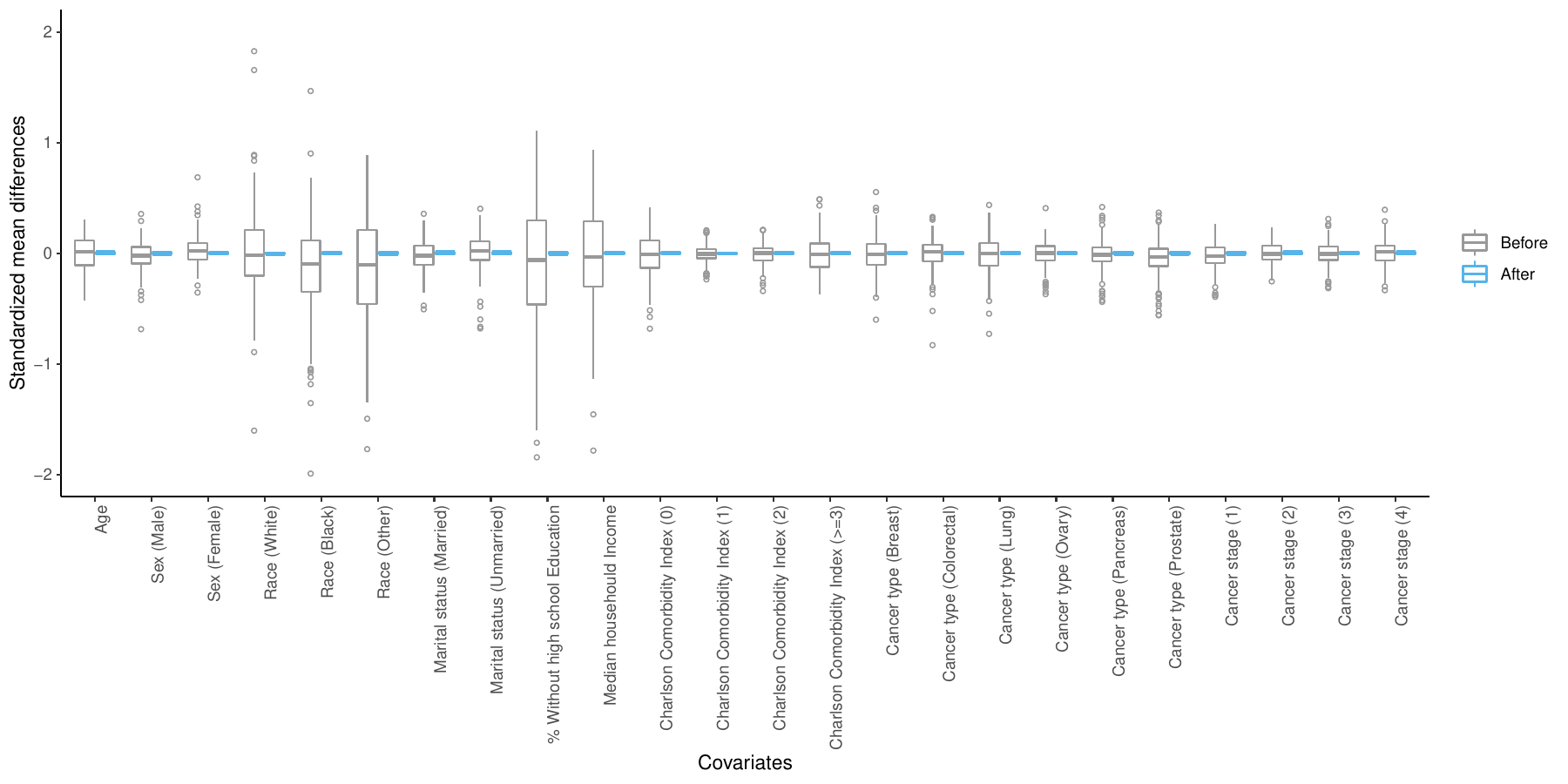}
\end{center}
\end{figure}

Figure \ref{fig_balance} shows covariate balance relative to the system profile before and after weighting using $\textrm{SBW}(\mathbb{R}^+_0, \boldsymbol{X})$.
The results are similar for other profiles.
Before weighting, there are substantial differences in covariates means between practices and the profile (most notably in race, education, and income) yet after weighting these differences are reduced, showing the case-mix adjustments and comparability of the patient samples after weighting.

\begin{figure}[!htbp]
\begin{center}
\caption{Estimated survival rates after one year for the system profile using $\textrm{SBW}(\mathbb{R}^+_0, \boldsymbol{X})$ (left) and $\{\textrm{SBW}(\mathbb{R}_0^+, \boldsymbol{X}_{\textrm{null}}^{\mathsf{c}}), \textrm{WR}(\boldsymbol{X}; \boldsymbol{w})\}$ (right).}
\label{fig_estimated_survival}
\includegraphics[scale = 0.4]{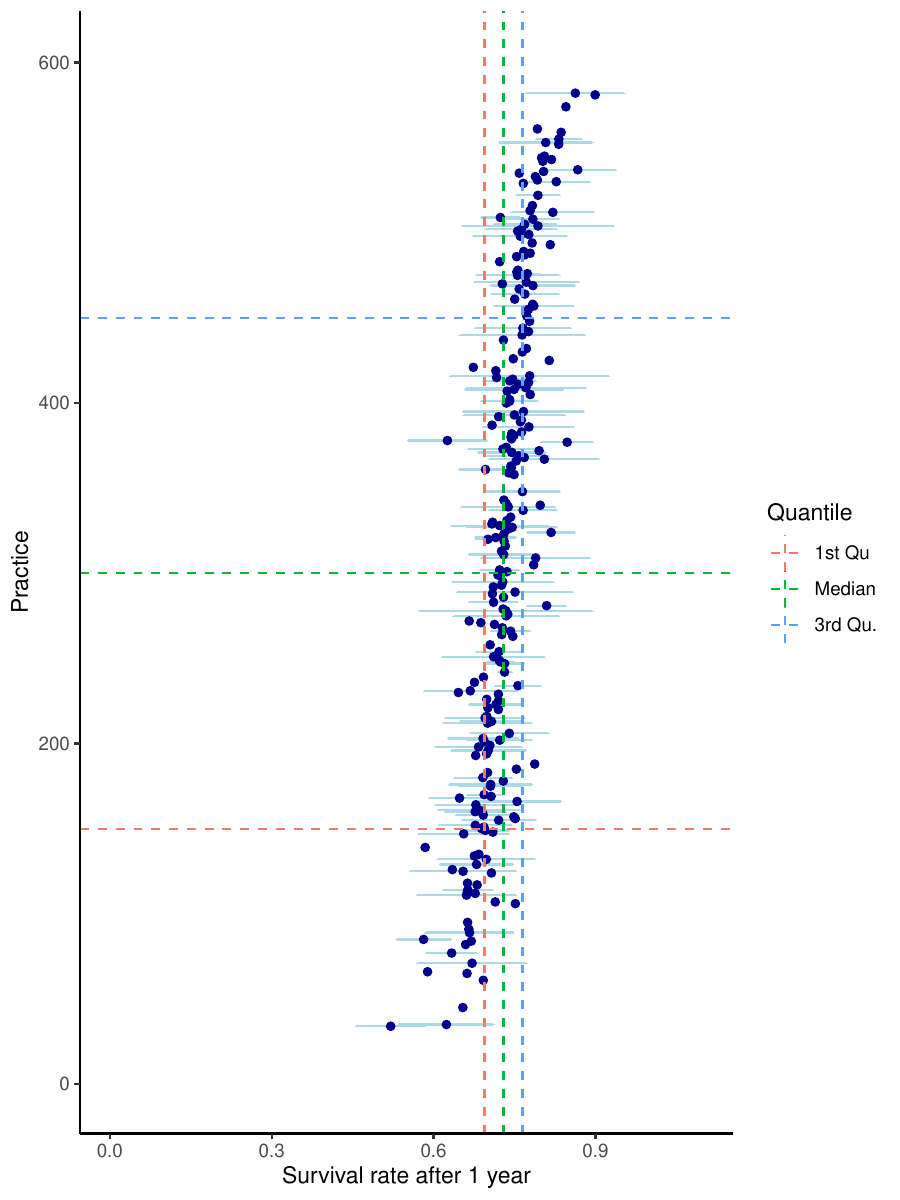}
\includegraphics[scale = 0.4]{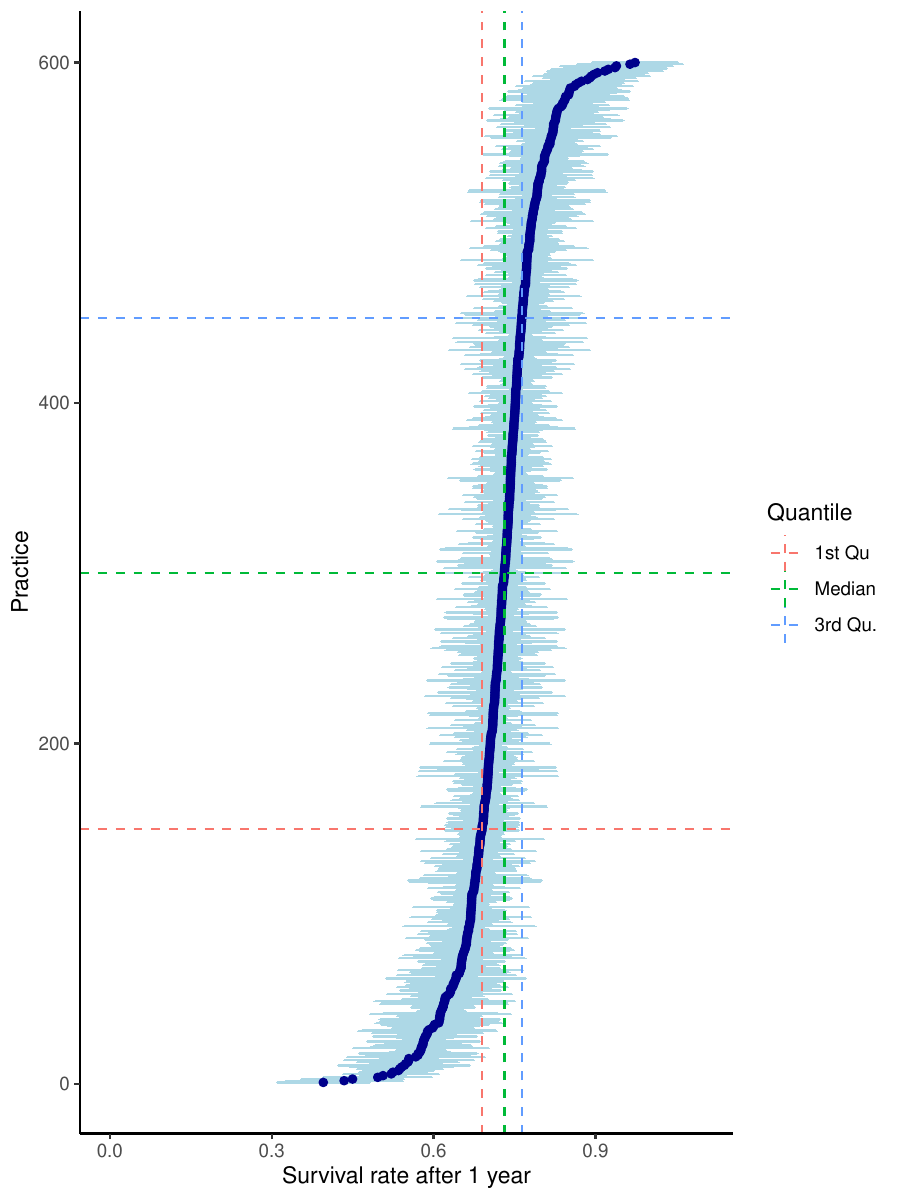}
\end{center}
\end{figure}

Figure \ref{fig_estimated_survival} follows \cite{keele2023hospital} and shows the estimated survival rate after one year using $\textrm{SBW}(\mathbb{R}^+_0, \boldsymbol{X})$ (left) and $\{\textrm{SBW}(\mathbb{R}^+_0, \boldsymbol{X}_{\textrm{null}}^{\mathsf{c}}), \textrm{WR}(\boldsymbol{X})\}$ (right) for the system profile.
Point estimates and confidence intervals are depicted in blue and light blue, respectively.
In both plots, practices are indexed by the order of their estimated 1-year survival rates using $\{\textrm{SBW}(\mathbb{R}_0^+, \boldsymbol{X}_{\textrm{null}}^{\mathsf{c}}),$ $\textrm{WR}(\boldsymbol{X})\}$, so in the right plot Practice 1 has the lowest survival rate and Practice 600, the highest.
In the left plot, we observe that $\textrm{SBW}(\mathbb{R}^+_0, \boldsymbol{X})$ often does not produce point estimates, leaving blank spaces for the practices where forms of extrapolation and information borrowing are needed for case-mix adjustment.
Comparing the two plots, we note that while both sets of estimates follow a roughly similar ``S'' shape in aggregate, the individual practice point estimates can differ substantially. 

To quantify uncertainty, the figure displays confidence intervals for the point estimates $\{\widehat{\mu}_{p}(\boldsymbol{x}^*)\}_{p = 1}^P$. 
Their contrast $\widehat{\mu}_{p'}(\boldsymbol{x}^*)-\widehat{\mu}_{p''}(\boldsymbol{x}^*)$ estimates the conditional average treatment effect  $\mathrm{E}[Y(p') - Y(p'') \mid \boldsymbol{X}= \boldsymbol{x}^*]$. 
Using weighted linear regression, $\widehat{\mu}_{p'}(\boldsymbol{x}^*)-\widehat{\mu}_{p''}(\boldsymbol{x}^*)$ corresponds to the difference between two indicators coefficients, $\widehat{\alpha}_{p'}-\widehat{\alpha}_{p''}$, which can be tested using a t-test for the null hypothesis ${\mu}_{p'}(\boldsymbol{x}^*)={\mu}_{p''}(\boldsymbol{x}^*)$. 
Additionally, if the null hypothesis is ${\mu}_{1}(\boldsymbol{x}^*)=...={\mu}_{P}(\boldsymbol{x}^*)$, we can conduct a Rao's Score test (\citealt{rao1948large}; specifically, an F-test in this linear regression scenario) using the coefficients $\widehat{\alpha}_{1},...,\widehat{\alpha}_{P}$ and robust estimates of their variances and covariances. 
A major challenge arises when simultaneously testing all contrasts and rankings due to multiple testing, which is an ongoing challenge and will be explored in future research. 
This simulation study shows that the proposed approach performs well in estimating rankings, which is an important milestone at this stage.

Figure \ref{fig_rankings} 
in the Online Supplementary Materials shows the changes in rankings for different profiles.
The left plot compares the rankings for the system profile and that of Patient 1. 
There are substantial differences: 367 or 61.1\% of the practices change their ranking by 10\% (60 ranking places) or more.
The center plot makes a similar comparison between profiles 2 and 3, when only one covariate value (race) changes between profiles.
426 or 71\% of the practices change their ranking by 10\% (60 ranking places) or more.
To better grasp this variation, the right plot shows the changes in relative positions after grouping the practices by quintiles.
Table \ref{tab_rankings} counts the number of practices for each of these changes: 176 or 29.3\% of the practices remain in the same quintile when the profile changes (diagonal of the matrix), 215 or 35.8\% of the practices move up or down by one quintile, 209 or 34.8\% move by two or more quintiles, and 42 or 7\% of the practices move from the top to the bottom quintile or vice versa.
This exemplifies that the profile --- the ``test'' on which each practice is evaluated --- matters.
Among other reasons, this is important for quality measurement that is used for high-stakes purposes, such as public reporting or payment.

%%%%%%%%%%%%%%%%%%%%%%%%%%%%
%%%%%%%%%%%%%%%%%%%%%%%%%%%%
%%%%%%%%%%%%%%%%%%%%%%%%%%%%

\section{Discussion}
\label{sec_discussion}

Motivated by the challenge of assessing the quality of cancer care delivered by healthcare providers throughout the United States, we have presented a framework for institutional quality measurement that accounts for the heterogeneity of the patient populations these providers serve.
Central to this framework is the concept of a target covariate profile, which can represent the specific patient populations of interest for quality measurement.
Leveraging this profile, we have assessed the empirical performance of several estimators that combine weighting and regression modeling adjustments in a layered manner in order to reduce parametric extrapolation.
We have implemented a form of adjustment that approximately balances the covariance matrices of the observed covariates.
In particular, we have highlighted the practical utility of weighting methods that openly describe the target population and warn the investigator when case-mix adjustments and performance estimates are infeasible without extrapolation beyond the support of the data of a given practice.

Addressing settings with null-case covariates, our examination highlights that case-mix adjustment methods must rely on data from other practices to generate performance estimates due to the absence of relevant patient data in these instances. 
As we have discussed, if the investigator wishes to obtain estimates for practices with null-case covariates, there is no other option than to rely on a model fitted using data from other practices. 
As we have shown, it is important to be aware of these different types of adjustments, as they can meaningfully impact assessments and recommendations for both practices and patients.
In this context, we have studied a constrained optimization approach to weighting, which, unlike existing unconstrained weighting approaches, provides an explicit diagnostic for extrapolation through a feasibility check. 

We have underscored the empirical performance of two approaches for case-mix adjustments.
The first approach uses balancing weights and conducts the adjustments by interpolation.
Importantly, because it does not require using outcome information, it is part of the design stage of the study \citep{rubin2007design}.
The second approach combines balancing weights and regression modeling, allowing for extrapolation beyond the data support but in a more robust and constrained manner compared to traditional regression modeling approaches for quality measurement.
While the first approach is limited by sparse data --- producing estimates only when the case mix-adjustments are feasible by interpolation --- it allows for effect modification for all covariates in the target covariate profile and produces a sample-bounded estimator (i.e., one that is restricted to lie inside the range of the observed data).
The latter approach can extrapolate but is applicable to the entirety of health providers in the data.

Our methods assume the availability of common data across different practices.
When data-sharing between practices is restricted, some of our methods can be adapted following \cite{han2024privacy} allowing each practice to share only their balancing weights and summary statistics (such as weighted variance-covariance matrices) to the investigators. 

Starting from a covariate profile, the proposed methodology serves as a tool for assessing the quality of healthcare providers, enabling us to develop more personalized recommendations for specific patient groups. 
In a nutshell, with this methodology each patient has a covariate or risk profile, and asks: for a given outcome (e.g., survival after one year), what is the typical performance of providers for patients with a similar covariate profile to mine?
The methodology first finds providers who have data treating patients like them.
Then, for the relevant providers, it produces performance estimates.
While some providers may not have sufficient data to offer personalized estimates without extrapolation, the methodology still can produce estimates for those providers with extrapolation upon the patient's request.

Our work underscores the practical challenges inherent in evaluating all healthcare providers in a system with respect to a single covariate profile. We illustrate how performance estimates can vary depending on the profile used for assessment. Looking forward, this emphasizes that different healthcare providers should be evaluated more holistically with respect to diverse profiles --- also recognizing that not all providers can be evaluated simultaneously with respect to the same profile because of violations to the positivity assumption.

Future work entails addressing structural violations of the positivity assumption, which occur when certain treatments are unattainable in particular practices due to medical guidelines or geographic constraints. 
One promising approach for addressing this issue is the use of incremental propensity score interventions \citep{kennedy2019nonparametric}.
Additionally, future research should consider covariate mismatches, as discussed in \cite{han2023multiply}, allowing hospitals to incorporate practice-specific covariate information more effectively.
This effort also includes integrating non-parametric balancing approaches for general covariate profiles $\boldsymbol{x}^*$ and statistical learning methods for the mean potential outcome function $\mu_p$, such as ensemble learners \citep{van2011targeted}, as in \cite{hirshberg2021augmented}.
Finally, an essential consideration moving forward involves careful quantification of uncertainty through multiple testing adjustments, where ideas from statistical genetics may be helpful.
Our analysis applies to other settings besides health care, including business and education, where instead of practices, we have companies and schools. 
Furthermore, provided there are adequate data, this framework can be applied at the individual level, evaluating the performance of professionals such as physicians, school teachers, and executive officers, rather than entire institutions.

%%%%%%%%%%%%%%%%%%%%%%%%%%%%%%%%%%%%%%%%%%%
%%%%%%%%%%%%%%%%%%%%%%%%%%%%%%%%%%%%%%%%%%%
%%%%%%%%%%%%%%%%%%%%%%%%%%%%%%%%%%%%%%%%%%%

%\section{Summary and remarks}
%\vspace{1cm}
%\pagebreak
\onehalfspacing
\bibliographystyle{asa}
\bibliography{mybibliography19}

\begin{thebibliography}{35}
\newcommand{\enquote}[1]{``#1''}
\expandafter\ifx\csname natexlab\endcsname\relax\def\natexlab#1{#1}\fi

\bibitem[{Ash et~al.(2012)Ash, Fienberg, Louis, Normand, Stukel, and Utts}]{ash2013statistical}
Ash, A.~S., Fienberg, S.~F., Louis, T.~A., Normand, S.-L.~T., Stukel, T.~A., and Utts, J. (2012), \enquote{Statistical issues in assessing hospital performance,} \textit{COPPS-CMS White Paper. 2012}.

\bibitem[{Ben-Michael et~al.(2020)Ben-Michael, Feller, and Rothstein}]{ben2018augmented}
Ben-Michael, E., Feller, A., and Rothstein, J. (2020), \enquote{The augmented synthetic control method,} \textit{arXiv preprint arXiv:1811.04170}.

\bibitem[{Chattopadhyay et~al.(2020)Chattopadhyay, Hase, and Zubizarreta}]{chattopadhyay2020balancing}
Chattopadhyay, A., Hase, C.~H., and Zubizarreta, J.~R. (2020), \enquote{Balancing versus modeling approaches to weighting in practice,} \textit{Statistics in Medicine}, 39, 3227--3254.

\bibitem[{Chattopadhyay and Zubizarreta(2021)}]{chattopadhyay2021implied}
Chattopadhyay, A. and Zubizarreta, J.~R. (2021), \enquote{On the implied weights of linear regression in causal inference,} \textit{arXiv preprint arXiv:2104.06581}.

\bibitem[{Chattopadhyay and Zubizarreta(2023)}]{chattopadhyay2023implied}
--- (2023), \enquote{On the implied weights of linear regression for causal inference,} \textit{Biometrika}, 110, 615--629.

\bibitem[{Daignault and Saarela(2017)}]{daignault2017doubly}
Daignault, K. and Saarela, O. (2017), \enquote{Doubly Robust Estimator for Indirectly Standardized Mortality Ratios,} \textit{Epidemiologic Methods}, 6.

\bibitem[{George et~al.(2017)George, Ro{\v{c}}kov{\'a}, Rosenbaum, Satop{\"a}{\"a}, and Silber}]{george2017mortality}
George, E.~I., Ro{\v{c}}kov{\'a}, V., Rosenbaum, P.~R., Satop{\"a}{\"a}, V.~A., and Silber, J.~H. (2017), \enquote{Mortality rate estimation and standardization for public reporting: Medicare's hospital compare,} \textit{Journal of the American Statistical Association}, 112, 933--947.

\bibitem[{Goldstein and Spiegelhalter(1996)}]{goldstein1996league}
Goldstein, H. and Spiegelhalter, D.~J. (1996), \enquote{League tables and their limitations: statistical issues in comparisons of institutional performance,} \textit{Journal of the Royal Statistical Society: Series A (Statistics in Society)}, 159, 385--409.

\bibitem[{Gondi et~al.(2021)Gondi, Wright, Landrum, Meneades, Zubizarreta, Chernew, and Keating}]{gondi2021assessment}
Gondi, S., Wright, A.~A., Landrum, M.~B., Meneades, L., Zubizarreta, J.~R., Chernew, M.~E., and Keating, N.~L. (2021), \enquote{Assessment of Patient Attribution to Care from Medical Oncologists, Surgeons or Radiation Oncologists Following a New Cancer Diagnosis,} \textit{Journal of the American Medical Association, Network Open}, in press.

\bibitem[{H{\'a}jek(1971)}]{hajek1971comment}
H{\'a}jek, J. (1971), \enquote{Comment on An essay on the logical foundations of survey sampling by Basu, D,} \textit{Foundations of Statistical Inference}, 236.

\bibitem[{Han et~al.(2024)Han, Li, Niknam, and Zubizarreta}]{han2024privacy}
Han, L., Li, Y., Niknam, B., and Zubizarreta, J.~R. (2024), \enquote{{Privacy-preserving, communication-efficient, and target-flexible hospital quality measurement},} \textit{The Annals of Applied Statistics}, 18, 1337 -- 1359.

\bibitem[{Han et~al.(2023)Han, Shen, and Zubizarreta}]{han2023multiply}
Han, L., Shen, Z., and Zubizarreta, J. (2023), \enquote{Multiply robust federated estimation of targeted average treatment effects,} \textit{Advances in Neural Information Processing Systems}, 36, 70453--70482.

\bibitem[{Hellevik(2009)}]{Hellevik2009Linear}
Hellevik, O. (2009), \enquote{Linear versus logistic regression when the dependent variable is a dichotomy,} \textit{Quality \& Quantity}, 43, 59--74.

\bibitem[{Hirshberg and Wager(2021)}]{hirshberg2021augmented}
Hirshberg, D.~A. and Wager, S. (2021), \enquote{{Augmented minimax linear estimation},} \textit{The Annals of Statistics}, 49, 3206 -- 3227.

\bibitem[{Imbens(2000)}]{imbens2000role}
Imbens, G.~W. (2000), \enquote{The role of the propensity score in estimating dose-response functions,} \textit{Biometrika}, 87, 706--710.

\bibitem[{Jones and Spiegelhalter(2011)}]{jones2011identification}
Jones, H.~E. and Spiegelhalter, D.~J. (2011), \enquote{The Identification of ``Unusual'' Health-Care Providers From a Hierarchical Model,} \textit{The American Statistician}, 65, 154 -- 163.

\bibitem[{Keele et~al.(2023)Keele, Ben-Michael, Feller, Kelz, and Miratrix}]{keele2023hospital}
Keele, L.~J., Ben-Michael, E., Feller, A., Kelz, R., and Miratrix, L. (2023), \enquote{Hospital quality risk standardization via approximate balancing weights,} \textit{The Annals of Applied Statistics}, 17, 901--928.

\bibitem[{Kennedy(2019)}]{kennedy2019nonparametric}
Kennedy, E.~H. (2019), \enquote{Nonparametric causal effects based on incremental propensity score interventions,} \textit{Journal of the American Statistical Association}, 114, 645--656.

\bibitem[{Krumholz et~al.(2011)Krumholz, Lin, Drye, Desai, Han, Rapp, Mattera, and Normand}]{krumholz2011administrative}
Krumholz, H.~M., Lin, Z., Drye, E.~E., Desai, M.~M., Han, L.~F., Rapp, M.~T., Mattera, J.~A., and Normand, S.-L.~T. (2011), \enquote{An administrative claims measure suitable for profiling hospital performance based on 30-day all-cause readmission rates among patients with acute myocardial infarction,} \textit{Circulation: Cardiovascular Quality and Outcomes}, 4, 243--252.

\bibitem[{Longford(2020)}]{longford2020performance}
Longford, N.~T. (2020), \enquote{Performance assessment as an application of causal inference,} \textit{Journal of the Royal Statistical Society: Series A (Statistics in Society)}, 183, 1363--1385.

\bibitem[{Neyman(1923, 1990)}]{neyman1923application}
Neyman, J. (1923, 1990), \enquote{On the application of probability theory to agricultural experiments,} \textit{Statistical Science}, 5, 463--480.

\bibitem[{Normand et~al.(2016)Normand, Ash, Fienberg, Stukel, Utts, and Louis}]{normand2016league}
Normand, S.-L.~T., Ash, A.~S., Fienberg, S.~E., Stukel, T.~A., Utts, J., and Louis, T.~A. (2016), \enquote{League tables for hospital comparisons,} \textit{Annual Review of Statistics and Its Application}, 3, 21--50.

\bibitem[{Rao(1948)}]{rao1948large}
Rao, C.~R. (1948), \enquote{Large sample tests of statistical hypotheses concerning several parameters with applications to problems of estimation,} \textit{Mathematical Proceedings of the Cambridge Philosophical Society}, 44, 50 -- 57.

\bibitem[{Robins et~al.(2007)Robins, Sued, Lei-Gomez, and Rotnitzky}]{robins2007comment}
Robins, J., Sued, M., Lei-Gomez, Q., and Rotnitzky, A. (2007), \enquote{Comment: performance of double-robust estimators when ``inverse probability'' weights are highly variable,} \textit{Statistical Science}, 22, 544--559.

\bibitem[{Rubin(1974)}]{rubin1974estimating}
Rubin, D.~B. (1974), \enquote{Estimating causal effects of treatments in randomized and nonrandomized studies.} \textit{Journal of Educational Psychology}, 66, 688.

\bibitem[{Rubin(1980)}]{rubin1980randomization}
--- (1980), \enquote{Randomization analysis of experimental data: the Fisher randomization test comment,} \textit{Journal of the American Statistical Association}, 75, 591--593.

\bibitem[{Rubin(2007)}]{rubin2007design}
--- (2007), \enquote{The design versus the analysis of observational studies for causal effects: parallels with the design of randomized trials,} \textit{Statistics in Medicine}, 26, 20--36.

\bibitem[{Rubin(2008)}]{rubin2008objective}
--- (2008), \enquote{For objective causal inference, design trumps analysis,} \textit{Annals of Applied Statistics}, 2, 808--840.

\bibitem[{Silber et~al.(2014)Silber, Rosenbaum, Ross, Ludwig, Wang, Niknam, Mukherjee, Saynisch, Even-Shoshan, Kelz, et~al.}]{silber2014template}
Silber, J.~H., Rosenbaum, P.~R., Ross, R.~N., Ludwig, J.~M., Wang, W., Niknam, B.~A., Mukherjee, N., Saynisch, P.~A., Even-Shoshan, O., Kelz, R.~R., et~al. (2014), \enquote{Template matching for auditing hospital cost and quality,} \textit{Health Services Research}, 49, 1446--1474.

\bibitem[{Tang et~al.(2020)Tang, Austin, Lawson, Finelli, and Saarela}]{tang2020construct}
Tang, T.-S., Austin, P.~C., Lawson, K.~A., Finelli, A., and Saarela, O. (2020), \enquote{Constructing inverse probability weights for institutional comparisons in healthcare,} \textit{Statistics in Medicine}, 39, 3156 -- 3172.

\bibitem[{Van~der Laan et~al.(2011)Van~der Laan, Rose, et~al.}]{van2011targeted}
Van~der Laan, M.~J., Rose, S., et~al. (2011), \textit{Targeted learning: causal inference for observational and experimental data}, vol.~4, Springer.

\bibitem[{Varewyck et~al.(2014)Varewyck, Goetghebeur, Eriksson, and Vansteelandt}]{varewyck2014shrinkage}
Varewyck, M., Goetghebeur, E., Eriksson, M., and Vansteelandt, S. (2014), \enquote{On shrinkage and model extrapolation in the evaluation of clinical center performance,} \textit{Biostatistics}, 15, 651--664.

\bibitem[{Wang and Zubizarreta(2020)}]{wang2020minimal}
Wang, Y. and Zubizarreta, J.~R. (2020), \enquote{Minimal dispersion approximately balancing weights: asymptotic properties and practical considerations,} \textit{Biometrika}, 107, 93--105.

\bibitem[{Yang et~al.(2016)Yang, Imbens, Cui, Faries, and Kadziola}]{yang2016propensity}
Yang, S., Imbens, G.~W., Cui, Z., Faries, D.~E., and Kadziola, Z. (2016), \enquote{Propensity score matching and subclassification in observational studies with multi-level treatments,} \textit{Biometrics}, 72, 1055--1065.

\bibitem[{Zubizarreta(2015)}]{zubizarreta2015stable}
Zubizarreta, J.~R. (2015), \enquote{Stable weights that balance covariates for estimation with incomplete outcome data,} \textit{Journal of the American Statistical Association}, 110, 910--922.

\end{thebibliography}
%\nobibliography{append}

\clearpage
% \begin{figure}[!htbp]
% \begin{center}
% \caption{Extrapolation with ten covariates for three profiles (Large, Small, System) and five regression and weighting that balance different functions of the covariates (LR1, original covariates; LR2, second-order transformations; SBW1+, original covariates without extrapolation; SBW1, original covariates with extrapolation; SBW2, second-order transformations with extrapolation). In each subfigure, each (row, column) intersection (i.e., ``pixel'') corresponds to a (practice, data set) in the simulation study. In each subfigure, there are 1000 rows corresponding to the 1000 simulated data sets, and 100 columns corresponding to the 100 practices in each data set.
% The color grey represents practices for which an estimate was produced without extrapolation; the color white represents practices for which no estimate was produced because it required extrapolation; and the color orange represents practices for which an estimate was produced with extrapolation.
% This allows us to visualize the frequency of extrapolation in case-mix adjustments for different profiles with regression and weighting methods.}
% \label{fig_extent_of_extrapolation1}
% \includegraphics[scale=0.275]{graphics/a_plot_flag_10_2.png}
% \end{center}
% \end{figure}

\begin{center}
% \begin{landscape}
\begin{table}[htbp]
\caption{Absolute bias and RMSE based on 1,000 simulated data sets, averaged over 100 practices, for three target samples: all patients in the system, and patients from the practices near the center or at the boundary of the population. Two forms of covariates ($\boldsymbol{X}$ and $\widetilde{\boldsymbol{X}}$) are considered: $\boldsymbol{X}$ exact fits when the outcome is linear and under fits when the outcome is non-linear; $\widetilde{\boldsymbol{X}}$ over fits when the outcome is linear and exact fits when the outcome is quadratic.}
\label{tab_the_system_perspective}
\centering
\scriptsize
\setlength{\tabcolsep}{2pt}
\begin{tabular}{lrrrrrrrrrrrr}
  \hline
   & \multicolumn{12}{c}{Simulation setting} \\
  \cmidrule(lr){2-13} 
  &  \multicolumn{3}{c}{Linear} &  \multicolumn{3}{c}{Lowly non-linear} &  \multicolumn{3}{c}{Moderately non-linear} &  \multicolumn{3}{c}{Highly non-linear} \\
  \cmidrule(lr){2-4}  \cmidrule(lr){5-7} \cmidrule(lr){8-10} \cmidrule(lr){11-13}
     &  \multicolumn{3}{c}{Target} &  \multicolumn{3}{c}{Target} &  \multicolumn{3}{c}{Target} &  \multicolumn{3}{c}{Target} \\
Method     & System & Center & ~Edge & System & Center & ~Edge & System & Center & ~Edge & System & Center & ~Edge \\
   \cmidrule(lr){1-1}  \cmidrule(lr){2-4}  \cmidrule(lr){5-7} \cmidrule(lr){8-10} \cmidrule(lr){11-13}
Bias (parsimonious) & Exact &  & & &  & &  & &  & Under\\
\hspace{.05cm} $\{\cdot, \textrm{FE}(\boldsymbol{X})\}$ & 0.88 & 0.91 & 1.91 & 0.88 & 0.91 & 1.91 & 0.88 & 0.91 & 1.91 & 0.89 & 0.91 & 1.91 \\ 
\hspace{.05cm} $\{\cdot, \textrm{SR}(\boldsymbol{X})\}$ & {\bf 0.00} & {\bf 0.00} & 0.01 & 0.16 & 0.08 & 0.30 & 0.32 & 0.16 & 0.61 & 0.64 & 0.32 & 1.22 \\ 
\hspace{.05cm} $\{\cdot, \textrm{PR}(\boldsymbol{X})\}$ & {\bf 0.00} & {\bf 0.00} & {\bf 0.00} & 0.16 & 0.08 & 0.30 & 0.31 & 0.16 & 0.60 & 0.63 & 0.31 & 1.20 \\ 
% \hspace{.05cm} $\{\textrm{IPW}(\mathbb{R}^+, \boldsymbol{X}), \cdot\}$ \\ 
% \hspace{.05cm} $\{\textrm{IPW}(\mathbb{R}^+, \boldsymbol{X}), \textrm{FE}({\boldsymbol{X}}; w)\}$ \\ 
% \hspace{.05cm} $\{\textrm{IPW}(\mathbb{R}^+, \boldsymbol{X}), \textrm{WR}({\boldsymbol{X}}; w)\}$ \\ 
\hspace{.05cm} $\{\textrm{LW}( \boldsymbol{X}_{\textrm{null}}^{\mathsf{c}}), \cdot\}$ & 0.27 & 0.24 & 0.66 & 0.33 & 0.22 & 0.71 & 0.39 & 0.21 & 0.79 & 0.52 & 0.21 & 0.97 \\ 
\hspace{.05cm} $\{\textrm{LW}(\boldsymbol{X}_{\textrm{null}}^{\mathsf{c}}), \textrm{FE}(\boldsymbol{X})\}$ & 0.24 & 0.21 & 0.65 & 0.24 & 0.21 & 0.65 & 0.27 & 0.21 & 0.66 & 0.39 & 0.21 & 0.67\\
\hspace{.05cm} $\{\textrm{LW}( \boldsymbol{X}_{\textrm{null}}^{\mathsf{c}}), \textrm{WR}({\boldsymbol{X}})\}$ & 0.24 & 0.21 & 0.66 & 0.25 & 0.21 & 0.66 & 0.27 & 0.21 & 0.66 & {\bf 0.38} & 0.21 & 0.66 \\ 
\hspace{.05cm} $\{\textrm{SBW}(\boldsymbol{X}_{\textrm{null}}^{\mathsf{c}}), \textrm{WR}({\boldsymbol{X}})\}$ & 0.01 & 0.01 & 0.17 & {\bf 0.12} &  {\bf 0.04} & {\bf 0.18} & {\bf 0.24} & {\bf 0.09} & {\bf 0.24} & 0.47 & {\bf 0.17} & {\bf 0.41} \\ 
\cmidrule(lr){1-1}  \cmidrule(lr){2-4}  \cmidrule(lr){5-7} \cmidrule(lr){8-10} \cmidrule(lr){11-13}
Bias (comprehensive) & Over &  & & &  & &  & &  & Exact\\
\hspace{.05cm} $\{\cdot, \textrm{FE}(\widetilde{\boldsymbol{X}})\}$ & 0.88 & 0.91 & 1.91 & 0.88 & 0.91 & 1.91 & 0.88 & 0.91 & 1.91 & 0.88 & 0.91 & 1.91 \\ 
\hspace{.05cm} $\{\cdot, \textrm{SR}(\widetilde{\boldsymbol{X}})\}$ & 0.01 & 0.01 & {\bf 0.01} & {\bf 0.02} & 0.03 & 0.04 & {\bf 0.04} & 0.06 & 0.07 & {\bf 0.08} & 0.12 & 0.13 \\ 
\hspace{.05cm} $\{\cdot, \textrm{PR}(\widetilde{\boldsymbol{X}})\}$ & {\bf 0.00} & {\bf 0.00} & {\bf 0.01} & {\bf 0.02} & 0.03 & {\bf 0.03} & {\bf 0.04} & 0.06 & {\bf 0.06} & 0.07 & 0.12 & {\bf 0.12} \\ 
% \hspace{.05cm} $\{\textrm{IPW}(\mathbb{R}^+, \widetilde{\boldsymbol{X}}), \cdot\}$ \\ 
% \hspace{.05cm} $\{\textrm{IPW}(\mathbb{R}^+, \widetilde{\boldsymbol{X}}), \textrm{DR}(\widetilde{\boldsymbol{X}}; w)\}$ \\ 
% \hspace{.05cm} $\{\textrm{IPW}(\mathbb{R}^+, \widetilde{\boldsymbol{X}}), \textrm{WR}(\widetilde{\boldsymbol{X}}; w)\}$ \\ 
\hspace{.05cm} $\{\textrm{LW}( \widetilde{\boldsymbol{X}}_{\textrm{null}}^{\mathsf{c}}), \cdot\}$ & 0.24 & 0.20 & 0.61 & 0.25 & 0.18 & 0.64 & 0.26 & 0.18 & 0.70 & 0.28 & 0.18 & 0.83 \\ 
\hspace{.05cm} $\{\textrm{LW}( \widetilde{\boldsymbol{X}}_{\textrm{null}}^{\mathsf{c}}), \textrm{FE}(\widetilde{\boldsymbol{X}})\}$ & 0.22 & 0.17 & 0.61 & 0.22 & 0.17 & 0.61 & 0.22 & 0.17 & 0.61 & 0.22 & 0.17 & 0.61 \\ 
\hspace{.05cm} $\{\textrm{LW}( \widetilde{\boldsymbol{X}}_{\textrm{null}}^{\mathsf{c}}), \textrm{WR}(\widetilde{\boldsymbol{X}})\}$ & 0.23 & 0.17 & 0.62 & 0.23 & 0.17 & 0.62 & 0.23 & 0.17 & 0.62 & 0.23 & 0.17 & 0.62 \\ 
\hspace{.05cm} $\{\textrm{SBW}( \widetilde{\boldsymbol{X}}_{\textrm{null}}^{\mathsf{c}}), \textrm{WR}(\widetilde{\boldsymbol{X}})\}$ & 0.06 & 0.02 & 0.55 & 0.06 & {\bf 0.02} & 0.56 & 0.06 & {\bf 0.04} & 0.56 & 0.09 & {\bf 0.07} & 0.57 \\ 
\cmidrule(lr){1-1}  \cmidrule(lr){2-4}  \cmidrule(lr){5-7} \cmidrule(lr){8-10} \cmidrule(lr){11-13}
RMSE (parsimonious) & Exact &  & & &  & &  & &  & Under\\
\hspace{.05cm} $\{\cdot, \textrm{FE}(\boldsymbol{X})\}$ & 0.94 & 1.04 & 1.95 & 0.96 & 1.07 & 1.96 & 1.01 & 1.16 & 1.98 & 1.17 & 1.44 & 2.10 \\ 
\hspace{.05cm} $\{\cdot, \textrm{SR}(\boldsymbol{X})\}$ & 0.18 & 0.19 & 0.28 & 0.35 & 0.38 & 0.62 & 0.63 & 0.68 & 1.14 & 1.23 & 1.31 & 2.21 \\ 
\hspace{.05cm} $\{\cdot, \textrm{PR}(\boldsymbol{X})\}$ & {\bf 0.14} & {\bf 0.14} & {\bf 0.19} & 0.29 & 0.30 & {\bf 0.50} & 0.53 & 0.56 & 0.93 & 1.03 & 1.09 & 1.83 \\
% \hspace{.05cm} $\{\textrm{IPW}(\mathbb{R}^+, \boldsymbol{X}), \cdot\}$ \\
% \hspace{.05cm} $\{\textrm{IPW}(\mathbb{R}^+, \boldsymbol{X}), \textrm{DR}({\boldsymbol{X}}; w)\}$ \\ 
% \hspace{.05cm} $\{\textrm{IPW}(\mathbb{R}^+, \boldsymbol{X}), \textrm{WR}({\boldsymbol{X}}; w)\}$ \\ 
\hspace{.05cm} $\{\textrm{LW}(\boldsymbol{X}_{\textrm{null}}^{\mathsf{c}}), \cdot\}$ & 0.49 & 0.52 & 1.03 & 0.56 & 0.56 & 1.19 & 0.68 & 0.69 & 1.40 & 0.98 & 1.08 & 1.91 \\ 
\hspace{.05cm} $\{\textrm{LW}(\boldsymbol{X}_{\textrm{null}}^{\mathsf{c}}), \textrm{FE}(\boldsymbol{X})\}$ & 0.34 & 0.34 & 0.79 & 0.41 & 0.43 & 0.88 & 0.55 & 0.61 & 1.07 & 0.88 & 1.05 & 1.59 \\
\hspace{.05cm} $\{\textrm{LW}(\boldsymbol{X}_{\textrm{null}}^{\mathsf{c}}), \textrm{WR}({\boldsymbol{X}})\}$ & 0.34 & 0.34 & 0.80 & 0.41 & 0.43 & 0.86 & 0.54 & 0.60 & 1.00 & {\bf 0.87} & 1.02 & {\bf 1.39} \\ 
\hspace{.05cm} $\{\textrm{SBW}(\boldsymbol{X}_{\textrm{null}}^{\mathsf{c}}), \textrm{WR}({\boldsymbol{X}})\}$ & {\bf 0.14} & {\bf 0.14} & 0.36 & {\bf 0.27} & {\bf 0.29} & 0.53 & {\bf 0.48} & {\bf 0.52} & {\bf 0.83} & 0.92 & {\bf 1.00} & 1.52 \\ 
\cmidrule(lr){1-1}  \cmidrule(lr){2-4}  \cmidrule(lr){5-7} \cmidrule(lr){8-10} \cmidrule(lr){11-13}
RMSE (comprehensive) & Over &  & & &  & &  & &  & Exact\\
\hspace{.05cm} $\{\cdot, \textrm{FE}(\widetilde{\boldsymbol{X}})\}$ & 0.94 & 1.04 & 1.95 & 0.95 & 1.05 & 1.95 & 0.96 & 1.08 & 1.97 & 1.00 & 1.22 & 2.03 \\ 
\hspace{.05cm} $\{\cdot, \textrm{SR}(\widetilde{\boldsymbol{X}})\}$ & 0.26 & 0.25 & 0.43 & 0.31 & 0.35 & 0.64 & 0.42 & 0.56 & 1.04 & 0.72 & 1.03 & 1.94 \\ 
\hspace{.05cm} $\{\cdot, \textrm{PR}(\widetilde{\boldsymbol{X}})\}$ & {\bf 0.19} & {\bf 0.17} & {\bf 0.30} & {\bf 0.23} & {\bf 0.26} & {\bf 0.49} & {\bf 0.32} & {\bf 0.43} & {\bf 0.83} & 0.55 & 0.80 & 1.57 \\ 
% \hspace{.05cm} $\{\textrm{IPW}(\mathbb{R}^+, \widetilde{\boldsymbol{X}}), \cdot\}$ \\
% \hspace{.05cm} $\{\textrm{IPW}(\mathbb{R}^+, \widetilde{\boldsymbol{X}}), \textrm{DR}(\widetilde{\boldsymbol{X}}; w)\}$ \\ 
% \hspace{.05cm} $\{\textrm{IPW}(\mathbb{R}^+, \widetilde{\boldsymbol{X}}), \textrm{WR}(\widetilde{\boldsymbol{X}}; w)\}$ \\ 
\hspace{.05cm} $\{\textrm{LW}( \widetilde{\boldsymbol{X}}_{\textrm{null}}^{\mathsf{c}}), \cdot\}$ & 0.55 & 0.60 & 1.06 & 0.60 & 0.61 & 1.21 & 0.69 & 0.69 & 1.42 & 0.94 & 0.97 & 1.91 \\ 
\hspace{.05cm} $\{\textrm{LW}(\widetilde{\boldsymbol{X}}_{\textrm{null}}^{\mathsf{c}}), \textrm{FE}(\widetilde{\boldsymbol{X}})\}$ & 0.37 & 0.37 & 0.79 & 0.39 & 0.39 & 0.83 & 0.44 & 0.45 & 0.90 & 0.59 & 0.64 & 1.12 \\ 
\hspace{.05cm} $\{\textrm{LW}( \widetilde{\boldsymbol{X}}_{\textrm{null}}^{\mathsf{c}}), \textrm{WR}(\widetilde{\boldsymbol{X}})\}$ & 0.37 & 0.36 & 0.79 & 0.39 & 0.39 & 0.82 & 0.43 & 0.44 & 0.87 & 0.57 & {\bf 0.62} & {\bf 1.04} \\ 
\hspace{.05cm} $\{\textrm{SBW}(\widetilde{\boldsymbol{X}}_{\textrm{null}}^{\mathsf{c}}), \textrm{WR}(\widetilde{\boldsymbol{X}})\}$ & 0.24 & 0.23 & 0.71 & 0.26 & 0.30 & 0.76 & 0.34 & 0.46 & 0.88 & {\bf 0.54} & 0.82 & 1.26 \\ 
\hline
\end{tabular}
\vspace{.25cm}
\footnotesize{
\begin{flushleft}
\end{flushleft}
}
\end{table}
% \end{landscape}
\end{center}

\begin{center}
% \begin{landscape}
\begin{table}[htbp]
\caption{Absolute bias and RMSE based on 1,000 simulated data sets, averaged over 200 practices, for three target samples: all patients in the system, and patients from the practices near the center or at the boundary of the population. Two forms of covariates ($\boldsymbol{X}$ and $\widetilde{\boldsymbol{X}}$) are considered: $\boldsymbol{X}$ exact fits when the outcome is linear and under fits when the outcome is non-linear; $\widetilde{\boldsymbol{X}}$ over fits when the outcome is linear and exact fits when the outcome is quadratic.}
\label{tab_the_system_perspective_200}
\centering
\scriptsize
\setlength{\tabcolsep}{2pt}
\begin{tabular}{lrrrrrrrrrrrr}
  \hline
   & \multicolumn{12}{c}{Simulation setting} \\
  \cmidrule(lr){2-13} 
  &  \multicolumn{3}{c}{Linear} &  \multicolumn{3}{c}{Lowly non-linear} &  \multicolumn{3}{c}{Moderately non-linear} &  \multicolumn{3}{c}{Highly non-linear} \\
  \cmidrule(lr){2-4}  \cmidrule(lr){5-7} \cmidrule(lr){8-10} \cmidrule(lr){11-13}
     &  \multicolumn{3}{c}{Target} &  \multicolumn{3}{c}{Target} &  \multicolumn{3}{c}{Target} &  \multicolumn{3}{c}{Target} \\
Method & System & Center & ~Edge & System & Center & ~Edge & System & Center & ~Edge & System & Center & ~Edge \\
   \cmidrule(lr){1-1}  \cmidrule(lr){2-4}  \cmidrule(lr){5-7} \cmidrule(lr){8-10} \cmidrule(lr){11-13}
Bias (parsimonious) & Exact &  & & &  & &  & &  & Under \\
\hspace{.05cm} $\{\cdot, \textrm{FE}(\boldsymbol{X})\}$ & 0.88 & 0.93 & 1.93 & 0.87 & 0.93 & 1.93 & 0.88 & 0.92 & 1.92 & 0.89 & 0.92 & 1.92 \\ 
\hspace{.05cm} $\{\cdot, \textrm{SR}(\boldsymbol{X})\}$ & --  & -- & -- & -- & -- & -- & -- & -- & -- & -- & -- & --\\
\hspace{.05cm} $\{\cdot, \textrm{PR}(\boldsymbol{X})\}$ & {\bf 0.01} & {\bf 0.01} & {\bf 0.01} & 0.18 & 0.14 & {\bf 0.32} & 0.36 & 0.28 & 0.65 & 0.72 & 0.55 & 1.30 \\ 
% \hspace{.05cm} $\{\textrm{IPW}(\mathbb{R}^+, \boldsymbol{X}), \cdot\}$ \\
% \hspace{.05cm} $\{\textrm{IPW}(\mathbb{R}^+, \boldsymbol{X}), \textrm{DR}({\boldsymbol{X}}; w)\}$ \\ 
% \hspace{.05cm} $\{\textrm{IPW}(\mathbb{R}^+, \boldsymbol{X}), \textrm{WR}({\boldsymbol{X}}; w)\}$ \\ 
\hspace{.05cm} $\{\textrm{LW}( \boldsymbol{X}_{\textrm{null}}^{\mathsf{c}}), \cdot\}$ & 0.29 & 0.33 & 0.75 & 0.35 & 0.34 & 0.81 & 0.42 & 0.36 & 0.89 & 0.57 & 0.40 & 1.11 \\ 
\hspace{.05cm} $\{\textrm{LW}(\boldsymbol{X}_{\textrm{null}}^{\mathsf{c}}), \textrm{FE}(\boldsymbol{X})\}$ & 0.26 & 0.28 & 0.74 & 0.26 & 0.28 & 0.74 & 0.29 & 0.28 & 0.74 & 0.44 & 0.28 & 0.76 \\ 
\hspace{.05cm} $\{\textrm{LW}(\boldsymbol{X}_{\textrm{null}}^{\mathsf{c}}), \textrm{WR}({\boldsymbol{X}})\}$ & 0.26 & 0.28 & 0.74 & 0.26 & 0.28 & 0.74 & 0.29 & 0.28 & 0.74 & {\bf 0.43} & 0.28 & 0.74 \\ 
\hspace{.05cm} $\{\textrm{SBW}( \boldsymbol{X}_{\textrm{null}}^{\mathsf{c}}), \textrm{WR}({\boldsymbol{X}})\}$ & 0.03 & 0.07 & 0.47 & {\bf 0.14} & {\bf 0.08} & 0.47 & {\bf 0.27} & {\bf 0.14} & {\bf 0.49} & 0.54 & {\bf 0.26} & {\bf 0.58} \\ 
\cmidrule(lr){1-1}  \cmidrule(lr){2-4}  \cmidrule(lr){5-7} \cmidrule(lr){8-10} \cmidrule(lr){11-13}
Bias (comprehensive) & Over &  & & &  & &  & &  & Exact\\
\hspace{.05cm} $\{\cdot, \textrm{FE}(\widetilde{\boldsymbol{X}})\}$ & 0.87 & 0.93 & 1.93 & 0.87 & 0.93 & 1.93 & 0.87 & 0.92 & 1.93 & 0.87 & 0.92 & 1.93 \\ 
\hspace{.05cm} $\{\cdot, \textrm{SR}(\widetilde{\boldsymbol{X}})\}$ & --  & -- & -- & -- & -- & -- & -- & -- & -- & -- & -- & --\\
\hspace{.05cm} $\{\cdot, \textrm{PR}(\widetilde{\boldsymbol{X}})\}$  & {\bf 0.03} & {\bf 0.05} & {\bf 0.08} & {\bf 0.05} & {\bf 0.07} & {\bf 0.08} & {\bf 0.08} & {\bf 0.10} & {\bf 0.11} & {\bf 0.15} & {\bf 0.17} & {\bf 0.20} \\ 
% \hspace{.05cm} $\{\textrm{IPW}(\mathbb{R}^+, \widetilde{\boldsymbol{X}}), \cdot\}$ \\
% \hspace{.05cm} $\{\textrm{IPW}(\mathbb{R}^+, \widetilde{\boldsymbol{X}}), \textrm{DR}(\widetilde{\boldsymbol{X}}; w)\}$ \\ 
% \hspace{.05cm} $\{\textrm{IPW}(\mathbb{R}^+, \widetilde{\boldsymbol{X}}), \textrm{WR}(\widetilde{\boldsymbol{X}}; w)\}$ \\ 
\hspace{.05cm} $\{\textrm{LW}( \widetilde{\boldsymbol{X}}_{\textrm{null}}^{\mathsf{c}}), \cdot\}$ & 0.27 & 0.29 & 0.72 & 0.28 & 0.32 & 0.77 & 0.30 & 0.35 & 0.84 & 0.36 & 0.41 & 1.04 \\ 
\hspace{.05cm} $\{\textrm{LW}( \widetilde{\boldsymbol{X}}_{\textrm{null}}^{\mathsf{c}}), \textrm{FE}(\widetilde{\boldsymbol{X}})\}$ & 0.25 & 0.26 & 0.72 & 0.25 & 0.26 & 0.72 & 0.25 & 0.26 & 0.72 & 0.26 & 0.26 & 0.72 \\ 
\hspace{.05cm} $\{\textrm{LW}( \widetilde{\boldsymbol{X}}_{\textrm{null}}^{\mathsf{c}}), \textrm{WR}(\widetilde{\boldsymbol{X}})\}$  & 0.26 & 0.26 & 0.73 & 0.26 & 0.27 & 0.73 & 0.26 & 0.27 & 0.73 & 0.26 & 0.27 & 0.73 \\ 
\hspace{.05cm} $\{\textrm{SBW}( \widetilde{\boldsymbol{X}}_{\textrm{null}}^{\mathsf{c}}), \textrm{WR}(\widetilde{\boldsymbol{X}})\}$ & 0.22 & 0.31 & 0.94 & 0.22 & 0.31 & 0.94 & 0.22 & 0.31 & 0.94 & 0.22 & 0.31 & 0.94 \\  
\cmidrule(lr){1-1}  \cmidrule(lr){2-4}  \cmidrule(lr){5-7} \cmidrule(lr){8-10} \cmidrule(lr){11-13}
RMSE (parsimonious) & Exact &  & & &  & &  & &  & Under\\
\hspace{.05cm} $\{\cdot, \textrm{FE}(\boldsymbol{X})\}$ & 1.00 & 1.12 & 2.00 & 1.02 & 1.15 & 2.02 & 1.09 & 1.27 & 2.07 & 1.35 & 1.64 & 2.28 \\ 
\hspace{.05cm} $\{\cdot, \textrm{SR}(\boldsymbol{X})\}$ & --  & -- & -- & -- & -- & -- & -- & -- & -- & -- & -- & --\\
\hspace{.05cm} $\{\cdot, \textrm{PR}(\boldsymbol{X})\}$ & {\bf 0.21} & {\bf 0.23} & {\bf 0.31} & {\bf 0.42} & {\bf 0.50} & {\bf 0.69} & 0.75 & 0.91 & 1.27 & 1.44 & 1.77 & 2.48 \\ 
% \hspace{.05cm} $\{\textrm{IPW}(\mathbb{R}^+, \boldsymbol{X}), \cdot\}$ \\
% \hspace{.05cm} $\{\textrm{IPW}(\mathbb{R}^+, \boldsymbol{X}), \textrm{DR}({\boldsymbol{X}}; w)\}$ \\ 
% \hspace{.05cm} $\{\textrm{IPW}(\mathbb{R}^+, \boldsymbol{X}), \textrm{WR}({\boldsymbol{X}}; w)\}$ \\ 
\hspace{.05cm} $\{\textrm{LW}( \boldsymbol{X}_{\textrm{null}}^{\mathsf{c}}), \cdot\}$ & 0.68 & 0.86 & 1.30 & 0.77 & 0.98 & 1.51 & 0.93 & 1.21 & 1.80 & 1.36 & 1.80 & 2.49 \\ 
\hspace{.05cm} $\{\textrm{LW}(\boldsymbol{X}_{\textrm{null}}^{\mathsf{c}}), \textrm{FE}(\boldsymbol{X})\}$ & 0.45 & 0.54 & 0.96 & 0.55 & 0.68 & 1.09 & 0.75 & 0.95 & 1.36 & 1.23 & 1.62 & 2.08 \\ 
\hspace{.05cm} $\{\textrm{LW}( \boldsymbol{X}_{\textrm{null}}^{\mathsf{c}}), \textrm{WR}({\boldsymbol{X}})\}$ & 0.45 & 0.54 & 0.97 & 0.55 & 0.66 & 1.06 & 0.74 & 0.92 & 1.26 & {\bf 1.21} & {\bf 1.55} & {\bf 1.84} \\ 
\hspace{.05cm} $\{\textrm{SBW}(\boldsymbol{X}_{\textrm{null}}^{\mathsf{c}}), \textrm{WR}({\boldsymbol{X}})\}$ & 0.25 & 0.31 & 0.69 & {\bf 0.42} & 0.52 & 0.84 & {\bf 0.73} & {\bf 0.89} & {\bf 1.17} & 1.38 & 1.69 & 1.99 \\ 
\cmidrule(lr){1-1}  \cmidrule(lr){2-4}  \cmidrule(lr){5-7} \cmidrule(lr){8-10} \cmidrule(lr){11-13}
RMSE (comprehensive) & Over &  & & &  & &  & &  & Exact\\
\hspace{.05cm} $\{\cdot, \textrm{FE}(\widetilde{\boldsymbol{X}})\}$ & 1.00 & 1.12 & 2.00 & 1.00 & 1.13 & 2.01 & 1.02 & 1.18 & 2.04 & 1.09 & 1.35 & 2.16 \\ 
\hspace{.05cm} $\{\cdot, \textrm{SR}(\widetilde{\boldsymbol{X}})\}$ & --  & -- & -- & -- & -- & -- & -- & -- & -- & -- & -- & --\\
\hspace{.05cm} $\{\cdot, \textrm{PR}(\widetilde{\boldsymbol{X}})\}$ & 1.26 & 2.01 & 2.76 & 1.50 & 2.44 & 3.17 & 2.00 & 3.38 & 4.32 & 3.35 & 5.56 & 7.58 \\ 
% \hspace{.05cm} $\{\textrm{IPW}(\mathbb{R}^+, \widetilde{\boldsymbol{X}}), \cdot\}$ \\
% \hspace{.05cm} $\{\textrm{IPW}(\mathbb{R}^+, \widetilde{\boldsymbol{X}}), \textrm{DR}(\widetilde{\boldsymbol{X}}; w)\}$ \\ 
% \hspace{.05cm} $\{\textrm{IPW}(\mathbb{R}^+, \widetilde{\boldsymbol{X}}), \textrm{WR}(\widetilde{\boldsymbol{X}}; w)\}$ \\ 
\hspace{.05cm} $\{\textrm{LW}( \widetilde{\boldsymbol{X}}_{\textrm{null}}^{\mathsf{c}}), \cdot\}$ & 0.85 & 1.05 & 1.42 & 0.91 & 1.14 & 1.61 & 1.04 & 1.31 & 1.88 & 1.42 & 1.80 & 2.52 \\ 
\hspace{.05cm} $\{\textrm{LW}( \widetilde{\boldsymbol{X}}_{\textrm{null}}^{\mathsf{c}}), \textrm{FE}(\widetilde{\boldsymbol{X}})\}$ & 0.54 & 0.64 & 1.02 & 0.58 & 0.69 & 1.07 & 0.65 & 0.78 & 1.16 & 0.86 & 1.06 & 1.45 \\ 
\hspace{.05cm} $\{\textrm{LW}(\widetilde{\boldsymbol{X}}_{\textrm{null}}^{\mathsf{c}}), \textrm{WR}(\widetilde{\boldsymbol{X}})\}$ & 0.54 & 0.64 & {\bf 1.01} & 0.57 & 0.67 & {\bf 1.05} & 0.63 & {\bf 0.76} & {\bf 1.13} & 0.84 & {\bf 1.01} & {\bf 1.37} \\ 
\hspace{.05cm} $\{\textrm{SBW}( \widetilde{\boldsymbol{X}}_{\textrm{null}}^{\mathsf{c}}), \textrm{WR}(\widetilde{\boldsymbol{X}})\}$ & {\bf 0.44} & {\bf 0.54} & 1.09 & {\bf 0.47} & {\bf 0.61} & 1.14 & {\bf 0.54} & 0.78 & 1.26 & {\bf 0.78} & 1.25 & 1.65 \\ 
\hline
\end{tabular}
\vspace{.25cm}
\footnotesize{
\begin{flushleft}
\end{flushleft}
}
\end{table}
% \end{landscape}
\end{center}

%\singlespacing
\begin{table}[htbp]
\small
\setlength{\tabcolsep}{4pt}
\caption{Covariate profiles for the system and three hypothetical patients.}
\begin{center}
\label{tab_profiles}
\begin{tabular}{lrrrrrr}
\hline
& System & Patient 1 & Patient 2 & Patient 3\\ 
\hline
Age & 76.10 & 70.77 & 78.34 & 78.34\\ 
Sex (Male) & 0.34 & 1 & 0 & 0 \\ 
Sex (Female) & 0.66 & 0 & 1 & 1\\
Race (White) & 0.82 & 1 & 1 & 0\\ 
Race (Black) & 0.08 & 0 & 0 & 1\\ 
Race (Other) & 0.10 & 0 & 0 & 0\\  
Marital status (Married) & 0.52 & 0 & 1 & 1\\ 
Marital status (Unmarried)  & 0.44 & 1 & 0 & 0\\ 
Percentage of residents without a high school education & 13.09 & 27.75 & 13.57 &  13.57\\
Census-tract median household income & 1.08 & 0.60 &  0.89 &  0.89\\ 
Charlson Comorbidity Index (0) & 0.34 & 0 & 0 & 0\\ 
Charlson Comorbidity Index (1) & 0.26 & 0 & 0 & 0\\ 
Charlson Comorbidity Index (2) & 0.16 & 0 & 0 & 0\\ 
Charlson Comorbidity Index ($\geq$3) & 0.24 & 1 & 1 & 1\\ 
Cancer type (Breast) & 0.35 & 0 & 0 & 0\\ 
Cancer type (Colorectal) & 0.17 & 0 & 0 & 0\\ 
Cancer type (Lung) & 0.34 & 1 & 0 & 0\\ 
Cancer type (Ovary) & 0.02 & 0 & 0 & 0\\ 
Cancer type (Pancreas) & 0.07 & 0 & 1 & 1\\ 
Cancer type (Prostate)& 0.05 & 0 & 0 & 0\\ 
Cancer stage (1) & 0.27 & 0 & 1 & 1\\ 
Cancer stage (2) & 0.22 & 0 & 0 & 0\\ 
Cancer stage (3) & 0.20 & 1 & 0 & 0\\ 
Cancer stage (4) & 0.27 & 0 & 1 & 1\\ 
\hline
\end{tabular}
\end{center}
\end{table}

\begin{table}[!htbp]
\centering
\caption{Changes in quintile rankings of practices for the profiles of patients 2 and 3.}
\label{tab_rankings}
\begin{tabular}{lrrrrr}
\hline
Patient 2 / 3 & [1,120] & (120,240] & (240,361] & (361,480] & (480,600] \\ 
\hline
[1,120] &  33 &  32 &  20 &  13 &  22 \\ 
(120,240] &  23 &  36 &  29 &  15 &  17 \\ 
(240,361] &  23 &  24 &  39 &  17 &  17 \\ 
(361,480] &  21 &  18 &  19 &  33 &  29 \\ 
(480,600] &  20 &  10 &  13 &  42 &  35 \\ 
\hline
\end{tabular}
\end{table}

%%%%%%%%%%%%%%%%%%%%
%%%%%%%%%%%%%%%%%%%%
%%%%%%%%%%%%%%%%%%%%
% \clearpage
% \appendix
% \section*{Supplementary Materials} 

% \input{sec_supp}

%\bibliographystyle{asa}

\end{document}

\pagestyle{plain}

\newtheoremstyle{mystyle}% name
{\topsep}% Space above
{\topsep}% Space below
{\it}% Body font
{}% Indent amount
{\bf}% Theorem head font
{.}%Punctuation after theorem head
{.5em}%Space after theorem head
{}% theorem head spec
\theoremstyle{mystyle}
\newtheorem{assumptionex}{Assumption}
\newenvironment{assumption}
  {\pushQED{\qed}\renewcommand{\qedsymbol}{}\assumptionex}
  {\popQED\endassumptionex}
\newtheorem{assumptionexp}{Assumption}
\newenvironment{assumptionp}
  {\pushQED{\qed}\renewcommand{\qedsymbol}{}\assumptionexp}
  {\popQED\endassumptionexp}
\renewcommand{\theassumptionexp}{\arabic{assumptionexp}$'$}

\newtheorem{assumptionexpp}{Assumption}
\newenvironment{assumptionpp}
  {\pushQED{\qed}\renewcommand{\qedsymbol}{}\assumptionexpp}
  {\popQED\endassumptionexpp}
\renewcommand{\theassumptionexpp}{\arabic{assumptionexpp}$''$}

\newtheorem{assumptionexppp}{Assumption}
\newenvironment{assumptionppp}
  {\pushQED{\qed}\renewcommand{\qedsymbol}{}\assumptionexppp}
  {\popQED\endassumptionexppp}
\renewcommand{\theassumptionexppp}{\arabic{assumptionexppp}$'''$}

\renewcommand{\arraystretch}{1.3}

\newcommand\carl[1]{\cmnt{#1}{Carl}}
\newcommand\ambarish[1]{\cmnt{#1}{Ambarish}}
\newcommand\jose[1]{\cmnt{#1}{Jose}}

\newcommand{\argmin}{\mathop{\mathrm{argmin}}}
\makeatletter
\newcommand{\grande}{\bBigg@{2.25}}
\newcommand{\enorme}{\bBigg@{5}}

\newcommand{\blind}{0}

\newcommand{\tit}{%Revisiting and Extending the Finite Selection Model for Experimental Design
Supplementary Materials for\\
Targeted Quality Measurement of Health Care Providers}

\if0\blind

%{\title{\tit\thanks{For comments and suggestions, we thank...}}
{\title{\tit\thanks{
This research was supported by Arnold Ventures. Jos\'e Zubizarreta also acknowledges support by awards ME-2019C1-16172 and ME-2022C1- 25648 from the Patient-Centered Outcomes Research Institute (PCORI).}}
\author{Yige Li
\thanks{Department of Health Care Policy, Harvard University.
email: \url{yige_li@hcp.med.harvard.edu}.},
\and 
Nancy L. Keating
\thanks{Department of Health Care Policy, Harvard University; Division of General Internal Medicine, Brigham and Women's Hospital. email: \url{keating@hcp.med.harvard.edu}.},
\and
Mary Beth Landrum
\thanks{Department of Health Care Policy, Harvard University. 
email: \url{landrum@hcp.med.harvard.edu}.
}, 
\and 
Jos\'{e} R. Zubizarreta\thanks{Departments of Health Care Policy, Biostatistics, and Statistics, Harvard University, 180 Longwood Avenue, Office 307-D, Boston, MA 02115; email: \url{zubizarreta@hcp.med.harvard.edu}.}
}

\date{} 

\maketitle
}\fi

\if1\blind
\title{\tit}
\date{} 
\maketitle
\fi

%\vspace*{.3in}

%\begin{center}
%\noindent Keywords:
%{Causal inference; Covariate balance; Experimental design; Multi-valued treatments}
%\end{center}
\clearpage
\doublespacing

%\singlespacing
%\pagebreak
%\tableofcontents
%\pagebreak
%\doublespacing

%\section{Summary and remarks}
%\vspace{1cm}
%\pagebreak
%\onehalfspacing
%\bibliographystyle{asa}
%\bibliography{mybibliography21}
%\nobibliography{append}

%%%%%%%%%%%%%%%%%%%%
%%%%%%%%%%%%%%%%%%%%
%%%%%%%%%%%%%%%%%%%%
\clearpage
\appendix
%\section*{Supplementary Materials} 
\section*{}

\addcontentsline{toc}{section}{Supplementary Materials}
\renewcommand{\thesubsection}{\Alph{subsection}}
\setcounter{table}{0}
\renewcommand{\thetable}{A\arabic{table}}

\renewcommand{\theequation}{A\arabic{equation}}
\setcounter{figure}{0}
\renewcommand\thefigure{A\arabic{figure}}

\setcounter{theorem}{0}
\renewcommand\thetheorem{A\arabic{theorem}}

\setcounter{lemma}{0}
\renewcommand\thelemma{A\arabic{lemma}}

\subsection{Notation}

\begin{table}[!htbp]
\caption{Notation}
\label{tab_notation}
\begin{center}
\begin{tabular}{r c p{11cm} }
\hline
$i$ & $\triangleq$ & Index of patient $i = 1, ..., n$ \\
$p$ & $\triangleq$ & Practice assignment $p = 1, ..., P$	\\
$\boldsymbol{X}_{i}$ & $\triangleq$ & Observed covariates of patient $i$\\
%$\boldsymbol{X}_{\textrm{inter}}$ & $\triangleq$ & Covariates in $\boldsymbol{X}$ whose empirical convex hull contains the target covariate profile for all the practices	\\
%$\boldsymbol{X}_{\textrm{extra}}$ & $\triangleq$ & Covariates in $\boldsymbol{X}$ whose empirical convex hull does not contain the target covariate profile for all the practices	\\
$\boldsymbol{X}_{\textrm{null}}$ & $\triangleq$ & Covariates in $\boldsymbol{X}$ that are null for at least one practice	\\
$\boldsymbol{X}_{\textrm{null}}^{\mathsf{c}}$ & $\triangleq$ & Covariates that are not null for all practices, $\boldsymbol{X}_{\textrm{null}} \cup \boldsymbol{X}_{\textrm{null}}^{\mathsf{c}} = \boldsymbol{X}$	\\
$\widetilde{\boldsymbol{X}}$ & $\triangleq$ & Transformations of the observed covariates $\boldsymbol{X}$	\\
$\boldsymbol{U}$ & $\triangleq$ & Unobserved covariates	\\
$\indicator{p,i}$ & $\triangleq$ & Practice assignment indicator of patient $i$, $p = 1, ..., P$	\\
% $S_{p,i}$ & $\triangleq$ & Selection indicator of patient $i$ into practice $p = 1, ..., P$	\\
$Y_i(p)$ & $\triangleq$ & Potential outcome of patient $i$ under practice assignment $p$	\\
$Y_i$ & $\triangleq$ & Observed outcome of patient $i$, $Y_i = \sum_{p = 1}^{P} \indicator{p,i} Y_i(p)$	\\
{$\textrm{LW}(a,b)$} & $\triangleq$ & Inverse probability weighting with logistic regression model, domain of weights $a$ (e.g., $a = \mathbb{R}^+$), and adjustment for the covariates in $b$ (e.g., $b = \boldsymbol{X}_{\textrm{null}}^{\mathsf{c}}, \boldsymbol{X}$) \\
$\textrm{SBW}(a, b)$ & $\triangleq$ & Stable balancing weighting with constraints $a$ on the weights to require non-negative weights and interpolation adjustments ($a = \mathbb{R}^+_0$) or unrestricted weights and extrapolation adjustments ($a = \mathbb{R}$), and balance on the covariates in $b$ (e.g., $b = \boldsymbol{X}_{\textrm{null}}^{\mathsf{c}}, \boldsymbol{X}$)	\\
% $\textrm{H}$ & $\triangleq$ & Hajek estimator	\\
$\textrm{FE}(b)$ & $\triangleq$ & Fixed-effects regression with adjustment for the covariates in $b$ (e.g., $b = \boldsymbol{X}$) plus practice indicators	$\indicator{1},...,\indicator{P}$\\
$\textrm{SR}(b)$ & $\triangleq$ & Practice-stratified regression with adjustment for the covariates in $b$ (e.g., $b = \boldsymbol{X}$)	\\
{$\textrm{PR}(b)$} & $\triangleq$ & Practice-pooled regression with adjustment for the covariates in $b$ (e.g., $b = \boldsymbol{X}$) plus practice indicators $\indicator{1},...,\indicator{P}$ plus interactions of $\boldsymbol{X}_{\text{null}}^{\mathsf c}$ and practice indicators\\
% {$\textrm{FE}(b)$} & $\triangleq$ & Doubly robust estimator with adjustment for the covariates in $b$ (e.g., $b = \boldsymbol{X}$) plus practice indicators in the outcome model  \\
$\textrm{WR}(b)$ & $\triangleq$ & Weighted regression with adjustment for the covariates in $b$ (e.g., $b = \boldsymbol{X}$) plus practice indicators	$\indicator{1},...,\indicator{P}$ \\
\hline
\end{tabular}
\end{center}
\end{table}

%%%%%%%%%%%%%%%%%%%%%%%%%%%%%%%%%%%%%%%%%%%
%%%%%%%%%%%%%%%%%%%%%%%%%%%%%%%%%%%%%%%%%%%
%%%%%%%%%%%%%%%%%%%%%%%%%%%%%%%%%%%%%%%%%%%
\clearpage
\subsection{Methods considered in the empirical evaluations}
\begin{table}[!htbp]\caption{Methods considered in the empirical evaluations, $\{weighting, modeling\}$}
\label{tab_methods}
\begin{center}
\begin{tabular}{r c p{9cm} }
\hline
% $\{\cdot, \textrm{FE}(\boldsymbol{X})\}$ & $\triangleq$ & Fixed-effects regression of the outcome on $\boldsymbol{X}$ and the practice indicators; we fit one single model for all the practices to then predict the patient outcomes and average them within each of the practices	\\
$\{\cdot, \textrm{FE}(\widetilde{\boldsymbol{X}})\}$ & $\triangleq$ & Fixed-effects regression of the outcome on $\widetilde{\boldsymbol{X}}$ and the practice indicators; we fit one single model for all the practices, to then predict the patient outcomes and average them within each of the practices	\\
% $\{\cdot, \textrm{SR}(\boldsymbol{X})\}$ & $\triangleq$ & Stratified regression of the outcome on $\boldsymbol{X}$; we fit separate models for each practice, to then predict the patient outcomes and average them within each of the practices	\\
$\{\cdot, \textrm{SR}(\widetilde{\boldsymbol{X}})\}$ & $\triangleq$ & Stratified regression of the outcome on $\widetilde{\boldsymbol{X}}$; we fit separate models for each practice, to then predict the patient outcomes and average them within each of the practices	\\
{$\{\cdot, \textrm{PR}(\widetilde{\boldsymbol{X}})\}$} & $\triangleq$ & Pooled regression of the outcome on $\widetilde{\boldsymbol{X}}$; we fit one model for all practice at the same time, to then predict the patient outcomes and average them within each of the practices	\\
{$\{\textrm{LW}(\mathbb{R}^+, \widetilde{\boldsymbol{X}}_{\textrm{null}}^{\mathsf{c}}), \cdot\}$} & $\triangleq$ & Inverse probability weighting H\'ajek 
 estimator with the propensity scores estimated by logistic regression model on $\widetilde{\boldsymbol{X}}_{\textrm{null}}^{\mathsf{c}}$\\
{$\{\textrm{LW}(\mathbb{R}^+, \widetilde{\boldsymbol{X}}_{\textrm{null}}^{\mathsf{c}}), \textrm{FE}(\widetilde{\boldsymbol{X}})\}$} & $\triangleq$ & Bias-corrected estimator with propensity score model on $\widetilde{\boldsymbol{X}}_{\textrm{null}}^{\mathsf{c}}$ and fixed-effects outcome model on $\widetilde{\boldsymbol{X}}$\\
{$\{\textrm{LW}(\mathbb{R}^+, \widetilde{\boldsymbol{X}}_{\textrm{null}}^{\mathsf{c}}), \textrm{WR}(\widetilde{\boldsymbol{X}})\}$} & $\triangleq$ & Weighted regression of the outcome on $\widetilde{\boldsymbol{X}}$ and the practice indicators with the weights $\textrm{LW}(\mathbb{R}^+, \widetilde{\boldsymbol{X}}_{\textrm{null}}^{\mathsf{c}})$; we fit one single model for all the practices, to then predict the patient outcomes and average them within each of the practices\\
$\{\textrm{SBW}(\mathbb{R}^+_0, \widetilde{\boldsymbol{X}}_{\textrm{null}}^{\mathsf{c}}), \textrm{WR}(\widetilde{\boldsymbol{X}})\}$ & $\triangleq$ & Weighted regression of the outcome on $\widetilde{\boldsymbol{X}}$ and the practice indicators with the weights $\textrm{SBW}(\mathbb{R}_0^+, \widetilde{\boldsymbol{X}}_{\textrm{null}}^{\mathsf{c}})$; we fit one single model for all the practices, to then predict the patient outcomes and average them within each of the practices \\
{$\{\textrm{SBW}(\mathbb{R}^+_0, \widetilde{\boldsymbol{X}}), \cdot\}$} & $\triangleq$ & SBW with non-negative 
weights and balance conditions for $\widetilde{\boldsymbol{X}}$ \\
\hline
\end{tabular}
\end{center}
\end{table}

%%%%%%%%%%%%%%%%%%%%%%%%%%%%%%%%%%%%%%%%%%%
%%%%%%%%%%%%%%%%%%%%%%%%%%%%%%%%%%%%%%%%%%%
%%%%%%%%%%%%%%%%%%%%%%%%%%%%%%%%%%%%%%%%%%%
\clearpage
% \subsection{Extrapolation with 30 covariates}
% \begin{figure}[!htbp]
% \begin{center}
% \caption{Extrapolation with 30 covariates.}\label{fig_extent_of_extrapolation2}
% \includegraphics[scale=0.275]{graphics/a_plot_flag_30_2.png}
% \end{center}
% \end{figure}

%%%%%%%%%%%%%%%%%%%%%%%%%%%%%%%%%%%%%%%%%%%
%%%%%%%%%%%%%%%%%%%%%%%%%%%%%%%%%%%%%%%%%%%
%%%%%%%%%%%%%%%%%%%%%%%%%%%%%%%%%%%%%%%%%%%
{\subsection{Methods for layered case-mix adjustments}}

\subsubsection{Weighted linear regression}
\label{sec_estimation}

To accommodate covariates with null values in certain practices, we assume the following underlying structure for the outcomes,
\begin{equation}
     \mathrm{E}[Y|\boldsymbol{X}, p] = \boldsymbol{\alpha}^\top \widetilde{\boldsymbol{X}}_{\text{null}} +\alpha_p +  \boldsymbol{\beta}_p^\top \widetilde{\boldsymbol{X}}^{\mathsf{c}}_{\text{null}},
\end{equation}
where only non-null covariates $\widetilde{\boldsymbol X}^{\mathsf{c}}_{\text{null}}$ can be effect modifiers. 
Thus, the estimand  $\mu_{p}(\boldsymbol{x}^*) = \boldsymbol{\alpha}^\top \widetilde{\boldsymbol{x}}^{*}_{\text{null}}+ \alpha_{p}  + \boldsymbol{\beta}_{p}^\top \widetilde{\boldsymbol{x}}^{*\mathsf{c}}_{\text{null}}$; the contrast $\mu_{p'}(\boldsymbol{x}^*) - \mu_{p''}(\boldsymbol{x}^*) = \alpha_{p'} -\alpha_{p''} + (\beta_{p'}- \beta_{p''})^\top \widetilde{\boldsymbol{x}}^{*\mathsf{c}}_{\text{null}}$.

To increase the robustness and stability of the estimator, we employ a two-step approach.
    First, we balance $\widetilde{\boldsymbol{X}}^{\mathsf{c}}_{\text{null}}$ of each practice to its target $\widetilde{\boldsymbol{x}}^{*\mathsf{c}}_{\text{null}}$. 
    When considering the total population as the target, we set $\widetilde{\boldsymbol{x}}^{*\mathsf{c}}_{\text{null}} = \mathrm E_{n}[\widetilde{\boldsymbol{X}}^{\mathsf{c}}_{\text{null}}]$, which corresponds to the sample mean. 
    After obtaining the weights $\boldsymbol{w}_p$ for $p = 1,...,P$,  we then fit a weighted linear regression of the outcome vector $\boldsymbol{Y}$ on $n\times (L+P)$ design matrix $\textbf{Z}:=(\widetilde{\textbf{X}}_{\text{null}},\widetilde{\textbf{X}}^{\mathsf{c}}_{\text{null}},\indicatorbf{1},...,\indicatorbf{P})$.
    This yields estimates for the corresponding coefficients $\widehat{\boldsymbol{\alpha}}$,  $\widehat{\boldsymbol{\beta}}$, and $(\widehat{\alpha}_1,...,\widehat{\alpha}_P)$.
    Consequently, we compute $\widehat{\mu}_p(\boldsymbol{x}^*) := \widehat{\boldsymbol{\alpha}}^\top \widetilde{\boldsymbol{x}}^*_{\text{null}}  + \widehat{\boldsymbol{\beta}}^\top \widetilde{\boldsymbol{x}}^{*\mathsf{c}}_{\text{null}}+ \widehat{\alpha}_p$. 
Therefore, the estimator of the contrast $\mu_{p'}(\boldsymbol{x}^*) - \mu_{p''}(\boldsymbol{x}^*)$ is $\widehat{\alpha}_{p'} -\widehat{\alpha}_{p''}$.

With this procedure, the implied weights will be 
$\boldsymbol{\omega}_p^\top := (\widetilde{\boldsymbol{x}}_{\text{null}}^{*\top} ,\widetilde{\boldsymbol{x}}^{*\mathsf{c}\top}_{\text{null}},\boldsymbol{e}_p^\top)(\textbf{Z}^\top \textbf{W} \textbf{Z})^{-1}(\textbf{Z}^\top \textbf{W})$,
where $\boldsymbol{e}_p = (0,...,1,...,0)^\top$ is a length-$P$ vector with a 1 in the $p$-th element and 0's elsewhere, $\textbf{W}$ is an $n\times n$ matrix of normalized weights $\textbf{W} := \sum^{P}_{q = 1}\frac{n_q}{n}\text{diag}(\boldsymbol{w}_q)$. 
We are going to show the robustness and stability properties of this estimator.

The weighted estimator can be written as
\begin{align*}
    \boldsymbol{\omega}_p^\top \boldsymbol{Y} 
     &=  \boldsymbol{\omega}_p^\top \widetilde{\textbf{X}}_{\text{null}} \boldsymbol{\alpha} +\boldsymbol{\omega}_p^\top \big[\sum_{q = 1}^P \indicatorbf{q}\alpha_q+\sum_{q = 1}^P\textbf{1}_q \widetilde{\textbf{X}}^{\mathsf{c}}_{\text{null}}\boldsymbol{\beta}_q\big]+\boldsymbol{\omega}_p^\top \boldsymbol{\varepsilon} \\
     &= \boldsymbol{\omega}_p^\top \widetilde{\textbf{X}}_{\text{null}}\boldsymbol{\alpha} +\boldsymbol{\omega}_p^\top\big[\sum_{q = 1}^P \indicatorbf{q}\alpha_q+ \widetilde{\textbf{X}}^{\mathsf{c}}_{\text{null}}\boldsymbol{\beta}\big]+\boldsymbol{\omega}_p^\top \big[\sum_{q = 1}^P \textbf{1}_q \widetilde{\textbf{X}}^{\mathsf{c}}_{\text{null}}\boldsymbol{\beta}_q-\widetilde{\textbf{X}}^{\mathsf{c}}_{\text{null}} \boldsymbol{\beta} \big] +\boldsymbol{\omega}_p^\top \boldsymbol{\varepsilon}\\
     &= \boldsymbol{\alpha}^\top \widetilde{\boldsymbol{x}}^*_{\text{null}}+\alpha_p+\boldsymbol{\beta}^\top \widetilde{\boldsymbol{x}}^{*\mathsf{c}}_{\text{null}} +\boldsymbol{\omega}_p^\top \big[\sum_{q = 1}^P \textbf{1}_q \widetilde{\textbf{X}}^{\mathsf{c}}_{\text{null}}\boldsymbol{\beta}_q-\sum_{q = 1}^P \textbf{1}_q \widetilde{\textbf{X}}^{\mathsf{c}}_{\text{null}}  \boldsymbol{\beta}\big]+\boldsymbol{\omega}_p^\top \boldsymbol{\varepsilon},
\end{align*}
where $\textbf{1}_q := \text{diag}(\indicatorbf{q})$, $\boldsymbol{\beta} = \sum_{q = 1}^P \frac{n_q}{n} \boldsymbol{\beta}_q$.
Therefore, the estimation error is $\boldsymbol{\omega}_p^\top [\sum_{q = 1}^P \textbf{1}_q \widetilde{\textbf{X}}^{\mathsf{c}}_{\text{null}} (\boldsymbol{\beta}_q-\boldsymbol{\beta})] -  (\boldsymbol{\beta}_p-\boldsymbol{\beta})^\top \widetilde{\boldsymbol{x}}^{*\mathsf{c}}_{\text{null}} +\boldsymbol{\omega}_p^\top \boldsymbol{\varepsilon}$. To further decompose the error, $\boldsymbol{\omega}_p$ is decomposed into two components, balancing weights and leveraging weights, % interpolating extrapolating
\begin{align*}
    \boldsymbol{\omega}_p^\top 
    &= \big\{(\widetilde{\boldsymbol{x}}_{\text{null}}^{*\top},\widetilde{\boldsymbol{x}}_{\text{null}}^{*\mathsf{c}\top},\boldsymbol{e}_p^\top)(\textbf{Z}^\top \textbf{W} \textbf{Z})^{-1}\textbf{Z}^\top-\frac{n}{n_p}\indicatorbf{p}^\top+\frac{n}{n_p}\indicatorbf{p}^\top\big\} \textbf{W}\\
    &= \big\{\textbf{Z}(\textbf{Z}^\top \textbf{W} \textbf{Z})^{-1}(\widetilde{\boldsymbol{x}}_{\text{null}}^{*\top},\widetilde{\boldsymbol{x}}_{\text{null}}^{*\mathsf{c}\top},\boldsymbol{e}_p^\top)^\top-\frac{n}{n_p}\indicatorbf{p}\big\}^\top \textbf{W} + \frac{n}{n_p}\indicatorbf{p}^\top \textbf{W}.
\end{align*}
Since the balancing weights $\frac{n}{n_p}\indicatorbf{p}^\top \textbf{W}$ lets $(\frac{n}{n_p}\indicatorbf{p}^\top \textbf{W})^\top[\sum_{q = 1}^P \textbf{1}_q \widetilde{\textbf{X}}^{\mathsf{c}}_{\text{null}}(\boldsymbol{\beta}_q-\boldsymbol{\beta})] - (\boldsymbol{\beta}_p-\boldsymbol{\beta})^\top \widetilde{\boldsymbol{x}}^{*\mathsf{c}}_{\text{null}} = 0$, it remains to show the leveraging weights $\big\{\textbf{Z}(\textbf{Z}^\top \textbf{W} \textbf{Z})^{-1}(\widetilde{\boldsymbol{x}}_{\text{null}}^{*\top},\widetilde{\boldsymbol{x}}_{\text{null}}^{*\mathsf{c}\top},\boldsymbol{e}_p^\top)^\top-\frac{n}{n_p}\indicatorbf{p}\big\}^\top \textbf{W}$ can maintain balance at some extent. We let $\left[\begin{matrix}
(\textbf{M}_{11})_{L\times L} & (\textbf{M}_{12})_{L\times P} \\
(\textbf{M}_{21})_{P\times L} & (\textbf{M}_{22})_{P\times P} \\
\end{matrix}\right]$ denote $\textbf{Z}^\top \textbf{W} \textbf{Z}$.
\begin{align*}
& \textbf{Z}(\textbf{Z}^\top \textbf{W} \textbf{Z})^{-1}(\widetilde{\boldsymbol{x}}^{*\top}_{\text{null}} ,\widetilde{\boldsymbol{x}}^{*\mathsf{c}\top}_{\text{null}},\boldsymbol{e}_p^\top)^\top =[\widetilde{\textbf{X}}_{\text{null}},\widetilde{\textbf{X}}^{\mathsf{c}}_{\text{null}},\indicatorbf{1},...,\indicatorbf{P}]\left[\begin{matrix}
\textbf{M}_{11} & \textbf{M}_{12} \\
\textbf{M}_{21} & \textbf{M}_{22} \\
\end{matrix}\right]^{-1}(\widetilde{\boldsymbol{x}}^{*\top}_{\text{null}},\widetilde{\boldsymbol{x}}^{*\mathsf{c}\top}_{\text{null}},\boldsymbol{e}_p^\top)^\top 
\end{align*}
\begin{align*}
=& [\widetilde{\textbf{X}}_{\text{null}},\widetilde{\textbf{X}}^{\mathsf{c}}_{\text{null}},\indicatorbf{1},...,\indicatorbf{P}]
\left[\begin{matrix}
(\textbf{M}_{11}-\textbf{M}_{12}\textbf{M}_{22}^{-1}\textbf{M}_{21})^{-1} & 0 \\
0 & (\textbf{M}_{22}-\textbf{M}_{21}\textbf{M}_{11}^{-1}\textbf{M}_{12})^{-1} \\
\end{matrix}\right] \cdot \\
&
\left[\begin{matrix}
\textbf{I}_P & -\textbf{M}_{12}\textbf{M}_{22}^{-1} \\
-\textbf{M}_{21}\textbf{M}_{11}^{-1} & \textbf{I}_P \\
\end{matrix}\right]
\left[\begin{matrix}
\widetilde{\boldsymbol{x}}^*_{\text{null}} \\
\widetilde{\boldsymbol{x}}^{*\mathsf{c}}_{\text{null}} \\
\boldsymbol{e}_p \\
\end{matrix}\right] \\
=& 
\left[\begin{matrix}
[\widetilde{\textbf{X}}_{\text{null}},\widetilde{\textbf{X}}^{\mathsf{c}}_{\text{null}}](\textbf{M}_{11}-\textbf{M}_{12}\textbf{M}_{22}^{-1}\textbf{M}_{21})^{-1} & [\indicatorbf{1},...,\indicatorbf{P}](\textbf{M}_{22}-\textbf{M}_{21}\textbf{M}_{11}^{-1}\textbf{M}_{12})^{-1} \\
\end{matrix}\right]\left[\begin{matrix}
\left[\begin{matrix}
\widetilde{\boldsymbol{x}}^*_{\text{null}} \\
\widetilde{\boldsymbol{x}}^{*\mathsf{c}}_{\text{null}}
\end{matrix}\right] -\textbf{M}_{12}\textbf{M}_{22}^{-1}\boldsymbol{e}_p \\
\boldsymbol{e}_p-\textbf{M}_{21}\textbf{M}_{11}^{-1}\left[\begin{matrix}
\widetilde{\boldsymbol{x}}^*_{\text{null}} \\
\widetilde{\boldsymbol{x}}^{*\mathsf{c}}_{\text{null}}
\end{matrix}\right] \\
\end{matrix}\right] \\
=& 
[\widetilde{\textbf{X}}_{\text{null}},\widetilde{\textbf{X}}^{\mathsf{c}}_{\text{null}}](\textbf{M}_{11}-\textbf{M}_{12}\textbf{M}_{22}^{-1}\textbf{M}_{21})^{-1} \left(\left[\begin{matrix}
\widetilde{\boldsymbol{x}}^*_{\text{null}} \\
\widetilde{\boldsymbol{x}}^{*\mathsf{c}}_{\text{null}}
\end{matrix}\right] -\textbf{M}_{12}\frac{n}{n_p}\boldsymbol{e}_p\right) +\\
& [\indicatorbf{1},...,\indicatorbf{P}](\textbf{M}_{22}-\textbf{M}_{21}\textbf{M}_{11}^{-1}\textbf{M}_{12})^{-1}
\left(-\textbf{M}_{21}\textbf{M}_{11}^{-1}\left[\begin{matrix}
\widetilde{\boldsymbol{x}}^*_{\text{null}} \\
\widetilde{\boldsymbol{x}}^{*\mathsf{c}}_{\text{null}}
\end{matrix}\right]\right) +\\
& [\indicatorbf{1},...,\indicatorbf{P}]\{(\textbf{M}_{22}-\textbf{M}_{21}\textbf{M}_{11}^{-1}\textbf{M}_{12})^{-1}-\textbf{M}_{22}^{-1}\}\boldsymbol{e}_p+[\indicatorbf{1},...,\indicatorbf{P}]\frac{n}{n_p}\boldsymbol{e}_p \\
=& 
[\widetilde{\textbf{X}}_{\text{null}},\widetilde{\textbf{X}}^{\mathsf{c}}_{\text{null}}](\textbf{M}_{11}-\textbf{M}_{12}\textbf{M}_{22}^{-1}\textbf{M}_{21})^{-1}\left(\left[\begin{matrix}
\widetilde{\boldsymbol{x}}^*_{\text{null}} \\
\widetilde{\boldsymbol{x}}^{*\mathsf{c}}_{\text{null}}
\end{matrix}\right] -\left[\begin{matrix}
\mathrm{E}_{\frac{n}{n_p}\indicatorbf{p}^\top \textbf{W}}[\widetilde{\boldsymbol{X}}_{\text{null}}] \\
\mathrm{E}_{\frac{n}{n_p}\indicatorbf{p}^\top \textbf{W}}[\widetilde{\boldsymbol{X}}^{\mathsf{c}}_{\text{null}}]
\end{matrix}\right]\right) + \\
& [\indicatorbf{1},...,\indicatorbf{P}](\textbf{M}_{22}-\textbf{M}_{21}\textbf{M}_{11}^{-1}\textbf{M}_{12})^{-1}
\left(-\textbf{M}_{21}\textbf{M}_{11}^{-1}\left[\begin{matrix}
\widetilde{\boldsymbol{x}}^*_{\text{null}} \\
\widetilde{\boldsymbol{x}}^{*\mathsf{c}}_{\text{null}}
\end{matrix}\right]\right) + \\
&[\indicatorbf{1},...,\indicatorbf{P}]\{(\textbf{M}_{22}-\textbf{M}_{21}\textbf{M}_{11}^{-1}\textbf{M}_{12})^{-1}-\textbf{M}_{22}^{-1}\}\boldsymbol{e}_p+\frac{n}{n_p}\indicatorbf{p} 
\end{align*}

\begin{align*}
=& 
[\widetilde{\textbf{X}}_{\text{null}},\widetilde{\textbf{X}}^{\mathsf{c}}_{\text{null}}](\textbf{M}_{11}-\textbf{M}_{12}\textbf{M}_{22}^{-1}\textbf{M}_{21})^{-1}
\left[\begin{matrix}
\widetilde{\boldsymbol{x}}^*_{\text{null}} -\mathrm{E}_{\frac{n}{n_p}\indicatorbf{p}^\top \textbf{W}}[\widetilde{\boldsymbol{X}}_{\text{null}}] \\
0
\end{matrix}\right] + \\
& [\indicatorbf{1},...,\indicatorbf{P}]\textbf{M}_{22}^{-1}
\textbf{M}_{21}\textbf{M}_{11}^{-1}\left[\begin{matrix}
\widetilde{\boldsymbol{x}}^*_{\text{null}} \\
\widetilde{\boldsymbol{x}}^{*\mathsf{c}}_{\text{null}}
\end{matrix}\right] + [\indicatorbf{1},...,\indicatorbf{P}]\{(\textbf{M}_{22}-\textbf{M}_{21}\textbf{M}_{11}^{-1}\textbf{M}_{12})^{-1}-\textbf{M}_{22}^{-1}\}\boldsymbol{e}_p- \\
&[\indicatorbf{1},...,\indicatorbf{P}]\{(\textbf{M}_{22}-\textbf{M}_{21}\textbf{M}_{11}^{-1}\textbf{M}_{12})^{-1}-\textbf{M}_{22}^{-1}\}
\textbf{M}_{21}\textbf{M}_{11}^{-1}\left[\begin{matrix}
\widetilde{\boldsymbol{x}}^*_{\text{null}} \\
\widetilde{\boldsymbol{x}}^{*\mathsf{c}}_{\text{null}}
\end{matrix}\right] + \frac{n}{n_p}\indicatorbf{p} \\
=& 
\underbrace{[\widetilde{\textbf{X}}_{\text{null}},\widetilde{\textbf{X}}^{\mathsf{c}}_{\text{null}}](\textbf{M}_{11}-\textbf{M}_{12}\textbf{M}_{22}^{-1}\textbf{M}_{21})^{-1}
\left[\begin{matrix}
\widetilde{\boldsymbol{x}}^*_{\text{null}} -\mathrm{E}_{ \boldsymbol{w}_p}[\widetilde{\boldsymbol{X}}_{\text{null}}] \\
0
\end{matrix}\right]}_{\Delta_1} +\\
&\underbrace{[\indicatorbf{1},...,\indicatorbf{P}]
\left[\begin{matrix}
\mathrm{E}_{\boldsymbol{w}_1}[\widetilde{\boldsymbol{X}}_{\text{null}}]^\top & \widetilde{\boldsymbol{x}}^{*\mathsf{c}\top}_{\text{null}}\\
\vdots & \vdots \\
\mathrm{E}_{\boldsymbol{w}_P}[\widetilde{\boldsymbol{X}}_{\text{null}}]^\top & \widetilde{\boldsymbol{x}}^{*\mathsf{c}\top}_{\text{null}}\\
\end{matrix}\right]
\textbf{M}_{11}^{-1}\left[\begin{matrix}
\widetilde{\boldsymbol{x}}^*_{\text{null}} \\
\widetilde{\boldsymbol{x}}^{*\mathsf{c}}_{\text{null}}
\end{matrix}\right]}_{\Delta_2}+\\
& \underbrace{[\indicatorbf{1},...,\indicatorbf{P}]\{(\textbf{M}_{22}-\textbf{M}_{21}\textbf{M}_{11}^{-1}\textbf{M}_{12})^{-1}-\textbf{M}_{22}^{-1}\}\Bigg\{\boldsymbol{e}_p-
\textbf{M}_{21}\textbf{M}_{11}^{-1}\left[\begin{matrix}
\widetilde{\boldsymbol{x}}^*_{\text{null}} \\
\widetilde{\boldsymbol{x}}^{*\mathsf{c}}_{\text{null}}
\end{matrix}\right]\Bigg\}}_{\Delta_3} + \frac{n}{n_p}\indicatorbf{p}.
\end{align*}

% Condition 1: As $P\rightarrow \infty$, $\boldsymbol{\beta}_p$ is not confounded by the covariates or $a = 0$.

% Condition 2: $b = b'$ (outcome model is correctly specified) or the weights converge to the actual inverse propensity scores.

% We let $\beta := \sum_{q = 1}^P \frac{n_q}{n} \boldsymbol{\beta}_q$

% Multiple groups comparison.
Hence, the first component of the estimation error is
\begin{align*}
    &\Delta_1^\top \textbf{W} \Big[\sum_{q = 1}^P \textbf{1}_q \widetilde{\textbf{X}}^{\mathsf c}_{\text{null}}(\boldsymbol{\beta}_q-\boldsymbol{\beta})\Big]\\
    =& \left[\begin{matrix}
\widetilde{\boldsymbol{x}}^*_{\text{null}} -\mathrm{E}_{\boldsymbol{w}_p}[\widetilde{\boldsymbol{X}}_{\text{null}}]\\
0
\end{matrix}\right]^\top (\textbf{M}_{11}-\textbf{M}_{12}\textbf{M}_{22}^{-1}\textbf{M}_{21})^{-1}\left[\begin{matrix}
    \widetilde{\textbf{X}}_{\text{null}}^\top \textbf{W} [\sum\limits_{q = 1}^P \textbf{1}_q \widetilde{\textbf{X}}^{\mathsf c}_{\text{null}}(\boldsymbol{\beta}_q-\boldsymbol{\beta})] \\
    \widetilde{\textbf{X}}^{\mathsf c\top}_{\text{null}} \textbf{W} [\sum\limits_{q = 1}^P \textbf{1}_q \widetilde{\textbf{X}}^{\mathsf c}_{\text{null}}(\boldsymbol{\beta}_q-\boldsymbol{\beta})]
    \end{matrix}\right] \\
    =& \text{tr}\left(\left[\begin{matrix}
    \sum\limits_{q = 1}^P  \widetilde{\textbf{X}}_{\text{null}}^\top \textbf{W}\textbf{1}_q \widetilde{\textbf{X}}^{\mathsf c}_{\text{null}}(\boldsymbol{\beta}_q-\boldsymbol{\beta}) \\
    \sum\limits_{q = 1}^P\widetilde{\textbf{X}}^{\mathsf c \top }_{\text{null}}\textbf{W}\textbf{1}_q \widetilde{\textbf{X}}^{\mathsf c}_{\text{null}}(\boldsymbol{\beta}_q-\boldsymbol{\beta}) \\
    \end{matrix}\right] \left[\begin{matrix}
\widetilde{\boldsymbol{x}}^*_{\text{null}} -\mathrm{E}_{\boldsymbol{w}_p}[\widetilde{\boldsymbol{X}}_{\text{null}}] \\
0
\end{matrix}\right]^\top (\textbf{M}_{11}-\textbf{M}_{12}\textbf{M}_{22}^{-1}\textbf{M}_{21})^{-1}\right).
\end{align*} 
Under the following conditions, the first component converges to 0. If $\boldsymbol{w}_{p,i}$ converges to the true inverse propensity scores $\frac{\Pr(\indicator{*,i} = 1 |\boldsymbol{X}_i)}{\Pr(\indicator{p,i} = 1 | \boldsymbol{X}_i)}$, where $\indicator{*}$ is the indicator for the target population, then $\widetilde{\boldsymbol{X}}_{\text{null}}$ is already balanced, $
\mathrm{E}_{\boldsymbol{w}_p}[\widetilde{\boldsymbol{X}}_{\text{null}}]-\widetilde{\boldsymbol{x}}^*_{\text{null}}  = o_{\mathbb P}(1)$. If $\boldsymbol{w}_p$ balances the means and the covariance of $\widetilde{\boldsymbol{X}}^{\mathsf c}_{\text{null}}$, then $\sum_{q = 1}^P\widetilde{\textbf{X}}^{\mathsf c\top}_{\text{null}} \textbf{W} \textbf{1}_{q} \widetilde{\textbf{X}}^{\mathsf c}_{\text{null}} (\boldsymbol{\beta}_q-\boldsymbol{\beta}) = o_{\mathbb P}(1)$.
If $\boldsymbol{w}_p$ balances the cross product of $\widetilde{\boldsymbol{X}}_{\text{null}}$ and $\widetilde{\boldsymbol{X}}^{\mathsf c}_{\text{null}}$ or the coefficient $\{\boldsymbol{\beta}_q\}_{q = 1}^P$ is independent of imbalance, then $\sum_{q = 1}^P\widetilde{\textbf{X}}_{\text{null}}^\top \textbf{W} \textbf{1}_q \widetilde{\textbf{X}}^{\mathsf c}_{\text{null}} (\boldsymbol{\beta}_q-\boldsymbol{\beta}) = \text{cor}(\{\boldsymbol{\beta}_q-\boldsymbol{\beta}\}_{q = 1}^P, \{\mathrm{E}_{\boldsymbol{w}_q}[\widetilde{\boldsymbol{X}}_{\text{null}}^\top \widetilde{\boldsymbol{X}}^{\mathsf c}_{\text{null}}]\}_{q = 1}^P) =o_{\mathbb P}(1)$. 

The second component of the estimation error is uniform for all $p = 1,...,P$ and thus does not change the ranking. If $\boldsymbol{w}_p$, $p = 1,...,P$, further balances $\widetilde{\boldsymbol{X}}_{\text{null}}$ to the same value or converges to the true inverse propensity scores, then the second component is $o_{\mathbb P}(1)$. 
\begin{align*}
&\Delta_2^\top \textbf{W} \Big[\sum_{q = 1}^P \textbf{1}_q \widetilde{\textbf{X}}^{\mathsf c}_{\text{null}} (\boldsymbol{\beta}_q-\boldsymbol{\beta})\Big]\\
=& \left[\begin{matrix}
\widetilde{\boldsymbol{x}}^*_{\text{null}} \\
\widetilde{\boldsymbol{x}}^{*\mathsf{c}}_{\text{null}}
\end{matrix}\right]^\top \textbf{M}_{11}^{-1} \left[\begin{matrix}
\mathrm{E}_{\boldsymbol{w}_1}[\widetilde{\boldsymbol{X}}_{\text{null}}] &\cdots & \mathrm{E}_{\boldsymbol{w}_P}[\widetilde{\boldsymbol{X}}_{\text{null}}] \\
\widetilde{\boldsymbol{x}}^{*\mathsf{c}}_{\text{null}} & \cdots  & \widetilde{\boldsymbol{x}}^{*\mathsf{c}}_{\text{null}}\\
\end{matrix}\right] \left[\begin{matrix}
    \mathrm{E}_{\indicatorbf{1}^\top \textbf{W}}[\widetilde{\boldsymbol{X}}^{\mathsf c}_{\text{null}}]^\top(\boldsymbol{\beta}_1-\boldsymbol{\beta}) \\
    \vdots \\
    \mathrm{E}_{\indicatorbf{P}^\top \textbf{W}}[\widetilde{\boldsymbol{X}}^{\mathsf c}_{\text{null}}]^\top (\boldsymbol{\beta}_P-\boldsymbol{\beta})
    \end{matrix}\right] \\
=& \left[\begin{matrix}
\widetilde{\boldsymbol{x}}^*_{\text{null}} \\
\widetilde{\boldsymbol{x}}^{*\mathsf{c}}_{\text{null}}
\end{matrix}\right]^\top \textbf{M}_{11}^{-1} \left[\begin{matrix}
\sum_{q = 1}^P\frac{n_q}{n}\mathrm{E}_{\boldsymbol{w}_q}[\widetilde{\boldsymbol{X}}_{\text{null}}] \widetilde{\boldsymbol{x}}^{*\mathsf{c}\top}_{\text{null}} (\boldsymbol{\beta}_q-\boldsymbol{\beta})\\
0
\end{matrix}\right].
\end{align*}

The third component is
\begin{align*}
    &\Delta_3^\top \textbf{W} \Big[\sum_{q = 1}^P \textbf{1}_q \widetilde{\textbf{X}}^{\mathsf c}_{\text{null}} (\boldsymbol{\beta}_q-\boldsymbol{\beta})\Big]\\
    =& \left(\boldsymbol{e}_p-
\textbf{M}_{21}\textbf{M}_{11}^{-1}\left[\begin{matrix}
\widetilde{\boldsymbol{x}}^*_{\text{null}} \\
\widetilde{\boldsymbol{x}}^{*\mathsf{c}}_{\text{null}}
\end{matrix}\right]\right)^\top 
\Big\{(\textbf{M}_{22}-\textbf{M}_{21}\textbf{M}_{11}^{-1}\textbf{M}_{12})^{-1}-\textbf{M}_{22}^{-1}\Big\}
\left[\begin{matrix}
    \mathrm{E}_{\indicatorbf{1}^\top \textbf{W}}[\widetilde{\boldsymbol{X}}^{\mathsf c}_{\text{null}}]^\top (\boldsymbol{\beta}_1-\boldsymbol{\beta}) \\
    \vdots \\
    \mathrm{E}_{\indicatorbf{P}^\top \textbf{W}}[\widetilde{\boldsymbol{X}}^{\mathsf c}_{\text{null}}]^\top(\boldsymbol{\beta}_P-\boldsymbol{\beta})
    \end{matrix}\right] \\
     =& \left(\boldsymbol{e}_p-
\textbf{M}_{21}\textbf{M}_{11}^{-1}\left[\begin{matrix}
\widetilde{\boldsymbol{x}}^*_{\text{null}} \\
\widetilde{\boldsymbol{x}}^{*\mathsf{c}}_{\text{null}}
\end{matrix}\right]\right)^\top 
\Big\{(\textbf{I}_P-\text{diag}(\frac{n_1}{n},...,\frac{n_P}{n}))^{-1}-\textbf{I}_P\Big\}
\left[\begin{matrix}
    \widetilde{\boldsymbol{x}}^{*\mathsf{c}\top}_{\text{null}} (\boldsymbol{\beta}_1-\boldsymbol{\beta}) \\
    \vdots \\
    \widetilde{\boldsymbol{x}}^{*\mathsf{c}\top}_{\text{null}} (\boldsymbol{\beta}_P-\boldsymbol{\beta})
    \end{matrix}\right], 
\end{align*}
which is $o_{\mathbb P}(1)$ as each $n_p/n$ approaches 0.

% If cor, benefit.

% Converge to 0 is P is large or homogeneous.

% Will not increase bias

% Benefit very imbalanced practices.

% worst case.

Conditioning on knowing the weights, the variance of the estimator is
\begin{eqnarray*}
\mathrm{E}\|\boldsymbol{\omega}_p^\top\boldsymbol{\varepsilon}\|_2^2 = \sigma^2 (\widetilde{\boldsymbol{x}}^{*\top}_{\text{null}},\widetilde{\boldsymbol{x}}^{*\mathsf{c}\top}_{\text{null}},\boldsymbol{e}_p^\top)(\textbf{Z}^\top \textbf{W} \textbf{Z})^{-1}\textbf{Z}^\top \textbf{W}^2 \textbf{Z}(\textbf{Z}^\top \textbf{W} \textbf{Z})^{-1}(\widetilde{\boldsymbol{x}}^{*\top}_{\text{null}},\widetilde{\boldsymbol{x}}^{*\mathsf{c}\top}_{\text{null}},\boldsymbol{e}_p^\top)^\top.
\end{eqnarray*}
The variance of $\widehat{\mu}_{p}(\boldsymbol{x}^*)$ is $\mathrm{E}\|\boldsymbol{\omega}_p^\top\boldsymbol{\varepsilon}\|_2^2$. The variance of the contrast $\widehat{\mu}_{p'}(\boldsymbol{x}^*) - \widehat{\mu}_{p''}(\boldsymbol{x}^*)$ derived from the weighted linear regression is $\sigma^2(\boldsymbol{e}_{p'}-\boldsymbol{e}_{p''})^\top(\textbf{Z}^\top \textbf{W} \textbf{Z})^{-1}\textbf{Z}^\top \textbf{W}^2 \textbf{Z}(\textbf{Z}^\top \textbf{W} \textbf{Z})^{-1}(\boldsymbol{e}_{p'}-\boldsymbol{e}_{p''})$. %Similarly, the variance of the contrast $\widehat{\mu}_{\mathcal S'}(\boldsymbol{x}^*) - \widehat{\mu}_{\mathcal S''}(\boldsymbol{x}^*)$ is $\sigma^2(\frac{1}{|\mathcal S'|}\boldsymbol{e}_{\mathcal S'}-\frac{1}{|\mathcal S''|}\boldsymbol{e}_{\mathcal S''})^\top(\textbf{Z}^\top \textbf{W} \textbf{Z})^{-1}\textbf{Z}^\top \textbf{W}^2 \textbf{Z}(\textbf{Z}^\top \textbf{W} \textbf{Z})^{-1}(\frac{1}{|\mathcal S'|}\boldsymbol{e}_{\mathcal S'}-\frac{1}{|\mathcal S''|}\boldsymbol{e}_{\mathcal S''})$, where $\mathcal S'$ and $\mathcal S''$ are two disjoint sets of indexes. The questions we are interested in can be written as paired comparisons, for example, whether the upper quartile of $\{\widehat{\mu}_{p}(\boldsymbol{x}^*)\}_{p = 1}^P$ is significantly greater than the lower quartile. 
The null hypothesis $H_0: \widehat{\mu}_{(1)}(\boldsymbol{x}^*)  = \widehat{\mu}_{(2)}(\boldsymbol{x}^*) = ... = \widehat{\mu}_{(P)}(\boldsymbol{x}^*)$ is equivalent to the null hypothesis for the intercepts $H_0: \widehat{\alpha}_{(1)} = \widehat{\alpha}_{(2)} = ... = \widehat{\alpha}_{(P)}$, for which one can use the Rao's Score test (\citealt{rao1948large}; specifically, the F-test in this linear regression setting). Hypothesis testing for rankings will be one of our future works.

We minimize the variance of each $\boldsymbol{w}_p$ to increase stability. To solve for $\boldsymbol{w}_p$ during implementation, we utilize the stable balancing weighting (SBW; \citealp{zubizarreta2015stable}) approach.

\subsubsection{Fixed-effects regression}

In fixed-effects regression, the estimator fixes the coefficients of $\widetilde{\boldsymbol{X}}_{\text{null}}$ and $\widetilde{\boldsymbol{X}}^{\mathsf c}_{\text{null}}$, letting the intercepts vary across practices. Ignorance of effect modifier $\widetilde{\boldsymbol{X}}^{\mathsf c}_{\text{null}}$ results in bias. 
% \citet{imai2019when} suggest weighted fixed-effects regression models alternatively. 
\begin{eqnarray*}
    \boldsymbol{\omega}_p^{\text{FE}\top}\boldsymbol{Y}
    := \big[(\widetilde{\boldsymbol{x}}^{*\top}_{\text{null}},\widetilde{\boldsymbol{x}}^{*\mathsf{c}\top}_{\text{null}},\boldsymbol{e}_p^\top)(\textbf{Z}^\top \textbf{Z})^{-1}\textbf{Z}^\top\big]\boldsymbol{Y}.
\end{eqnarray*}

\subsubsection{Practice-stratified regression}

As stated before, when fitting a stratified model for each practice, one runs into singularity issues. To make fair comparisons among practices, we use the same set of variables for all the practices. 

\subsubsection{Practice-pooled regression}

Another way of covariate adjustments is to put the interaction terms in a pooled regression model. The variance of the following estimator can be considerable when the dimension of $\boldsymbol{X}^c_{\text{null}}$ is large.
\begin{eqnarray*}
    \boldsymbol{\omega}_p^{\text{PR}\top} \boldsymbol{Y}
    := (\widetilde{\boldsymbol{x}}^{*\top}_{\text{null}},\boldsymbol{e}_p^\top,0,...,\widetilde{\boldsymbol{x}}^{*\mathsf{c}\top}_{\text{null}},...,0)(\textbf{Z}_0^\top \textbf{Z}_0)^{-1}\textbf{Z}_0^\top \boldsymbol{Y},
\end{eqnarray*}
where the design matrix $\textbf{Z}_0 := (\widetilde{\textbf{X}}_{\text{null}},\indicatorbf{1},...,\indicatorbf{P}, \textbf{1}_1 \widetilde{\textbf{X}}^{\mathsf c}_{\text{null}},...,\textbf{1}_P \widetilde{\textbf{X}}^{\mathsf c}_{\text{null}})$.

% \begin{eqnarray*}
%     && \mathrm{E}\|\boldsymbol{\omega}_p^\top\varepsilon\|_2^2 \\
%     &=& \sigma^2 (a\{x^*_{\text{null}}\}^\top,\boldsymbol{e}_p^\top,0,...,b\{x^{*c}_{\text{null}}\}^\top,...,0)(\textbf{Z}_0^\top \textbf{Z}_0)^{-1}(a\{x^*_{\text{null}}\}^\top,\boldsymbol{e}_p^\top,0,...,b\{x^{*c}_{\text{null}}\}^\top,...,0)^\top
% \end{eqnarray*}

\subsubsection{Bias-corrected estimator}

We consider the following bias-corrected (BC) doubly robust estimator where the outcome model is the same as the fixed-effects model. Since the outcome model misspecifies the heterogeneous effects, the unbiasedness relies on the correct estimation of the practice assignment probabilities. Let $\boldsymbol{w}_p$ denote the estimated inverse propensity score weights. When $P$ is large and the covariate overlap is poor, the BC estimator is unstable.
\begin{align*}
    \boldsymbol{\omega}_p^{\text{BC}\top} \boldsymbol{Y} := &\boldsymbol{w}_p^\top(\boldsymbol{Y}-\textbf{Z}(\textbf{Z}^\top \textbf{Z})^{-1}\textbf{Z}^\top \boldsymbol{Y}) + (\widetilde{\boldsymbol{x}}^{*\top}_{\text{null}},\widetilde{\boldsymbol{x}}^{*\mathsf{c}\top}_{\text{null}},\boldsymbol{e}_p^\top)(\textbf{Z}^\top \textbf{Z})^{-1}\textbf{Z}^\top \boldsymbol{Y} \\
    =& [\boldsymbol{w}_p^\top-\boldsymbol{w}_p^\top\textbf{Z}(\textbf{Z}^\top \textbf{Z})^{-1}\textbf{Z}^\top + (\widetilde{\boldsymbol{x}}^{*\top}_{\text{null}},\widetilde{\boldsymbol{x}}^{*\mathsf{c}\top}_{\text{null}},\boldsymbol{e}_p^\top)(\textbf{Z}^\top \textbf{Z})^{-1}\textbf{Z}^\top] [\textbf{Z}(\boldsymbol{\alpha}^\top, \boldsymbol{\beta}^\top, \alpha_1,...,\alpha_P)^\top+\\
    & \sum_{q = 1}^P \textbf{1}_q \widetilde{\textbf{X}}^{\mathsf c}_{\text{null}} (\boldsymbol{\beta}_q-\boldsymbol{\beta})] +\boldsymbol{\omega}_p^\top \boldsymbol{\varepsilon} \\
    =&  0+\boldsymbol{\omega}_{p}^\top[\sum_{q = 1}^P \textbf{1}_q \widetilde{\textbf{X}}^{\mathsf c}_{\text{null}} (\boldsymbol{\beta}_q-\boldsymbol{\beta})] + \widetilde{\boldsymbol{x}}^{*\top}_{\text{null}}\boldsymbol{\alpha}+ \alpha_p+ \widetilde{\boldsymbol{x}}^{*\mathsf{c}\top}_{\text{null}} \boldsymbol{\beta}+\boldsymbol{\omega}_p^\top \boldsymbol{\varepsilon} 
\end{align*}
Therefore, the estimation error is 
\begin{align*}
& \big\{\boldsymbol{w}_p^\top+\big[-\boldsymbol{w}_p^\top\textbf{Z} + (\widetilde{\boldsymbol{x}}^{*\top}_{\text{null}},\widetilde{\boldsymbol{x}}^{*\mathsf{c}\top}_{\text{null}},\boldsymbol{e}_p^\top)\big](\textbf{Z}^\top \textbf{Z})^{-1}\textbf{Z}^\top\big\} \big[\sum_{q = 1}^P \textbf{1}_q \widetilde{\textbf{X}}^{\mathsf c}_{\text{null}} (\boldsymbol{\beta}_q-\boldsymbol{\beta})\big] - \widetilde{\boldsymbol{x}}^{*\mathsf{c}\top}_{\text{null}} (\boldsymbol{\beta}_p-\boldsymbol{\beta})+\boldsymbol{\omega}_p^\top \boldsymbol{\varepsilon} \\
=& \big[-\boldsymbol{w}_p^\top\textbf{Z} + (\widetilde{\boldsymbol{x}}^{*\top}_{\text{null}},\widetilde{\boldsymbol{x}}^{*\mathsf{c}\top}_{\text{null}},\boldsymbol{e}_p^\top)\big](\textbf{Z}^\top \textbf{Z})^{-1}\textbf{Z}^\top \big[\sum_{q = 1}^P \textbf{1}_q \widetilde{\textbf{X}}^{\mathsf c}_{\text{null}} (\boldsymbol{\beta}_q-\boldsymbol{\beta})\big] + \boldsymbol{w}_{p}^\top \big[\sum_{q = 1}^P \textbf{1}_q \widetilde{\textbf{X}}^{\mathsf c}_{\text{null}} (\boldsymbol{\beta}_q-\boldsymbol{\beta})\big] - \\
& \widetilde{\boldsymbol{x}}^{*\mathsf{c}\top}_{\text{null}} (\boldsymbol{\beta}_p-\boldsymbol{\beta})+\boldsymbol{\omega}_p^\top \boldsymbol{\varepsilon} \\
=& \big[-\boldsymbol{w}_p^\top\textbf{Z} + (\widetilde{\boldsymbol{x}}^{*\top}_{\text{null}},\widetilde{\boldsymbol{x}}^{*\mathsf{c}\top}_{\text{null}},\boldsymbol{e}_p^\top)\big](\textbf{Z}^\top \textbf{Z})^{-1}\textbf{Z}^\top \big[\sum_{q = 1}^P \textbf{1}_q \widetilde{\textbf{X}}^{\mathsf c}_{\text{null}} (\boldsymbol{\beta}_q-\boldsymbol{\beta})\big] + (\boldsymbol{w}_{p}^\top \widetilde{\textbf{X}}^{\mathsf c}_{\text{null}} - \widetilde{\boldsymbol{x}}^{*\mathsf{c}\top}_{\text{null}}) (\boldsymbol{\beta}_p-\boldsymbol{\beta})+\boldsymbol{\omega}_p^\top \boldsymbol{\varepsilon}.
\end{align*}
If $\boldsymbol{w}_{p,i}$ converges to the true inverse propensity scores $\frac{\Pr(\indicator{*,i} = 1 |\boldsymbol{X}_i)}{\Pr(\indicator{p,i} = 1 | \boldsymbol{X}_i)}$, then the error converges to 0.

\subsubsection{Other balancing estimators}

An alternative approach to our proposed one is replacing the balancing weights with the inverse propensity scores. The procedure, adapted to the null-case covariates, is a compromise version of the doubly robust estimation in \citet{uysal2015doubly}. It is less efficient than the direct balancing weights when linear models for the outcomes are good enough. In the simulation, we will show it performs slightly better when the outcomes are highly non-linear.

%  Extrapolation: FE weights, leverage across practices; MR weights, extrapolate within each practice; WR weights, combine interpolate (for b) and extrapolate (for a) (Interpolate and allow for heterogeneity when data is not sparse, extrapolate only when data does not support, performs well when there is a mild overlap for dense covariates, ridge regression with penalty for the $b$-component?)calibrate

% Outcome regression is sensitive to variable selection, underfitting VS overfitting, because of extrapolation

% Random effects model: may violate the assumption of drawing from a common distribution.

% Shrinkage estimates: not robust.

% Ridge: 

% If we can show the bias goes to zero. 

% 1: deviation, 3: shrinkage

% \begin{equation*}
% \frac{1}{\mathrm{P}(p|X^c_{\text{null}}, X_{\text{null}})} = \frac{\mathrm{P}(X_{\text{null}}|X^c_{\text{null}})}{\mathrm{P}(p,X_{\text{null}}|X^c_{\text{null}})}
% \end{equation*}

% We adjust for the square of principle component scores

%%%%%%%%%%%%%%%%%%%%%%%%%%%%%%%%%%%%%%%%%%%
%%%%%%%%%%%%%%%%%%%%%%%%%%%%%%%%%%%%%%%%%%%
%%%%%%%%%%%%%%%%%%%%%%%%%%%%%%%%%%%%%%%%%%%
\subsection{Accuracy of estimated rankings}

\begin{center}
% \begin{landscape}
\begin{table}[!ht]
\caption{Absolute difference between the true and estimated rankings across 100 practices and 1000 simulated data sets for three target samples: all patients in the system, and patients from the practices near the center or at the boundary of the population. The mean value is averaged across 100 practices and 1000 simulated data sets.  The maximum value is computed  for each data set and then averaged all over.}
\label{tab_the_practice_perspective}
\centering
\scriptsize
\setlength{\tabcolsep}{2pt}
\begin{tabular}{lrrrrrrrrrrrr}
  \hline
   & \multicolumn{12}{c}{Simulation setting} \\
     \cmidrule(lr){2-13} 
  &  \multicolumn{3}{c}{Linear} &  \multicolumn{3}{c}{Non-linear: low} &  \multicolumn{3}{c}{Non-linear: medium} &  \multicolumn{3}{c}{Non-linear: high} \\
  \cmidrule(lr){2-4}  \cmidrule(lr){5-7} \cmidrule(lr){8-10} \cmidrule(lr){11-13}
     &  \multicolumn{3}{c}{Target} &  \multicolumn{3}{c}{Target} &  \multicolumn{3}{c}{Target} &  \multicolumn{3}{c}{Target} \\
Method     & System & Center & Edge & System & Center & Edge & System & Center & Edge & System & Center & Edge \\
   \cmidrule(lr){1-1}  \cmidrule(lr){2-4}  \cmidrule(lr){5-7} \cmidrule(lr){8-10} \cmidrule(lr){11-13}
Mean (parsimonious)  & Exact &  & & &  & &  & &  & Under\\
\hspace{.05cm} $\{\cdot, \textrm{FE}(\boldsymbol{X})\}$ & 7.23 & 7.00 & 15.81 & 7.22 & 7.10 & 15.93 & 7.38 & 7.44 & 16.07 & 8.15 & 8.52 & 16.54 \\ 
\hspace{.05cm} $\{\cdot, \textrm{SR}({\boldsymbol{X}})\}$ & 1.26 & 1.39 & 2.06 & 2.39 & 2.43 & 4.21 & 4.21 & 4.10 & 7.74 & 8.03 & 7.38 & 15.97 \\ 
\hspace{.05cm} $\{\cdot, \textrm{PR}(\boldsymbol{X})\}$ & {\bf 0.93} & {\bf 0.94} & {\bf 1.36} & 1.84 & {\bf 1.76} & {\bf 3.02} & 3.33 & 3.06 & 5.83 & 6.56 & 5.68 & 12.74 \\ 
% \hspace{.05cm} $\{\textrm{IPW}(\mathbb{R}^+, \boldsymbol{X}), \cdot\}$ \\
% \hspace{.05cm} $\{\textrm{IPW}(\mathbb{R}^+, \boldsymbol{X}), \textrm{FE}({\boldsymbol{X}}; w)\}$ \\ 
% \hspace{.05cm} $\{\textrm{IPW}(\mathbb{R}^+, \boldsymbol{X}), \textrm{WR}({\boldsymbol{X}}; w)\}$ \\ 
\hspace{.05cm} $\{\textrm{LW}( \boldsymbol{X}_{\textrm{null}}^{\mathsf{c}}), \cdot\}$ & 3.57 & 3.68 & 7.93 & 3.93 & 3.84 & 8.58 & 4.64 & 4.50 & 9.54 & 6.59 & 6.42 & 12.18 \\ 
\hspace{.05cm} $\{\textrm{LW}(\boldsymbol{X}_{\textrm{null}}^{\mathsf{c}}), \textrm{FE}(\boldsymbol{X})\}$& 2.68 & 2.53 & 6.57 & 3.03 & 2.95 & 6.97 & 3.80 & 3.84 & 7.76 & 5.81 & 5.91 & 10.21 \\ 
\hspace{.05cm} $\{\textrm{LW}( \boldsymbol{X}_{\textrm{null}}^{\mathsf{c}}), \textrm{WR}({\boldsymbol{X}})\}$ & 2.68 & 2.52 & 6.60 & 3.02 & 2.94 & 6.85 & 3.77 & 3.80 & 7.39 & {\bf 5.74} & 5.83 & {\bf 9.14} \\ 
\hspace{.05cm} $\{\textrm{SBW}(\boldsymbol{X}_{\textrm{null}}^{\mathsf{c}}), \textrm{WR}({\boldsymbol{X}})\}$ & 0.96 & 0.99 & 2.63 & {\bf 1.76} & {\bf 1.75} & 3.61 & {\bf 3.06} & {\bf 3.00} & {\bf 5.40} & 5.77 & {\bf 5.56} & 9.29 \\ 
\cmidrule(lr){1-1}  \cmidrule(lr){2-4}  \cmidrule(lr){5-7} \cmidrule(lr){8-10} \cmidrule(lr){11-13}
Mean (comprehensive) & Over &  & & &  & &  & &  & Exact\\
\hspace{.05cm} $\{\cdot, \textrm{FE}(\widetilde{\boldsymbol{X}})\}$ & 7.20 & 6.97 & 15.76 & 7.23 & 7.04 & 15.99 & 7.30 & 7.26 & 16.24 & 7.56 & 7.95 & 16.83 \\
\hspace{.05cm} $\{\cdot, \textrm{SR}(\widetilde{\boldsymbol{X}})\}$ & 1.76 & 1.82 & 2.99 & 2.09 & 2.50 & 4.22 & 2.82 & 3.78 & 6.41 & 4.59 & 6.53 & 10.98 \\ 
\hspace{.05cm} $\{\cdot, \textrm{PR}(\widetilde{\boldsymbol{X}})\}$ & {\bf 1.28} & {\bf 1.23} & {\bf 2.05} & {\bf 1.56} & {\bf 1.78} & {\bf 3.14} & {\bf 2.16} & {\bf 2.80} & {\bf 5.00} & {\bf 3.62} & 4.99 & 8.86 \\ 
% \hspace{.05cm} $\{\textrm{IPW}(\mathbb{R}^+, \widetilde{\boldsymbol{X}}), \cdot\}$ \\
% \hspace{.05cm} $\{\textrm{IPW}(\mathbb{R}^+, \widetilde{\boldsymbol{X}}), \textrm{FE}(\widetilde{\boldsymbol{X}}; w)\}$ \\ 
% \hspace{.05cm} $\{\textrm{IPW}(\mathbb{R}^+, \widetilde{\boldsymbol{X}}), \textrm{WR}(\widetilde{\boldsymbol{X}}; w)\}$ \\ 
\hspace{.05cm} $\{\textrm{LW}(\widetilde{\boldsymbol{X}}_{\textrm{null}}^{\mathsf{c}}), \cdot\}$ & 4.01 & 4.27 & 7.94 & 4.21 & 4.24 & 8.57 & 4.66 & 4.62 & 9.46 & 6.01 & 6.04 & 11.79 \\ 
\hspace{.05cm} $\{\textrm{LW}( \widetilde{\boldsymbol{X}}_{\textrm{null}}^{\mathsf{c}}), \textrm{FE}(\widetilde{\boldsymbol{X}}\}$& 2.83 & 2.68 & 6.42 & 2.95 & 2.82 & 6.61 & 3.23 & 3.13 & 6.93 & 4.07 & 4.06 & 7.88 \\ 
\hspace{.05cm} $\{\textrm{LW}( \widetilde{\boldsymbol{X}}_{\textrm{null}}^{\mathsf{c}}), \textrm{WR}(\widetilde{\boldsymbol{X}})\}$ & 2.81 & 2.67 & 6.42 & 2.93 & 2.78 & 6.58 & 3.18 & 3.07 & 6.82 & 3.97 & {\bf 3.93} & {\bf 7.58} \\ 
\hspace{.05cm} $\{\textrm{SBW}( \widetilde{\boldsymbol{X}}_{\textrm{null}}^{\mathsf{c}}), \textrm{WR}(\widetilde{\boldsymbol{X}})\}$ & 1.62 & 1.61 & 5.49 & 1.85 & 2.13 & 5.82 & 2.37 & 3.14 & 6.57 & 3.72 & 5.39 & 8.78 \\ 
% \hspace{.25cm} $\{\cdot, \textrm{PR}({\boldsymbol{X}})\}$ & 0.94 & 0.94 & 1.35 & 1.85 & 1.77 & 3.06 & 3.34 & 3.04 & 5.84 & 6.56 & 5.67 & 12.73 \\ 
% \hspace{.25cm} $\{\cdot, \textrm{PR}(\widetilde{\boldsymbol{X}})\}$ & 1.27 & 1.25 & 2.03 & 1.58 & 1.83 & 3.07 & 2.18 & 2.85 & 4.87 & 3.63 & 5.05 & 8.55 \\ 
\cmidrule(lr){1-1}  \cmidrule(lr){2-4}  \cmidrule(lr){5-7} \cmidrule(lr){8-10} \cmidrule(lr){11-13}
Maximum (parsimonious) & Exact &  & & &  & &  & &  & Under\\ 
\hspace{.05cm} $\{\cdot, \textrm{FE}(\boldsymbol{X})\}$ & 20.77 & 17.95 & 36.03 & 20.45 & 18.55 & 35.84 & 22.10 & 19.98 & 37.91 & 27.23 & 24.92 & 44.54 \\ 
\hspace{.05cm} $\{\cdot, \textrm{SR}({\boldsymbol{X}})\}$ & 5.81 & 5.68 & 10.21 & 12.29 & 10.54 & 22.31 & 21.88 & 18.06 & 41.72 & 41.79 & 31.68 & 81.35 \\ 
\hspace{.05cm} $\{\cdot, \textrm{PR}(\boldsymbol{X})\}$ & {\bf 4.12} & {\bf 3.94} & {\bf 6.47} & 9.22 & 7.56 & {\bf 16.09} & 16.99 & 13.32 & 31.36 & 33.35 & 24.27 & 69.11 \\ 
% \hspace{.05cm} $\{\textrm{IPW}(\mathbb{R}^+, \boldsymbol{X}), \cdot\}$ \\
% \hspace{.05cm} $\{\textrm{IPW}(\mathbb{R}^+, \boldsymbol{X}), \textrm{FE}({\boldsymbol{X}}; w)\}$ \\ 
% \hspace{.05cm} $\{\textrm{IPW}(\mathbb{R}^+, \boldsymbol{X}), \textrm{WR}({\boldsymbol{X}}; w)\}$ \\ 
\hspace{.05cm} $\{\textrm{LW}(\boldsymbol{X}_{\textrm{null}}^{\mathsf{c}}), \cdot\}$ & 16.49 & 13.94 & 33.12 & 18.38 & 14.50 & 36.67 & 21.48 & 17.49 & 41.37 & 28.65 & 26.07 & 52.86 \\ 
\hspace{.05cm} $\{\textrm{LW}( \boldsymbol{X}_{\textrm{null}}^{\mathsf{c}}), \textrm{FE}(\boldsymbol{X})\}$ & 10.66 & 8.58 & 22.21 & 12.99 & 10.40 & 25.86 & 16.85 & 14.80 & 31.13 & 25.04 & 24.47 & 42.47 \\ 
\hspace{.05cm} $\{\textrm{LW}(\boldsymbol{X}_{\textrm{null}}^{\mathsf{c}}), \textrm{WR}({\boldsymbol{X}})\}$  & 10.62 & 8.57 & 22.17 & 12.84 & 10.24 & 24.60 & 16.60 & 14.38 & 28.67 & {\bf 24.67} & 23.75 & {\bf 37.60} \\ 
\hspace{.05cm} $\{\textrm{SBW}(\boldsymbol{X}_{\textrm{null}}^{\mathsf{c}}), \textrm{WR}({\boldsymbol{X}})\}$ & 4.45 & 4.14 & 16.61 & {\bf 8.53} & {\bf 7.39} & 21.20 & {\bf 14.74} & {\bf 12.71} & {\bf 28.58} & 26.00 & {\bf 22.86} & 43.65 \\ 
\cmidrule(lr){1-1}  \cmidrule(lr){2-4}  \cmidrule(lr){5-7} \cmidrule(lr){8-10} \cmidrule(lr){11-13} 
Maximum (comprehensive) & Over &  & & &  & &  & &  & Exact\\
\hspace{.05cm} $\{\cdot, \textrm{FE}(\widetilde{\boldsymbol{X}})\}$ & 20.42 & 18.03 & 35.30 & 20.40 & 18.24 & 35.79 & 21.13 & 18.93 & 37.68 & 23.68 & 21.98 & 43.44 \\ 
\hspace{.05cm} $\{\cdot, \textrm{SR}(\widetilde{\boldsymbol{X}})\}$ & 10.79 & 8.37 & 17.97 & 12.83 & 11.55 & 24.59 & 17.08 & 17.66 & 37.03 & 27.33 & 29.54 & 61.07 \\ 
\hspace{.05cm} $\{\cdot, \textrm{PR}(\widetilde{\boldsymbol{X}})\}$ & {\bf 7.31} & {\bf 5.22} & {\bf 11.82} & {\bf 8.97} & {\bf 7.64} & {\bf 18.39} & {\bf 12.52} & 12.22 & 28.92 & 20.75 & 21.63 & 49.97 \\ 
% \hspace{.05cm} $\{\textrm{IPW}(\mathbb{R}^+, \widetilde{\boldsymbol{X}}), \cdot\}$ \\
% \hspace{.05cm} $\{\textrm{IPW}(\mathbb{R}^+, \widetilde{\boldsymbol{X}}), \textrm{FE}(\widetilde{\boldsymbol{X}}; w)\}$ \\ 
% \hspace{.05cm} $\{\textrm{IPW}(\mathbb{R}^+, \widetilde{\boldsymbol{X}}), \textrm{WR}(\widetilde{\boldsymbol{X}}; w)\}$ \\ 
\hspace{.05cm} $\{\textrm{LW}(\widetilde{\boldsymbol{X}}_{\textrm{null}}^{\mathsf{c}}), \cdot\}$ & 18.94 & 18.79 & 33.93 & 20.30 & 18.57 & 37.15 & 23.01 & 20.96 & 41.59 & 29.44 & 28.23 & 52.42 \\ 
\hspace{.05cm} $\{\textrm{LW}(\widetilde{\boldsymbol{X}}_{\textrm{null}}^{\mathsf{c}}), \textrm{FE}(\widetilde{\boldsymbol{X}})\}$& 11.66 & 10.37 & 22.57 & 12.79 & 11.16 & 24.31 & 14.73 & 12.83 & 26.93 & 19.48 & 17.61 & 33.53 \\ 
\hspace{.05cm} $\{\textrm{LW}(\widetilde{\boldsymbol{X}}_{\textrm{null}}^{\mathsf{c}}), \textrm{WR}(\widetilde{\boldsymbol{X}})\}$ & 11.41 & 9.97 & 22.19 & 12.31 & 10.66 & 23.57 & 14.13 & {\bf 12.17} & {\bf 25.71} & {\bf 18.58} & {\bf 16.55} & {\bf 31.32} \\ 
\hspace{.05cm} $\{\textrm{SBW}( \widetilde{\boldsymbol{X}}_{\textrm{null}}^{\mathsf{c}}), \textrm{WR}(\widetilde{\boldsymbol{X}})\}$ & 10.61 & 6.87 & 23.33 & 11.48 & 9.02 & 25.72 & 13.58 & 13.22 & 30.57 & 19.67 & 22.31 & 42.13 \\ 
\hline
\end{tabular}
\vspace{.25cm}
\footnotesize{
\begin{flushleft}
\end{flushleft}
}
\end{table}
% \end{landscape}
\end{center}

\begin{center}
% \begin{landscape}
\begin{table}[!ht]
\caption{Absolute difference between the true and estimated rankings across 200 practices and 1000 simulated data sets for three target samples: all patients in the system, and patients from the practices near the center or at the boundary of the population. The mean value is averaged across 200 practices and 1000 simulated data sets.  The maximum value is computed  for each data set and then averaged all over.}
\label{tab_the_practice_perspective_200}
\centering
\scriptsize
\setlength{\tabcolsep}{2pt}
\begin{tabular}{lrrrrrrrrrrrr}
  \hline
   & \multicolumn{12}{c}{Simulation setting} \\
     \cmidrule(lr){2-13} 
  &  \multicolumn{3}{c}{Linear} &  \multicolumn{3}{c}{Non-linear: low} &  \multicolumn{3}{c}{Non-linear: medium} &  \multicolumn{3}{c}{Non-linear: high} \\
  \cmidrule(lr){2-4}  \cmidrule(lr){5-7} \cmidrule(lr){8-10} \cmidrule(lr){11-13}
     &  \multicolumn{3}{c}{Target} &  \multicolumn{3}{c}{Target} &  \multicolumn{3}{c}{Target} &  \multicolumn{3}{c}{Target} \\
Method & System & Center & Edge & System & Center & Edge & System & Center & Edge & System & Center & Edge \\
   \cmidrule(lr){1-1}  \cmidrule(lr){2-4}  \cmidrule(lr){5-7} \cmidrule(lr){8-10} \cmidrule(lr){11-13}
Mean (parsimonious)  & Exact &  & & &  & &  & &  & Under\\
\hspace{.05cm} $\{\cdot, \textrm{FE}(\boldsymbol{X})\}$ & 8.19 & 9.01 & 17.93 & 8.23 & 9.00 & 17.90 & 8.54 & 9.23 & 17.89 & 9.90 & 10.38 & 18.10 \\
\hspace{.05cm} $\{\cdot, \textrm{SR}(\boldsymbol{X})\}$ & -- & -- & -- & -- & -- & --  & -- & -- & --  & -- & -- & --\\
\hspace{.05cm} $\{\cdot, \textrm{PR}(\boldsymbol{X})\}$ & {\bf 1.56} & {\bf 1.73} & {\bf 2.32} & {\bf 2.86} & {\bf 3.22} & {\bf 4.54} & 4.99 & 5.64 & 8.15 & 9.49 & 10.77 & 16.13 \\ 
% \hspace{.05cm} $\{\textrm{IPW}(\mathbb{R}^+, \boldsymbol{X}), \cdot\}$ \\
% \hspace{.05cm} $\{\textrm{IPW}(\mathbb{R}^+, \boldsymbol{X}), \textrm{FE}({\boldsymbol{X}}; w)\}$ \\ 
% \hspace{.05cm} $\{\textrm{IPW}(\mathbb{R}^+, \boldsymbol{X}), \textrm{WR}({\boldsymbol{X}}; w)\}$ \\ 
\hspace{.05cm} $\{\textrm{LW}( \boldsymbol{X}_{\textrm{null}}^{\mathsf{c}}), \cdot\}$  & 5.02 & 6.08 & 9.83 & 5.48 & 6.60 & 10.66 & 6.46 & 7.68 & 11.92 & 9.16 & 10.63 & 15.28 \\ 
\hspace{.05cm} $\{\textrm{LW}(\boldsymbol{X}_{\textrm{null}}^{\mathsf{c}}), \textrm{FE}(\boldsymbol{X})\}$ & 3.53 & 4.14 & 8.01 & 4.09 & 4.77 & 8.56 & 5.26 & 6.08 & 9.70 & 8.26 & 9.37 & 13.12 \\ 
\hspace{.05cm} $\{\textrm{LW}(\boldsymbol{X}_{\textrm{null}}^{\mathsf{c}}), \textrm{WR}({\boldsymbol{X}})\}$ & 3.54 & 4.14 & 8.05 & 4.08 & 4.69 & 8.42 & 5.24 & 5.89 & 9.27 & {\bf 8.19} & {\bf 8.95} & {\bf 11.93} \\ 
\hspace{.05cm} $\{\textrm{SBW}( \boldsymbol{X}_{\textrm{null}}^{\mathsf{c}}), \textrm{WR}({\boldsymbol{X}})\}$ & 1.80 & 2.25 & 5.44 & 2.95 & 3.45 & 6.28 & {\bf 4.90} & {\bf 5.55} & {\bf 8.14} & 9.00 & 9.97 & 12.88 \\ 
\cmidrule(lr){1-1}  \cmidrule(lr){2-4}  \cmidrule(lr){5-7} \cmidrule(lr){8-10} \cmidrule(lr){11-13}
Mean (comprehensive) & Over &  & & &  & &  & &  & Exact\\
\hspace{.05cm} $\{\cdot, \textrm{FE}(\widetilde{\boldsymbol{X}})\}$ & 8.18 & 8.97 & 17.87 & 8.19 & 8.98 & 17.98 & 8.26 & 9.12 & 18.10 & 8.59 & 9.78 & 18.44 \\ 
\hspace{.05cm} $\{\cdot, \textrm{SR}(\widetilde{\boldsymbol{X}})\}$ & -- & -- & -- & -- & -- & --  & -- & -- & --  & -- & -- & --\\
\hspace{.05cm} $\{\cdot, \textrm{PR}(\widetilde{\boldsymbol{X}})\}$ & 3.46 & {\bf 3.28} & {\bf 5.27} & 3.98 & {\bf 4.56} & {\bf 7.20} & 5.15 & 6.88 & 10.59 & 8.04 & 11.79 & 17.43 \\ 
% \hspace{.05cm} $\{\textrm{IPW}(\mathbb{R}^+, \widetilde{\boldsymbol{X}}), \cdot\}$ \\
% \hspace{.05cm} $\{\textrm{IPW}(\mathbb{R}^+, \widetilde{\boldsymbol{X}}), \textrm{FE}(\widetilde{\boldsymbol{X}}; w)\}$ \\ 
% \hspace{.05cm} $\{\textrm{IPW}(\mathbb{R}^+, \widetilde{\boldsymbol{X}}), \textrm{WR}(\widetilde{\boldsymbol{X}}; w)\}$ \\ 
\hspace{.05cm} $\{\textrm{LW}( \widetilde{\boldsymbol{X}}_{\textrm{null}}^{\mathsf{c}}), \cdot\}$ & 6.21 & 7.45 & 10.55 & 6.45 & 7.79 & 11.33 & 7.13 & 8.61 & 12.48 & 9.27 & 11.08 & 15.53 \\ 
\hspace{.05cm} $\{\textrm{LW}(\widetilde{\boldsymbol{X}}_{\textrm{null}}^{\mathsf{c}}), \textrm{FE}(\widetilde{\boldsymbol{X}})\}$ & 4.13 & 4.83 & 8.30 & 4.32 & 5.04 & 8.55 & 4.74 & 5.49 & 8.99 & 6.05 & 6.89 & 10.32 \\ 
\hspace{.05cm} $\{\textrm{LW}(\widetilde{\boldsymbol{X}}_{\textrm{null}}^{\mathsf{c}}), \textrm{WR}(\widetilde{\boldsymbol{X}})\}$ & 4.11 & 4.80 & 8.30 & 4.28 & 4.97 & 8.50 & 4.67 & {\bf 5.38} & {\bf 8.86} & 5.93 & {\bf 6.64} & {\bf 9.97} \\ 
\hspace{.05cm} $\{\textrm{SBW}( \widetilde{\boldsymbol{X}}_{\textrm{null}}^{\mathsf{c}}), \textrm{WR}(\widetilde{\boldsymbol{X}})\}$ & {\bf 3.33} & 4.21 & 9.25 & {\bf 3.55} & 4.65 & 9.49 & {\bf 4.10} & 5.77 & 10.09 & {\bf 5.78} & 8.83 & 12.31 \\ 
\cmidrule(lr){1-1}  \cmidrule(lr){2-4}  \cmidrule(lr){5-7} \cmidrule(lr){8-10} \cmidrule(lr){11-13}
Maximum (parsimonious) & Exact &  & & &  & &  & &  & Under\\
\hspace{.05cm} $\{\cdot, \textrm{FE}(\boldsymbol{X})\}$ & 24.64 & 26.64 & 41.23 & 25.11 & 27.04 & 41.64 & 28.73 & {\bf 30.57} & {\bf 45.03} & {\bf 38.22} & {\bf 39.75} & {\bf 54.22} \\ 
\hspace{.05cm} $\{\cdot, \textrm{SR}(\boldsymbol{X})\}$ & -- & -- & -- & -- & -- & --  & -- & -- & --  & -- & -- & --\\
\hspace{.05cm} $\{\cdot, \textrm{PR}(\boldsymbol{X})\}$ & {\bf 7.92} & {\bf 8.87} & {\bf 13.36} & {\bf 16.74} & {\bf 18.98} & {\bf 28.80} & 30.20 & 34.27 & 51.05 & 55.34 & 63.38 & 100.21 \\ 
% \hspace{.05cm} $\{\textrm{IPW}(\mathbb{R}^+, \boldsymbol{X}), \cdot\}$ \\
% \hspace{.05cm} $\{\textrm{IPW}(\mathbb{R}^+, \boldsymbol{X}), \textrm{FE}({\boldsymbol{X}}; w)\}$ \\ 
% \hspace{.05cm} $\{\textrm{IPW}(\mathbb{R}^+, \boldsymbol{X}), \textrm{WR}({\boldsymbol{X}}; w)\}$ \\ 
\hspace{.05cm} $\{\textrm{LW}( \boldsymbol{X}_{\textrm{null}}^{\mathsf{c}}), \cdot\}$ & 26.68 & 32.58 & 45.93 & 30.22 & 36.78 & 52.47 & 36.00 & 43.40 & 60.70 & 48.99 & 57.71 & 75.82 \\ 
\hspace{.05cm} $\{\textrm{LW}(\boldsymbol{X}_{\textrm{null}}^{\mathsf{c}}), \textrm{FE}(\boldsymbol{X})\}$ & 16.36 & 19.24 & 30.08 & 21.19 & 25.55 & 37.63 & 29.00 & 35.26 & 48.94 & 44.76 & 53.02 & 68.26 \\ 
\hspace{.05cm} $\{\textrm{LW}(\boldsymbol{X}_{\textrm{null}}^{\mathsf{c}}), \textrm{WR}({\boldsymbol{X}})\}$ & 16.28 & 19.15 & 30.08 & 20.85 & 24.74 & 35.74 & {\bf 28.59} & 33.80 & 45.20 & 44.21 & 50.72 & 62.63 \\ 
\hspace{.05cm} $\{\textrm{SBW}( \boldsymbol{X}_{\textrm{null}}^{\mathsf{c}}), \textrm{WR}({\boldsymbol{X}})\}$ & 11.16 & 13.81 & 27.16 & 18.51 & 21.30 & 33.03 & 30.31 & 34.17 & 45.29 & 49.54 & 54.73 & 67.03 \\ 
\cmidrule(lr){1-1}  \cmidrule(lr){2-4}  \cmidrule(lr){5-7} \cmidrule(lr){8-10} \cmidrule(lr){11-13}
Maximum (comprehensive) & Over &  & & &  & &  & &  & Exact\\
\hspace{.05cm} $\{\cdot, \textrm{FE}(\widetilde{\boldsymbol{X}})\}$ & 24.30 & 26.20 & 40.74 & 24.54 & 26.57 & 41.31 & 25.82 & 28.94 & 43.78 & 29.84 & 35.76 & 51.18 \\ 
\hspace{.05cm} $\{\cdot, \textrm{SR}(\widetilde{\boldsymbol{X}})\}$ & -- & -- & -- & -- & -- & --  & -- & -- & --  & -- & -- & --\\
\hspace{.05cm} $\{\cdot, \textrm{PR}(\widetilde{\boldsymbol{X}})\}$ & 53.70 & 38.97 & 61.35 & 58.12 & 46.72 & 72.10 & 67.18 & 60.86 & 90.75 & 88.36 & 86.73 & 123.12 \\
% \hspace{.05cm} $\{\textrm{IPW}(\mathbb{R}^+, \widetilde{\boldsymbol{X}}), \cdot\}$ \\
% \hspace{.05cm} $\{\textrm{IPW}(\mathbb{R}^+, \widetilde{\boldsymbol{X}}), \textrm{FE}(\widetilde{\boldsymbol{X}}; w)\}$ \\ 
% \hspace{.05cm} $\{\textrm{IPW}(\mathbb{R}^+, \widetilde{\boldsymbol{X}}), \textrm{WR}(\widetilde{\boldsymbol{X}}; w)\}$ \\ 
\hspace{.05cm} $\{\textrm{LW}( \widetilde{\boldsymbol{X}}_{\textrm{null}}^{\mathsf{c}}), \cdot\}$ & 35.75 & 41.75 & 51.38 & 38.66 & 45.07 & 58.43 & 44.73 & 51.38 & 67.11 & 58.13 & 65.49 & 83.16 \\ 
\hspace{.05cm} $\{\textrm{LW}( \widetilde{\boldsymbol{X}}_{\textrm{null}}^{\mathsf{c}}), \textrm{FE}(\widetilde{\boldsymbol{X}})\}$ & 20.80 & 24.26 & 32.27 & 22.96 & 26.49 & 35.72 & 26.94 & 30.52 & 40.81 & 36.50 & 40.60 & 52.23 \\ 
\hspace{.05cm} $\{\textrm{LW}( \widetilde{\boldsymbol{X}}_{\textrm{null}}^{\mathsf{c}}), \textrm{WR}(\widetilde{\boldsymbol{X}})\}$ & 20.31 & 23.64 & 31.73 & 22.07 & 25.38 & 34.47 & 25.74 & 28.98 & {\bf 38.82} & 35.06 & {\bf 38.20} & {\bf 48.98} \\  
\hspace{.05cm} $\{\textrm{SBW}( \widetilde{\boldsymbol{X}}_{\textrm{null}}^{\mathsf{c}}), \textrm{WR}(\widetilde{\boldsymbol{X}})\}$ & {\bf 18.92} & {\bf 18.33} & {\bf 31.71} & {\bf 19.92} & {\bf 20.83} & {\bf 34.33} & {\bf 22.33} & {\bf 26.45} & 40.09 & {\bf 29.44} & 39.79 & 54.17 \\ 
\hline
\end{tabular}
\vspace{.25cm}
\footnotesize{
\begin{flushleft}
\end{flushleft}
}
\end{table}
% \end{landscape}
\end{center}

%%%%%%%%%%%%%%%%%%%%%%%%%%%%%%%%%%%%%%%%%%%
%%%%%%%%%%%%%%%%%%%%%%%%%%%%%%%%%%%%%%%%%%%
%%%%%%%%%%%%%%%%%%%%%%%%%%%%%%%%%%%%%%%%%%%
\clearpage
\subsection{Distribution of null{-case} covariates}

\begin{table}[!htbp]
\centering
\caption{Number of practices with at least one null{-case} covariate} 
\label{tab_null_covariates}
\begingroup
\footnotesize
%\scriptsize
\begin{tabular}{lrr}
\hline
Covariate & Profile & Violations\\
\hline
Age	& 0.278 & 0\\
Age (65-74)	& 0.459 & 0\\
Age (75-84)	& 0.421 & 0\\
Age ($\geq$85)	& 0.120 & 2\\
Sex (Male)  & 0.339 & 6\\
Sex (Female)	& 0.661 & 3\\ 
Race (White)	& 0.822 & 1\\
Race (Black)	& 0.075 & 108\\
Race (Other)	& 0.102 & 69\\
Marital status (Married)	& 0.516 & 0\\
Marital status (Unmarried)	 & 0.438 & 0\\ 
Census-level prop. of residents without a high school ed. & 0.164 & 0\\
Census-level prop. of residents without a high school ed. (Quartile 1)	& 0.256 & 71\\
Census-level prop. of residents without a high school ed. (Quartile 2)	& 0.257 & 20\\
Census-level prop. of residents without a high school ed. (Quartile 3)	& 0.251 & 4\\
Census-level prop. of residents without a high school ed. (Quartile 4)	& 0.237 & 17\\
Census-tract median household income  & 0.230 & 0\\
Census-tract median household income (Quartile 1)	& 0.247 & 7\\
Census-tract median household income (Quartile 2)	& 0.250 & 0\\
Census-tract median household income (Quartile 3)	& 0.251 & 3\\
Census-tract median household income (Quartile 4)	& 0.252 & 60\\
Charlson Comorbidity Index (0)	& 0.345 & 0\\
Charlson Comorbidity Index (1)	& 0.259 & 0\\
Charlson Comorbidity Index (2)	& 0.159 & 0\\
Charlson Comorbidity Index ($\geq$3)	& 0.237 & 0\\
Cancer type (Breast)	& 0.347 & 4\\
Cancer type (Colorectal)	& 0.171 & 10\\
Cancer type (Lung)	& 0.337 & 7\\
Cancer type (Ovary)	& 0.023 & 141\\
Cancer type (Pancreas)	& 0.075 & 28\\
Cancer type (Prostate)	& 0.048 & 55\\
Cancer stage (1)	& 0.269 & 3\\
Cancer stage (2)	& 0.217 & 0\\ 
Cancer stage (3)	& 0.196 & 3\\
Cancer stage (4)	& 0.274 & 2\\
\hline
\end{tabular}
\endgroup
\end{table}

%%%%%%%%%%%%%%%%%%%%%%%%%%%%%%%%%%%%%%%%%%%
%%%%%%%%%%%%%%%%%%%%%%%%%%%%%%%%%%%%%%%%%%%
%%%%%%%%%%%%%%%%%%%%%%%%%%%%%%%%%%%%%%%%%%%
\clearpage
\subsection{Number of practices with at least one null{-case} covariate}

\begin{figure}[!htbp]
\begin{center}
\caption{Number of practices with at least one null{-case} covariate} 
\label{fig_null_covariates}
\includegraphics[scale=0.375]{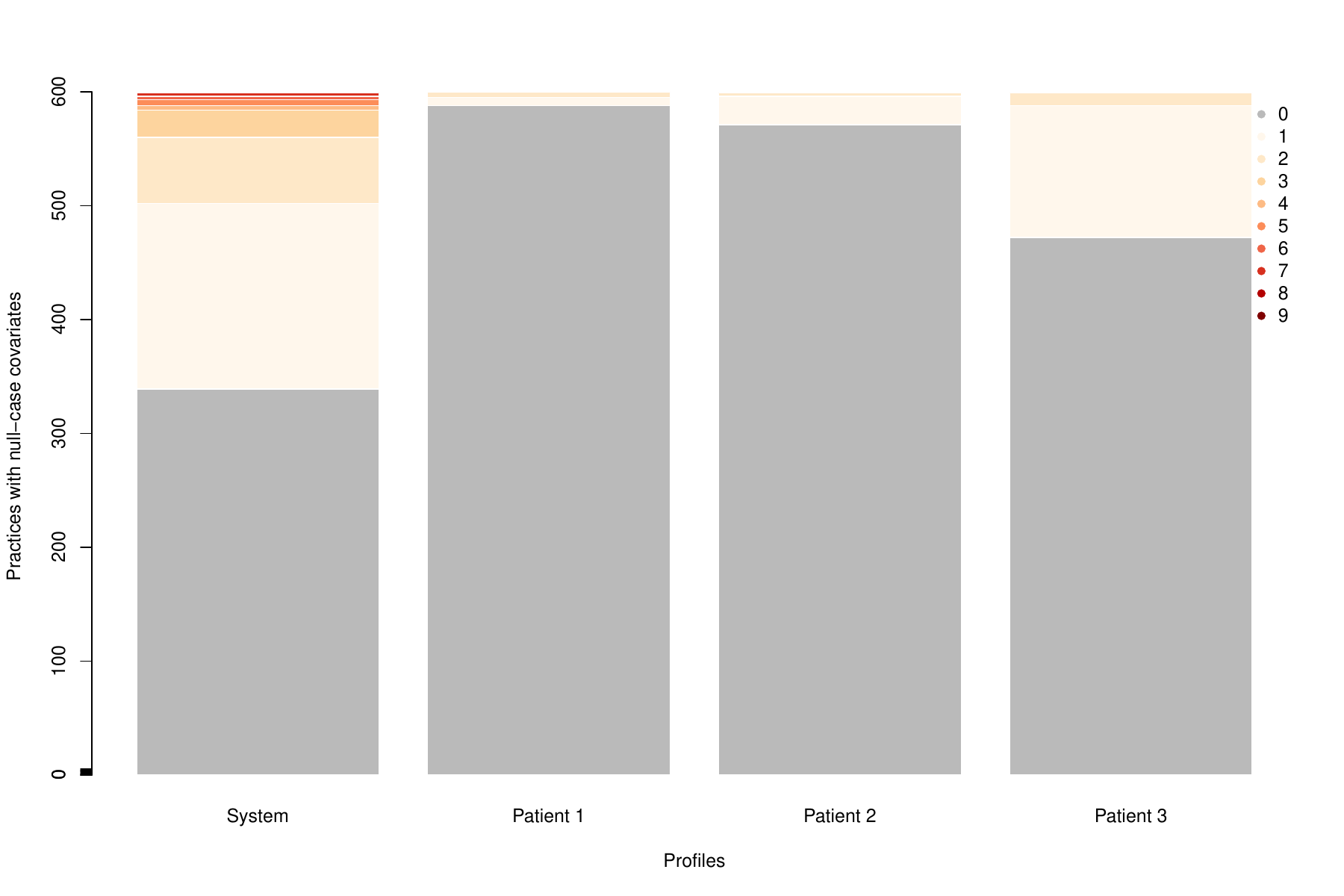}
\end{center}
\end{figure}

\begin{figure}[ht]
\begin{center}
\caption{Changes in practice rankings for different covariate profiles. The right plot groups the ranking by quintiles.}
\label{fig_rankings}
\includegraphics[scale = 0.35]{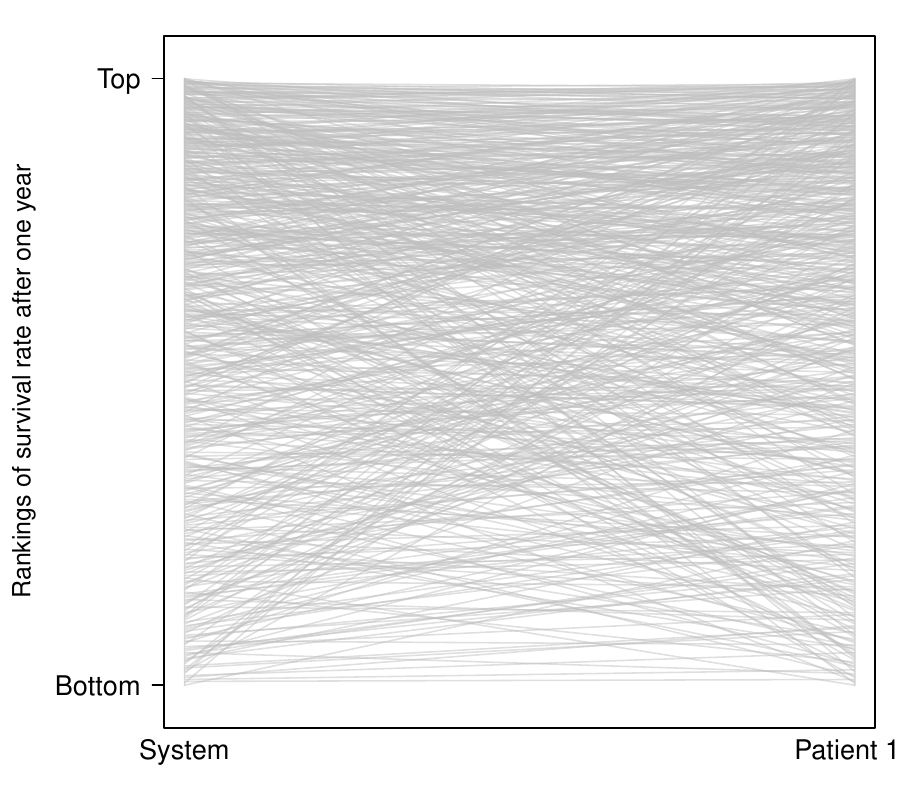}\includegraphics[scale = 0.35]{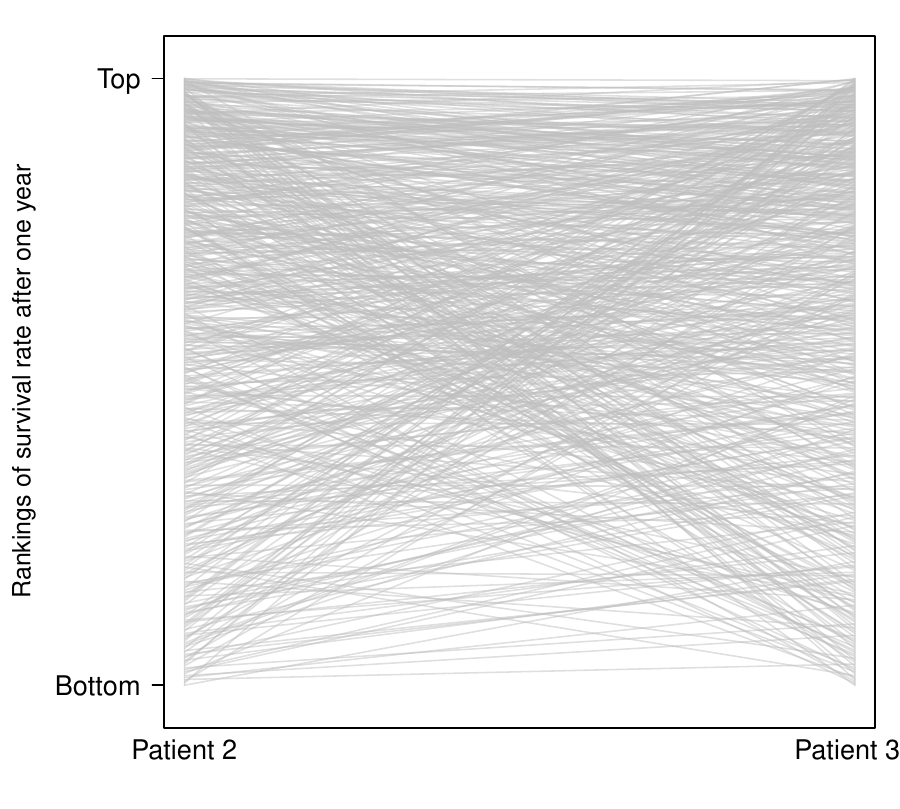}
\includegraphics[scale = 0.35]{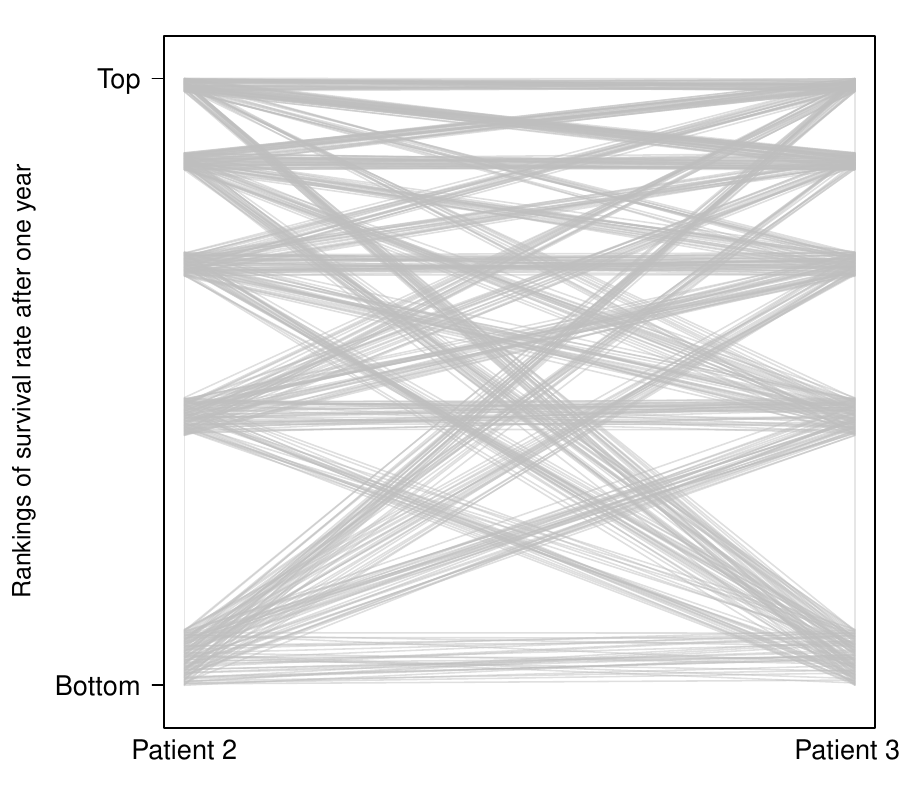}
\end{center}
\end{figure}

\bibliography{mybibliography19}
\bibliographystyle{rss}